\documentstyle{report}[10]
\title{Complex States of Simple Molecular Systems}
\author{
R. Englman$^{a,b}$ and A. Yahalom$^b$ \\
$^a$ Department of Physics and Applied Mathematics,\\
Soreq NRC,Yavne 81810,Israel\\
$^b$ College of Judea and Samaria, Ariel 44284, Israel\\
e-mail: englman@vms.huji.ac.il; asya@ycariel.yosh.ac.il;}
\newcommand{\beq} {\begin{equation}}
\newcommand{\enq} {\end{equation}}
\newcommand{\ber} {\begin {eqnarray}}
\newcommand{\enr} {\end {eqnarray}}
\newcommand{\eq} {equation }
\newcommand{\eqs} {equations }

\newcommand {\er}[1] {equation (\ref{#1}) }
\newcommand {\SE} {Schr\"{o}dinger equation}
\newcommand {\br} {{\bf r}}
\newcommand {\bR} {{\bf R}}
\newcommand {\bX} {{\bf X}}
\newcommand{\ce}  {continuity equation }
\newcommand{\ces} {continuity equations }
\newcommand{\hje} {Hamilton-Jacobi equation }
\newcommand{\hjes} {Hamilton-Jacobi equations }
\newcommand{\bp}  {\bar{\psi}}

\begin{document}

\maketitle

\begin  {abstract}
A review is given of phase properties in molecular wave functions,
 composed of a number of (and, at least, two) electronic states that become
 degenerate at some nearby values of the nuclear configuration. Apart from
 discussing phases and interference in classical (non-quantal) systems,
 including light-waves, the review looks at the
 constructability of complex wave functions from observable quantities
("the phase problem"), at the controversy regarding  quantum
mechanical
 phase-operators, at the modes of observability of phase and at the role of phases
in some non-demolition measurements. Advances in experimental and
(especially)
 theoretical aspects of Aharonov-Bohm and topological (Berry)
phases are described, including those involving two-electron and
relativistic systems. Several works in the phase control and
revivals of molecular
 wave-packets are cited as  developments and applications of
 complex-function theory. Further topics that this review touches on are: coherent
 states, semiclassical approximations and the Maslov index.  The interrelation
between time and the complex state is noted in the contexts of time
delays in scattering, of time-reversal invariance and of the
existence of a molecular time-arrow.

 When the stationary Born-Oppenheimer description for nearly degenerate state
is regarded as "embedded" in a broader dynamic formulation, namely
through solution of the time dependent {\SE},
 the wave function becomes necessarily complex. The analytic behavior of
 this function in the complex time-plane can be
 exploited to gain information about the connection between phases and
moduli of the component-amplitudes. One form of this connection
are a pair of reciprocal (Kramers-Kronig type) relations in the
complex time-domain.
 These show that certain phase changes in a component amplitude (such as,
e.g., lead to a non-zero Berry-phase) require changes in the
amplitude-moduli, too, and imply degeneracies of the electronic
state.

The subject of conical degeneracy, or intersection, of adjacent
potential
 surface is extended to nonlinear nuclear-electronic coupling, which can then
result in multiple conical degeneracies on a nuclear coordinate
plane. We treat cases of double and fourfold intersections, the
latter under
 trigonal symmetry, and employ an analytic-graphical phase tracing method
 to obtain the resulting Berry phases as the system circles around some
or all of the degeneracy points. These phases can take up values
that are all integral multiples of $\pi$, with integers that vary
with the physical situation (or with the model postulated for it).

 A further type of invariance,
with respect to gauge transformation, is also tied to the complex
form of the wave function. For a many-component state this leads
to a consideration of tensorial (Yang-Mills type) fields $F$ for molecular
systems. It is shown that it is the truncation of the Born-Oppenheimer
 electron-nuclear superposition that generates the molecular
 Yang-Mills fields.

The equations of motion for the nuclear degrees of freedom are
 derived from a Lagrangean density $\cal L$, representing a
non-Abelian situation.
 In $\cal L$ the non-adiabatic coupling terms (NACTs), that express
the electronic background on the nuclear states, enter as vector
potentials (or gauge fields). We demonstrate  a deep lying
 interrelation between two apparently distinct theoretical developments:
 that which led to the discovery of the Yang-Mills fields  from
 considerations of local gauge-invariance, and the use of an adiabatic-diabatic
transformation matrix for the molecular case due to Baer. A
generalized form of the NACTs is found, such that it makes the
field $F$ vanish for a complete electronic set.

Lastly, the use of phases and moduli as  variables was found to be very
 convenient to obtain variationally the equation of continuity and the
 Hamilton-Jacobi equations from the Lagrangian of an  electron following
 either the Schr\"{o}dinger or the Dirac equation. In the latter case, to a
good approximation (in the nearly non-relativistic limit) the phases
 in the different spinor components are not scrambled together and their
 Berry phases are unaffected.

\bigskip

PACS  numbers: 03.65.Bz, 03.65.Ca, 03.65.Ge, 31.15.-p, 31.30.-i,
 31.30.Gs, 31.70.-f

\bigskip
keywords: phases, Born-Oppenheimer states, nonadiabatic coupling

\end {abstract}

\tableofcontents
\pagestyle{plain}

\chapter{Complex States of Simple Molecular Systems}

\section {Introduction and Preview of the Chapter}

 In quantum theory physical systems move in
vector spaces that are, unlike those in classical physics, essentially complex.
 This  difference has had  considerable impact on the status, interpretation
 and mathematics  of the theory.  These aspects will be discussed in
this Chapter within the general context of simple molecular systems, while concentrating,
at the same time, on instances in which the electronic states of the molecule
are exactly or nearly degenerate. It is hoped that as the Chapter progresses,
the reader will obtain a clearer view of the relevance of the complex description
 of the state to the presence of a degeneracy.

 The difficulties that arose from the complex nature of the
wave function during the development of quantum theory are recorded by
 historians of science (\cite {Jammer} - \cite{Cao}). For some time during
 the early stages of
 the new quantum theory the existence of a complex state defied acceptance
(\cite {Jammer}, p.266). Thus,  both de Broglie and Schr\"{o}dinger  believed
that material  waves (or "matter" or "de Broglie" waves, as they were also called)
 are real (i.e., not complex) quantities, just as  electromagnetic
 waves are \cite {Cao}. The decisive step for the acceptance of the complex
wave came with the probabilistic interpretation
of the theory, also known as Born's probability postulate. This placed the
modulus of the wave function in the position of a (and, possibly, unique)
 connection between theory and experience. This development took place in the
 year 1926  and it is
 remarkable that already in the same year Dirac embraced the modulus based
interpretation wholeheartedly \cite {Dirac1}. [Oddly, it was
Schr\"{o}dinger who appears to have, in 1927, demurred at
accepting the probabilistic interpretation (\cite {Mehra}, p. 561,
footnote 350)]. Thus, the complex wave function was at last
 legitimated, but the modulus was
 and has remained for a considerable time the focal point of the formalism.

A somewhat different viewpoint motivates this article, which stresses the
added meaning that the complex nature of the wave function lends to our
understanding. Though it is only recently that this aspect has come to the
forefront, the essential point was affirmed already in 1972 by Wigner \cite
 {Wigner1} in his famous essay on the role of mathematics in physics.
 We quote from this here  at some length:

"The enormous usefulness of mathematics in the natural sciences is something
bordering on the mysterious and there is no rational explanation for ...
this uncanny usefulness of mathematical concepts...

The complex numbers provide a particularly striking example of the foregoing.
Certainly, nothing in our experience suggests the introducing of these
quantities... Let us not forget that the Hilbert space of quantum mechanics
 is the complex Hilbert space with a Hermitian scalar product. Surely to the
unpreoccupied mind, complex numbers ... cannot be suggested by physical
observations. Furthermore, the use of complex numbers is not a calculational
trick of applied mathematics, but comes close to being a necessity in the
 formulation of the laws of quantum mechanics. Finally, it now [1972] begins to
appear that not only complex numbers but analytic functions are destined to
play a decisive role in the formulation of quantum theory. I am referring
 to the rapidly developing theory of dispersion relations. It is difficult to
avoid the impression that a miracle confronts us here [i.e., in the agreement
between the properties of the hypernumber $\sqrt(-1)$ and those of the
 natural world]."

A shorter and more recent formulation is: "The concept of analyticity turns
 out to be astonishingly applicable"(\cite {Steiner}, p.37)

What is addressed by these sources is the ontology of quantal description.
 Wave functions (and other related quantities, like Green functions or
 density matrices), far from being mere compendia or short-hand listings
 of observational data, obtained
in the domain of real numbers, possess an actuality of their own. From a
 knowledge of the wave functions for real values
 of the variables and by  relying on their analytical behavior for complex
 values,  new properties
come to the open, in a way that one can perhaps view, echoing the quotations above,
 as "miraculous".

A term that is nearly synonymous with complex numbers or functions is
 their "phase". The rising preoccupation with the wave function phase
 in the last  few decades
is beyond doubt, to the extent that the importance of phases has of
 late become comparable to that of the
 moduli. (We use Dirac's terminology \cite{Dirac2}, that writes a wave
 function by a set of  coefficients, the ``amplitudes'',
  each expressible in terms of its absolute value, its ``modulus'',
 and its "phase"). There is a related growth of literature on
 interference effects, associated with Aharonov-Bohm and Berry phases
(\cite{AB} - \cite{Pati}). In parallel, one has witnessed in recent years a
 trend to construct selectively and to
manipulate wave functions.The necessary techniques to achieve these are also
 anchored in the phases of the
 wave function components. This trend is manifest in such diverse areas
as coherent or squeezed states \cite{Averbukh,Shapiro2}, electron transport
 in mesoscopic
systems \cite{Heiblum}, sculpting of Rydberg-atom wave-packets
\cite{Shapiro,Buchsbaum}, repeated and non-demolition
quantum measurements \cite{Nogues}, wave-packet collapse
\cite{GRW} and quantum computations
 \cite{Averin,Grover}.  Experimentally, the
 determination of phases frequently utilize measurement of Ramsey fringes
 \cite{Ramsey} or similar methods \cite{NoelS}.

The {\it status} of the phase in quantum mechanics has been the subject of debate.
Insomuch as classical mechanics  has successfully formulated and
solved problems using action-angle variables
\cite {Goldstein}, one would have expected to see in the phase of the
wave-function  a fully  "observable" quantity,
 equivalent to and and having a status similar to the modulus, or to the equivalent
concept of the
"number variable". This is not the case and, in fact, no exact, well behaved
 Hermitean
 phase operator conjugate to the number is known to exist. [An article by
Nieto \cite {Nieto} describes the early history of the phase operator question,
 and gives a feeling of the problematics of the field. An alternative
 discussion, primarily related to phases
 in the electromagnetic field, is available in \cite {Mandel}].
In  section 2  a brief review is provided of the various ways that phase
is linked to molecular properties.

Section 3 presents results that the analytic properties of  the
 wave function as a function of time $t$ imply and summarizes previous
 publications of the authors and of their collaborators (\cite
{EYB1} -\cite {EYBM4MCI}).  While the earlier quote from Wigner has prepared us
to expect from the analytic behavior of the wave function some general insight,
the equations in this section yield the specific result that, due to
 the analytic properties of the {\it logarithm} of wave function amplitudes,
certain forms of phase changes  lead
immediately to the logical necessity of enlarging the electronic set or, in
 other words, to the presence of an (otherwise) unsuspected state.

In the same section we also see that the
 source of the appropriate analytic behavior of the wave function is outside  its
defining equation, (the \SE), and is in general
 the consequence of either some very basic consideration or of the
 way  that experiments are conducted. The analytic behavior in question
can be in the frequency or in the time domain and leads in either case
to  a Kramers-Kronig type of reciprocal relations. We propose that behind
these relations there may be an "equation of restriction", but while in the former
case (where the variable is the frequency) the equation of restriction
expresses causality ("no effect before cause"), for the latter case (when the
variable is the time),  the restriction is in several instances the basic requirement
of lower boundedness of energies in (no-relativistic) spectra (\cite {Khalfin,
 PE}). In a previous work it has been shown  that  analyticity plays further
roles in these reciprocal relations, in that it
ensures that time-causality is not violated in the conjugate relations and
 that (ordinary) gauge invariance is observed \cite {PE}.

As already remarked, the results in section 3 are based on dispersions
 relations in the complex time domain.
 A complex time is not a new concept. It features  in wave optics
\cite{Mandel} for "complex analytic signals" (which is an electromagnetic field
with only positive frequencies) and in non-demolition measurements
performed on photons \cite{ScullyZubairy}. For transitions between adiabatic
 states (which is also discussed in this review), it was
previously introduced in several works (\cite {Miller71} - \cite{ZhuNN}).

Interestingly, the need  for a multiple electronic set, which we connect with
the reciprocal relations, was also a keynote of a recent review
(\cite {Bersuker1} and previous publications cited there and in \cite
{BersukerPolinger}). Though the considerations relevant to this effect
are not linked to complex nature of the states (but rather to the {\it
 stability} of the adiabatic states in the real domain),
 we have included in section 3 a mention of, and some elaboration on, this topic.

In further detail,  section 3 stakes out the following claims:
For time dependent wave functions  rigorous conjugate relations are derived
between analytic decompositions (in the complex t-plane) of phases and of
$\log \ moduli$. This entails a reciprocity, taking the form of Kramers-Kronig
integral relations (but in the time domain), holding  between observable
 phases and moduli in several physically important cases. These cases
include the nearly
adiabatic (slowly varying) case, a class of cyclic wave-functions, wave
packets and non-cyclic states in an "expanding potential". The results define a
unique phase through its analyticity properties and exhibit the
interdependence of geometric-phases and related decay probabilities. It turns
out that the reciprocity property obtained in this section holds for several
 textbook quantum mechanical applications (like the minimum width wave packet).

The multiple nature of electronic set becomes especially important when the
 potential energy surfaces of two (or more) electronic states come close,
 namely, near a "conical
 intersection" ({\it ci}). This is also the point in the space of nuclear
configurations at which  the phase of wave function components becomes anomalous.
 The basics of this situation have been extensively
studied and have been reviewed in various sources (\cite {Sidis}- \cite {DRY1}).
 Recent works
(\cite {Thummel} - \cite {AvoryBaerBilling}) have focused attention on a new
contingency: when there may be several {\it ci}'s between two adiabatic
 surfaces,
 their combined presence needs to be taken into account for calculations
 of the non-adiabatic corrections of the states and can have tangible
consequences in chemical reactions.
 Section 4 presents an analytic modeling of the multiple {\it ci} model, based on
the superlinear terms in the coupling between electronic and nuclear motion.
The section  describes
in detail a tracing method that keeps track of the phases, even when these
possess singular behavior (namely, at points where the moduli vanish or
become singular). The continuous tracing method is applicable to real states
 (including stationary ones). In these the phases are either zero
or $\pi$. [At this point, it might be objected that in so far that numerous properties
 of molecular systems (e.g., those
relating to questions of stability and, in general, to static situations
 and not involving a magnetic field) are well described in terms of {\it real}
 wave functions,  the {\it complex} form of the wave function
need, after all, not be regarded as  a fundamental property. However,
 it will be shown in Section 4 that wave functions
 that are real but are subject to a sign change, can be best treated as
 limiting cases in complex variable theory. In fact, the "phase tracing"
 method is logically connected to the time dependent wave-functions (and
 represents a case of mathematical "embedding")].

 A specific result in  Section 4 is the construction of highly non-linear
vibronic couplings near a {\it ci}. The construction shows, {\it inter alia},
 that the
 connection between the Berry (or "topological", or "geometrical") phase,
 acquired during cycling in a parameter space, and the number of {\it ci}'s
 circled  depends on the details of the case that is studied and can vary
 from one situation to another.
Though the subject of Berry phase is reviewed in a companion chapter in this
volume \cite{ChildRev}, we note here some recent extensions in the subject
(\cite{BEV} - \cite{YE1}). In these works, the phase changes were calculated
for {\it two-electron} wave functions, such that are subject to inter-electronic
forces . An added complication was also considered, for the case in
which the two electrons are acted upon by different fields. This can occur
 when the two electrons are placed
in different environments, as in asymmetric dimers. By and large, intuitively
 understandable
results are found for the combined phase factor but, under conditions of
accidental degeneracies, surprising jumps ( named "switching") are noted. Some
applications to quantum computations seem to be possible \cite {YE1}.

The theory of Born-Oppenheimer (abbreviated to BO) \cite {BO, BH}
has been hailed (in an authoritative but
unfortunately unidentified source) as one of the greatest advances in
 theoretical physics. Its power is in
disentangling the problem of two kinds of interacting particles into two separate
problems, ordered according to some property of the two kinds. In its most frequently
 encountered form, it is the nuclei and electrons that interact
(in a molecule or in a
solid) and the ordering of the treatment is based on the large difference
between their masses. However, other particle pairs can  be similarly handled, like
  hadronic mesons and baryons, except that a relativistic or field theoretical
version of the Born-Oppenheimer theory is not known. The price that is paid for
the strength of the method is that the remaining coupling between the two kinds
of particles is dynamic. This coupling is expressed by the so called
 Non-Adiabatic Coupling Terms (abbreviated  as NACTs), which involve derivatives of
(the electronic) states rather than the states themselves.  "Correction terms"
of this form are difficult to handle by conventional perturbation theory. For atomic collisions the method of " Perturbed Stationary States" was
 designed to overcome this difficulty \cite {BatesMcCarroll, MottMassey}, but
this is accurate only under restrictive conditions. On the other hand, the
 circumstance that this coupling is independent of the potential, indicates
 that a general procedure can be used to take care of the NACTs [68].
Such general procedure was developed by Yang and
Mills in 1954 \cite {YM} and has led to far reaching consequences in the
theory of weak and strong interactions between elementary particles.

 The interesting history of the Yang-Mills field  belongs essentially to
particle physics (\cite {Jackiw} - \cite {Weinberg}). The reason
 for mentioning it here in a chemical physics setting,
is to note that an apparently entirely different
 procedure was proposed for the equivalent problem arising in the molecular
 context, namely, for the elimination
 of the derivative terms (the NACTs) from the nuclear part of the BO
 {\SE} through an Adiabatic-Diabatic Transformation (ADT) matrix \cite
{Smith, Baer75}. It turns out that the quantity known as the tensorial
field  (or
covariant, or Yang-Mills, or YM field, with some further names also in use)
 enters also into the ADT description, though from a
 completely different viewpoint, namely through ensuring the validity
of the ADT matrix method by satisfaction of what is known as
the "curl cndition".  Formally, when the
"curl condition" holds, the (classical) Yang-Mills field is zero and this is
also the requirement for the strict validity of the ADT method. [A review
of the ADT and alternative methods is  available in, e.g., \cite {Sidis,PCK},
the latter of which  discusses also
the Yang-Mills field in the context of the BO treatment.]
However, it has recently been shown by a
formal proof, that an {\it approximate} construction of the ADT matrix (using
only a finite, and in practice small, number of BO, adiabatic states) is
possible even though the "curl condition" may be {\it formally} invalid
\cite {BE2}. An  example for such an approximate construction in a systematic way
was provided in a model that uses Mathieu functions for the BO
electronic states \cite {EYBMathieu}.

As has been noted some time ago, the non-adiabatic coupling terms (NACTs, mentioned above)
 can be incorporated in the nuclear part of the  {\SE} as a
 vector potential \cite {Mead87,Zygelman1}. The
 question of a possible magnetic field, associated with this vector potential
 has also been considered (\cite {MeadTruhlar79} - \cite{BaerMagnetic}).
 For an electron occupying an admixture of  two or more states (a case
that is commonly designated as non-commutative, "non-Abelian"), the fields
of physical interest are not only the magnetic field, being the curl of the
 "vector potential", but also
 tensorial (Yang-Mills) fields. The latter is the sum of the curl field and
 of a vector-product term of the NACTs. Physically these field  represent
 the  reaction of the electron on the nuclear motion via the NACTs.

In a situation characteristic of molecular systems, a conical
 intersection {\it ci}  arises from the degeneracy  point
of adiabatic potential energy surfaces in a plane of nuclear displacement
coordinates. There are also a number of orthogonal directions, representing a
so-called "seam" direction. In this setting, it emerges
 that both kinds of fields are aligned with the seam direction of the
 {\it ci} and are zero everywhere outside the seam,
 but they differ as regards the flux that they produce. Already in a
two-state situation, the fields are representation dependent
 and the values of the fluxes depend
 on the states the electron occupies. (This evidently differs from conventional
 electro-magnetism, in which the magnetic field and the flux are unchanged
 under a gauge transformation.)

Another subject in which there are implications of phase is the time
 evolution of atomic or molecular wave-packets. In some recently studied cases
 these might be a superposition of a good ten or so energy eigen-states.
Thanks to the availability of short, femtosecond laser pulses  both the control
of reactions by coherent light (\cite{Shapiro2}, \cite {BrumerS}-
\cite {BlanchetNBG})  and the probing of
phases in a wave packet are now experimental possibilities (\cite {Buchsbaum},
\cite{BuchsbaumNature} - \cite {SchleichNature}). With
short duration excitations the initial form of the wave packet is
a {\it real} "doorway state" (\cite {FeshbachKL}- \cite {Englman79})
and this develops phases for each of its component amplitudes as
the wave-packet evolves. It has recently been shown that the
phases of these components are signposts of a time arrow
 (\cite {Zeh}-\cite{Savitt}) and of
the irreversibility; both of these are inherent in the quantum mechanical
 process of
preparation and evolution \cite {EY3}. It was further shown in
\cite {EY3} (for systems
that are invariant under time-reversal, e.g. in the absence of a
 magnetic field) that the preparation
of an initially {\it complex} wave-packet requires finite times for its
construction (and cannot be achieved instantaneously).

The quantum phase factor is  the exponential
of an imaginary quantity ($i$ times the phase) which multiplies into a wave function.
Historically, a natural extension of this was
proposed  in the form of a gauge transformation, which both multiplies into
and admixes different components of
a multi-component wave function \cite {Weyl}. The resulting "gauge theories"
 have become an essential tool of quantum field theories and provide
(as already noted in the discussion of the YM field, above) the modern
 rationale of basic forces
between elementary particles (\cite {Jackiw} -\cite {Weinberg}).
 It has already been noted in an earlier paragraph, that gauge theories
 have also made notable impact on molecular
 properties, especially under conditions that the electronic state basis in the
molecule consists of more than one component. This situation also characterizes
 the conical intersections between potential surfaces, as already mentioned.
 In section 5 we show how an important theorem,
 originally due to Baer \cite {Baer75}, and subsequently used in several
equivalent forms, gives some new insight to the nature and source of these YM
fields in a molecular (and perhaps also in a particle-field) context. What the above
theorem shows is that it is the {\it truncation} of the BO set that leads to
the YM fields, whereas for a complete BO set the field is inoperative for
 molecular vector potentials.

Section 6 shows the power of the modulus-phase formalism and is included
in this review partly for methodological purposes. In this
formalism the equations of continuity and the Hamilton-Jacobi equations can
be naturally derived in both the non-relativistic and the relativistic
(Dirac) theories
of the electron. It is shown that in the four-component (spinor)
 theory of electrons, the two extra components in the spinor wave-function
will have only a minor effect on the topological phase, provided certain
 conditions are met ({\it nearly} non-relativistic velocities and external
 fields that are not excessively large).

So as to make the individual sections self-contained, we have found it
advisable to give some definitions and statements more than once.

\section {Aspects of Phase in Molecules}
\label{Some Aspects of Phases in Molecules}

This section attempts
a brief review of several areas of research on the significance of
phases, mainly for quantum phenomena in molecular systems.
 Evidently, due to limitation of  space, one cannot do justice
to the breadth of the subject and numerous important
works will go unmentioned. It is hoped that the several cited papers (some of
 which have been chosen from quite recent publications)
will lead the reader to other, related and earlier, publications.
It is essential to state at the outset that the {\it overall} phase of the
 wave-function
 is arbitrary and only the relative phases of its components are observable in
any meaningful sense. Throughout, we concentrate on the relative phases of the
components.  (In a coordinate representation of the state function,
 the "phases of the
components" are none other than the coordinate-dependent parts of the phase,
so it is also true that {\it this} part  is susceptible to measurement.
Similar statements can be made in  momentum, energy, etc. representations.)

 A further preliminary statement to this section
 would be that, somewhat analogously to classical physics or mechanics
 where positions and momenta (or velocities) are the two conjugate variables
that determine the motion, moduli and phases play similar roles. But the analogy is
not perfect. Indeed, early on it was questioned, apparently first by
 Pauli \cite {Pauli},
whether a wave function can be constructed from the knowledge of a set of
moduli alone. It was then argued by Lamb \cite {Lamb} that from a set of values
 of wave function moduli and of their rates of change, the wave function,
 including its phase, is uniquely found. Counter-examples were then given \cite {GaleGT, Royer}
and it now appears  that the knowledge of the moduli and some
 information on the analytic properties of the wave function
are both required for the construction of a state. (The following section
contains a formal treatment, based partly on \cite {EYB2} - \cite {EYB3} and
 \cite {BE1}-
\cite{EB1}.) In a recent research effort, states with definite phases were generated
for either stationary or traveling type of fields \cite {GuimareBVM}.

Recalling for a start phases in {\it classical} waves, these have already been
 the subject of consideration by Lord Rayleigh \cite {Rayleigh}, who noted
 that through interference between the probed and a probing wave the magnitude
and phase of acoustic waves can be separately determined, e.g., by
finding surfaces of minimum and of zero magnitudes. A recent review on
   classical waves is
given by Klyshko \cite {Klyshko}. The work of Pancharatnam on polarized light beams
(\cite  {Pancharatnam, RamaRama})  is regarded as the
  precursor of later studies of topological phases in quantum systems \cite
{Berry84}. This work contained a formal expression for the
relative phase  between beams in different elliptic polarizations of light,
 as well as a construction (employing
 the so-named "Poincare sphere") that related the phase
difference to a geometrical, area concept. (For experimental realizations with
polarized light beams we quote \cite{SchmitzerKD1, SchmitzerKD2}; the issue of
any arbitrariness in experimentally pinning down the topological part of the
phase was raised in \cite{GiavariniGRT}.)
Regarding the interesting question of any common ground  between classical and quantal phases,
 the relation between the adiabatic (Hannay's) angle in mechanics and the phase in
wave functions was  the subject of \cite {Berry85}. The difference in
 two-particle interference patterns of electromagnetic and matter waves
was noted, rather more recently, in \cite{ BruknerZ}.  The two phases, belonging
to light and to the particle wave-function, are expected
 to enter on an equal footing
when the material system is in strong interaction
with an electromagnetic field (as in the Jaynes-Cummings model). An example of
 this case was provided
in a study of a two level atom, which was placed in a cavity containing an
 electromagnetic field. Using  one or two photon excitations, it was found possible
 to obtain from the  Pancharatnam phase an indication  of  the statistics of
the quantized field \cite {LawandeLJ}.

Several essential basic properties of phases in optics are contained in \cite {Mandel,
ScullyZubairy, ShapiroShepard}. It was noted in \cite {Mandel}, with reference to the
"complex analytic signal"
(an electromagnetic field with positive frequency components), that the position of zeros
 (from which the phase can be determined) and the intensity represent two
sets of information which are intetwined by the analytic property of the
 wave. In the next section we shall again encounter this finding, in the context
of complex matter (Schr\"{o}dinger) waves.
Experimentally,  observations  in wave guide structures of the positions
of amplitude zeros (which are just the "phase singularities") were made in
 {\cite {BalisteriKKH}. An alternative way for the determination of phase
 is from location  of maxima in interference fringes (\cite {Mandel}, section 7.3.2).

{\it Interference} in optical waves is clearly a phase phenomenon;
in classical systems it arises from the signed superposition of positive and negative
 {\it real} wave amplitudes.
To single out some special results in the extremely broad field of interference,
 we point to recent observations using two-photon pulse transition \cite {BlanchetNBG}
 in which  a differentiation was achieved between interferences due to temporal
 overlap (with finite pulse-width) and quantum interference caused by delay.
 The (component-specific) topological phase in
wave functions has been measured, following the proposal of Berry in \cite{Berry84},
 by neutron interferometry in a number
 of works, e.g., \cite {WaghRFI,WaghBRBS} with continual improvements in the
 technique. The difficulties in the use of coherent neutron beams and the possibility
of using conventional neutron sources for phase-sensitive neutron radiography
have been noted in a recent review \cite {NugentPG}.

Phase interference in optical or material systems can be utilized to achieve
 a type of quantum measurement, known as non-demolition measurements (\cite
 {ScullyZubairy}, Chapter 19). The general objective is to make a measurement
that does not change some property of the system at the expense of some other
other property(s) that is (are) changed. In optics it is  the
phase that may act as a probe for determining the intensity (or photon number).
The phase can change in the course of the measurement, while the photon
number does not \cite {GrangierLP}.

 In an intriguing and potentially important proposal (apparently
not further followed up), a filtering method was suggested for image
 reconstruction (including phases) from the modulus of the correlation
 function \cite {KohlerM}.
[In mathematical terms this amounts to deriving the behavior of a function
in the full complex (frequency) plane from the knowledge of the absolute
value of the function on the real axis, utilizing some physically realizable
kernel function.] A different spectral filtering method was discussed in
\cite{VijayWB}.

Before concluding this sketch of optical phases and passing on to our next
 topic,
 the status of the "phase" in the representation of observables
as quantum mechanical operators, we wish to call attention to
 the theoretical demonstration, provided in  \cite {ReckZBB},
 that any (discrete, finite dimensional) operator can be constructed through
use of optical devices only.

The appropriate quantum mechanical {\it operator} form of the phase has been
the subject of numerous efforts.  At present, one can only speak
of the best approximate operator, and  this also is the subject of debate.
 A  personal-historical
 account by Nieto of  various operator definitions for the phase (and of its
probability distribution) is in \cite {Nieto} and in companion articles,
 e.g. \cite {NohFM} - \cite{Vourdas} and others, that have appeared in volume
 {\bf 48} of Physica Scripta T (1993) devoted to this subject.
(For an introduction to the unitarity requirements placed on a phase operator,
one can refer to \cite {Loudon}). In 1927 Dirac proposed a quantum mechanical
 operator $ \hat{\phi} $, defined in terms of the creation and destruction
operators  \cite {Dirac1927}, but London \cite {London} showed that this
is not  Hermitean. (A further source is \cite {Louisell}.) Another
candidate, $ e^{i\hat{\phi}}$ is not unitary, as was demonstrated, e.g.,
in  \cite {Mandel}, section 10.7. Following that,
Susskind and Glogower proposed a pair of operators $\hat{cos}$ and
 $\hat{sin}$
\cite {SusskindG}, but it emerged these do not commute with the number operator $\hat{n}$. In 1988
Pegg  and Barnett  introduced a Hermitean phase operator
through a limiting procedure based on the state with a definite phase in a
truncated Hilbert space \cite {PeggB}. Some time ago a comparison
was made between different  phase operators when used on  squeezed states
\cite {GronbechCR}. Unfortunately, there is as yet non consensus on the
status of the Pegg-Barnett operators (\cite {ShapiroShepard},
\cite {VorontsovR1} - \cite{VorontsovR2}). It may be that, at least,
 part of the difficulties are rooted in problems that arise from the coupling
 between the quantum system and the measuring device. However, this difficulty is
a moot point in quantum mechanical measurement theory, in general.

(For the special case of a two-state systems, a hermitean phase operator
was proposed, \cite {AMuller}. This was said to provide a quantitative measure
 for "phase information".)

A related issue is the experimental {\it accessibility} of phases: It is now
widely accepted  that there are essentially two experimental ways to observe
 phases  \cite {Berry84, Berry90, WaghBRBS}:
(1) through a two-Hamiltonian, one state method, interferometrically (namely, by
sending two identically prepared rays across two regions having different fields),
 (2) a one Hamiltonian, two-state method (meaning, a difference in the preparation of the
rays), e.g., \cite {UbernaKWPG, ZucchettiVWW}. (One recalls that already
several years ago it was noted that there are the two ways
  for measuring the phase of a four-component
state, a spinor \cite {AharonovS1967}.)
One can also note a further distinction proposed more recently, namely, that
 between  "observabilities" of bosonic and fermionic phases \cite {AharonovV2000}:
Boson phases are observable both {\it locally} (at one point) and nonlocally
(at
extended distances, which the wave reaches as it progresses). The former can
 lead to phase values that are
 incompatible with the Bell inequalities, while fermion phases are
 only nonlocally observable (i.e., by
interference)  and do not violate Bell's inequalities. The difference resides in
that only the former type of particles gives rise to a coherent state with arbitrarily large
 occupation number $n$, whereas for the latter the exclusion principle
allows only $n=0$ or $1$.

The question of determination of the phase of a field (classical or quantal,
 as of a wave-function)  from the modulus (absolute value) of the field
along a {\it real} parameter (for which alone experimental
determination is possible) is known as {\it "the phase problem"}
\cite {Mandel}. (So, also in crystallography.) The reciprocal
 relations derived in the next section represent a formal
 scheme for the determination of phase given the modulus, and {\it vice versa}.
 The physical basis of these singular integral relations was described in
 \cite  {Floissart} and in several companion articles in that volume; a more recent
 account can be found in
\cite {PeiponenVA}. Thus, the reciprocal relations in the time domain provide, under
certain conditions of analyticity, solutions to the phase problem. For
 electromagnetic fields, these were derived in \cite {Toll, BurgeFGR,
 ShapiroShepard}
 and reviewed in \cite {Mandel, PeiponenVA}. Matter or Schr\"odinger waves
were considered in a general manner in \cite{Khalfin}. The more complete treatment,
presented in the next section applies the results to several
 situations in molecular and solid state physics. It is likely that the full
 scope and meaning
of the modulus-phase relationship await further and deeper-going analyses.

In 1984 Berry made his striking discovery of time-scale-independent phase
 changes in many-component states \cite {Berry84} (now variously known as
 Berry
or topological or {\it geometric phase}) . This followed a line of
 important developments regarding the role of phases and phase factors in
quantum mechanics. The starting point of these may be taken with Aharonov and
 Bohm's
discovery of the topologically acquired phase \cite {AB}, named after them.
(As a curiosity, it is recorded that David Bohm himself referred to the
 "ESAB effect" \cite {Peat, EhrenbergS}.) The achievement,
stressed by the authors of \cite {AB}, was to have been able to show that when an
 electron traverses a closed path along which the magnetic field is zero, it
acquires an observable phase change, which is proportional to the "vector
potential".  The "topological" aspect, namely that the path is inside
 a multiply connected  portion of space (or that, in physical terms,
 the closed path cannot be shrunk without encountering an infinite barrier),
has subsequently turned out to be also of considerable importance
 \cite {PeshkinTT, Simon}, especially through later extensions and applications of
 the Aharonov-Bohm phase-change \cite {PeshkinT}. (Cf. the paper by Wu
 and Yang  \cite {WY} that showed the
  importance of the phase {\it factor} in quantum mechanics, which has, in turn,
 led to several developments in many domains of physics.)

 In molecular physics, the  "topological" aspect has met its analogue
in the Jahn-Teller effect \cite {BersukerPolinger, Englman72} and, indeed,
 in any situation where a degeneracy of electronic states is encountered.
 The phase-change was change was discussedfrom various viewpoints in
 \cite {Longuet} - \cite {Ham87} and  \cite{Mead92}.

 For the Berry-phase we shall quote a definition given in \cite {ChanceyO}:
{\it "The phase that can be acquired by a state moving adiabatically (slowly)
around a closed path in the parameter space of the system."} There is a further,
somewhat more general phase, that appears in any cyclic motion, not necessarily
slow) in the Hilbert space, which is  the Aharonov-Anandan phase \cite {AA}.
Other developments and applications are abundant. An interim summary was published
in 1989 in \cite {ShapereW}. A further, more up-to-date summary, especially
 on progress
in experimental developments, is much needed. (In section 4 of the present review
 we list some publications that report on the experimental determinations of the Berry phase.)
 Regarding theoretical advances, we note (in a somewhat subjective and
selective mode) some clarifications regarding parallel transport,
 e.g., \cite{Anandan90}. This paper discusses the "projective Hilbert space"
and its metric (the Fubini-Study metric). The projective Hilbert space arises
 from the Hilbert space of the electronic manifold by the removal of
the overall phase and is therefore a central geometrical concept in any
 treatment of the component phases, such as the present review.

The term "Open-path phase" was coined  for a
 non-fully-cyclic evolution \cite {JainPati, Pati}. This,
 unlike the Berry-phase proper, is not gauge-invariant, but is, nevertheless
 (partially) accessible by experiments (\cite {EYB2} - \cite {EYB3}).
 The Berry phase
for non-stationary states was given in \cite{MooreStedman}, the interchange
 between dynamic and geometric phases is treated in \cite{GiavariniGRT}.
 A geometrical interpretation is provided in \cite{Kohler98} and a
 simple proof for Berry's area formula in \cite{YangY}. The phases
 in off-diagonal terms form the basis of generalizations of the Berry-phase
in \cite{ManiniP,ManolopoulosC}; an experimental detection by neutron
interferometry was recently accomplished \cite{HasegawaLBBR}. The treatment
by Garrison and Wright of complex topological phases for non-Hermitean
Hamiltonians \cite {GarrisonW} was extended in \cite {MiniaturaSBB} - \cite{GeC2}.
 Further advances on Berry-phases are corrections due to non-adiabatic effects
(resulting, mainly, in a {\it decrease} from the value
of the phase in the adiabatic, infinitely slow limit) \cite {EYB2, BaiCG,
 WhitneyG}. In \cite {AharonovR} the complementarity between local
and nonlocal effects is studied by means of some examples. For more general
time dependent Hamiltonians than the cyclic one, the method of the Lewis
 and Riesenfeld invariant spectral operator is in use. This is discussed in
 \cite {Cervero}.

As already noted, the Berry-phase and the open path phase designate changes
in the phases of the state-components, rather than the total phase change
of the wave function, which belongs to the so-called "Dynamic phase" \cite
{Berry84, AA}. The existence of more than one  component in the state
function is a topological effect.
This assertion is based on a theorem by Longuet-Higgins (\cite {Longuet}, "Topological
test for intersections"), which states that, if the wave function of a given electronic state
state changes sign when transported around a loop in nuclear configuration space,
then the state must become degenerate with another at some point within the loop.

>From this theorem it  follows that, close to the point of intersection and slightly away
 from it, the corresponding adiabatic or BO electronic wave functions will be
given (to a good approximation) by a superposition of the two degenerate states,
 with coefficients that are functions of the nuclear coordinates.
(For a formal proof of this statement, one has to assume,
 as is done in \cite {Longuet}, that the state is continuous function of
 the nuclear coordinates.) Moreover, the coefficients of the
two states have to differ from each other, otherwise they can be made
 to disappear from the normalized electronic state. Necessarily, there is also
a second "superposition state", with coefficients such that it is orthogonal
to the first at all points in the configuration space. (If more than two states
happen to be codegenerate at a point, then the adiabatic states are mutually
 orthogonal superpositions of all these states, again with coefficients that
 are functions of the nuclear coordinates.)

If now the nuclear coordinates are regarded as dynamical variables, rather than
parameters, then in the vicinity of the intersection point, the energy
eigenfunction, which is a combined
{\it electronic-nuclear} wave function, will contain a superposition of the two
adiabatic, superposition-states, with nuclear wave functions as cofactors.
We thus see that the topological phase change leads, first, to the adiabatic electronic
state being a multi-component superposition (of diabatic states) and,
 secondly, to the full solution being a multi-component superposition (of
adiabatic states), in each case with nuclear-coordinate dependent coefficients.

The  design  and {\it control of molecular processes} has of late become possible
thanks to advances in laser technology, at first through the appearance
 of femtosecond
laser pulses and of pump-probe techniques \cite{FelkerZ} and, more recently,
 through the realization of
more advanced ideas, including feedback and automated control
 (\cite {GaspardB} - \cite {ZeidlerFKM}). In a typical procedure,
 the pump pulse prepares a coherent superposition of energy eigenstates, and a second
delayed pulse probes the the time-dependent transition between an excited and a
lower potential energy surface. When the desired outcome is a particular reaction
product, this can be promoted by the control of the {\it relative phases}
 of two fast pulses emanating from the same  coherent laser source. One of
the earliest works to achieve this is \cite{SchererRDF}. A recent study focuses on several basic questions, e.g.,
those regarding pulsed preparation of an excited state \cite {ZucchettiVWW}.
 In between the two, numerous
 works have seen light in this fast-expanding and technologically interesting
field. The purpose of mentioning them here is to single out this field as
an application of phases in atomic (\cite{NoelS},
 \cite{BuchsbaumNature} - \cite{AraujoWS}) and molecular (\cite{BrumerS}-
\cite{LeichtleSAS}) spectroscopies. In spite of the achievements in photochemistry,
 summarized e.g., in \cite {Tannor}, one
hardly expects phases to play a role in ordinary (that is, not state-selective
or photon-induced)
chemical reactions. Still, interference (of the kind seen in double-slit
experiments) has been observed between different pathways during the dissociation
 of water \cite {DixonHYHLY,
Clary}. Moreover, several theoretical ideas have also been
put forward to produce favored reaction products through the involvement of phase
effects (\cite{AvronB} -\cite {AdhikariBALB}). Calculations for  the scattering
cross sections in the four-atom
reaction $OH + H_2 \to H_2 O +H $ showed a few percent change due to the effect
of phase \cite {BillingK}.

{\it Wave-packet reconstruction}, or imaging from observed data, requires the
 derivation of a {\it complex} function from a set of real quantities. Again, this
 is essentially the "phase problem", well known also from crystallography and
 noted above in a different context than the present one \cite {Mandel}.
  An experimental study yielded the Wigner position-momemtum distribution
 function \cite {DunnWM}. This approach was named a "tomographic" method, since a single
beam scans the whole phase-space and is distinct from another approach,
 in which two different laser pulses create two wave-packets: an object and
 a reference.  When the two states are superimposed, as in a conventional
 holographic arrangement, the cross term in the modulus squared retains
 the phase information (\cite {Shapiro3, LeichtleSAS, Shapiro2}). Computer
 simulations have shown the theoretical proposal to be feasible.
In a different work, the preparation of a long lived atomic electron wave packet
in a Rydberg state, with principal quantum numbers around $n = 30 $, was
  achieved \cite {BromageS}.

Rydberg states, as well as others, can provide an illustration for another,
 spectacular phenomenon:  wave-packet
revivals \cite{Averbukh}. In this, a superposition of about ten energy-states
first spreads out in phase-space (due to phase decoherence), only to
return to its original shape after a time which is of the order of the
deviation of the spacing of the energy levels from a uniform one \cite
{EberlySN, AverbukhP2}. Not only is the theory firmly based, and simulations
convincing, but even an application, based on this phenomenon and  aimed
 at separation of isotopes, has been proposed \cite{AverbukhVVS}. Elsewhere,
 it was shown that the effect of slow cycling on the evolving wave-packet
 is to leave the revival period unchanged, but to cause a shift in the
 position of the revived wave packet \cite {Jarzynski}.

{\it Coherent states} and diverse {\it semiclassical approximations} to
molecular wave-packets are essentially dependent on the relative phases
between the wave-components. Due to the need to keep this review to a reasonable
size, we can mention here only a sample of original works, (e.g., \cite
 {Heller} - \cite{Kay}) and some summaries (\cite {Percival} - \cite {Childbook}). In these the reader will come across the {\it Maslov index}
 \cite {MaslovF}, which we pause to mention here, since it links up in a natural way
to the modulus-phase relations described in the next section and with the
phase tracing method in section 4. The Maslov index relates to the phase
 acquired when  the semi-classical wave function  traverses a zero
(or a singularity, if there be one) and it (and, particularly, its sign) is the consequence of
 the analytic behavior of the wave  function in the complex time-plane.

The subject of {\it time} connects with the complex nature of the wave
function in a straight forward way, through the definition in quantum mechanics
of the Wigner time-reversal operator \cite {Davydov, Levine}. In a rough way,
the definition implies that the conjugate of the complex wave function
describes (in several instances) the behavior of the system with the time
running backwards. Given, on one hand, "the time-reversal invariant" structure of accepted
 physical theories and, on the other hand, the
experience of passing time and the successes of non-equilibrium statistical mechanics
and thermodynamics, the question that is being asked is: when and
where does a physical theory pick out a preferred direction of time (or a
"time arrow")? From the numerous sources that discuss  this
 subject, we call attention to some early controversies (\cite{Popper} -\cite
 {Costa}) and to more recent accounts (\cite{Zeh}, \cite{Lebowitz}
 -\cite {Schulman}),  as well as to a volume
 with philosophical orientation \cite {Savitt}. Several attempts have been
 made recently to change the original formalism of quantum mechanics
by adding non-Hermitean terms (\cite {BanksSP} - \cite {Gheorghiu}), or by
 extending ("rigging")
 the Hilbert space of admissible wave-functions \cite {ABohmAK, ABohm}. The
last two papers emphasize the preparation process as part of the wave-evolution.
 By an extension of this idea, it has recently been shown
 that the {\it relative} phases in a wave-packet, brought to life by fast
laser pulses, constitute a uni-directional clock for the passage of time (at least
for the initial stages of the wave-packet) \cite{EY3}. Thus, developing
 phases in real life are hallmarks of both a time-arrow and of
 irreversibility. It also emerged that, in a setting that is invariant
under time reversal, the preparation of an "initially" complex wave-packet needs finite
times to accomplish, i.e. it is not instanteneous \cite {ZucchettiVWW, EY3}.

{\it Time shifts} or delays in scattering processes are present in areas as
 diverse as particle, molecular and solid state phenomena, all of which are
due to the complex nature of
the wave function. For a considerable time it was thought that the instance
 of formation of a particle or of an excited  state is restricted only by the
 time-energy uncertainty relation. The time {\it delay} $\tau$ was first recognized by David Bohm
\cite {DBohm} and by Eisenbud and Wigner \cite {Wigner}, and was then given
by F. Smith  \cite {Smith} a unifying
expression in terms of the frequency ($\omega$)-derivative of the scattering-
(or $S$-) matrix, as
\beq
\tau = Re \frac{\partial lnS}{i\partial\omega}
\label {taushift}
\enq
The $Re$ pre-symbol signifies that essentially it is the phase part of the scattering
matrix that is involved. A conjugate quantity, in which the imaginary part is
 taken, was later identifed as the particle formation time (\cite {PollakM} -
 \cite {Perel'man2}).
Real and imaginary parts of derivatives were associated with
the delay time in tunnelling processes across a potential barrier in the
 Buttiker-Landauer approach. (A review is in \cite {LandauerM}.) Experimentally, an
example of time delay in reflection was found recently \cite {ChauvatEBL}.
The question of time reversal invariance, or of its default, is naturally a
 matter of great and continued interest for theories of interaction
between the fundamental constituents of matter.
A summary that provides an updating, good to its time of printing, is found
 in \cite {Debu}.

Another type of invariance, namely with respect to unitary or gauge transformation
 of the wave-functions (without change of norm)  is a corner-stone of modern physical
 theories \cite {YM}. Such transformations can be  global (i.e., coordinate
independent)  or local (coordinate dependent). Some of the observational
 aspects arising from gauge transformation have caused some controversy;
 e.g., what is the effect of a gauge transformation on an observable \cite {
AharonovAu, FeuchtwangKGC}. The justification for gauge-invariance
goes back to an argument due to Dirac \cite {Dirac1927},
reformulated more recently in \cite {Kaempffer}. This is based on
the observability of the moduli of overlaps between different
 wave-function, which then leads to a definite phase difference between
 any two coordinate values, the same for all wave functions. From this Dirac
goes on to deduce the invariance of Abelian systems under an arbitrary
local phase change, but the same
argument holds true also for the local gauge-invariance of non-Abelian,
 multi-component cases \cite {Weinberg}.

We end this section of phase effects in complex states by reflecting on
how, in the first place, we have arrived at a complex description
of phenomena that take place in a real world. There are actually two ways to
come by this:

 First, the time-dependent wave function is necessarily complex and
 this is due to the form of the time-dependent  {\SE} for
{\it real} times, which contains $i$. This equation will be the starting point
 of the next section, where we derive some consequence arising from
the analytic properties of the complex wave function.
But, secondly,  there are also defining equations that do not contain $i$ (like the
  time-independent {\SE}). Here also the wave-function can be made
complex through making some or other of the variables take complex values.
The advantage lies frequently in removing possible ambiguities that arise in the
solution  at a singular point (which may be an infinity). Complex times have been
 considered in several theoretical works, e.g., \cite {Miller71,HwangPechukas}).
 It is possible to associate a purely imaginary time with {\it temperature}.
Then, recognizing
 that negative temperatures are unphysical in an unrestricted Hilbert space,
 we immediately see that the upper and lower halves of the complex t-plane are
non-equivalent. Specifically, regions of non-analytic behavior are expected
 to be found in the upper half, which is the one that corresponds to negative
 temperatures, and analytic behavior is expected in the lower half plane,
 that corresponds to positive temperatures.
  The formal extension of the nuclear coordinate space onto a complex
 plane, as is done in  \cite {Nikitin, ZhuNN}, is an {\it essentially}
 equivalent procedure, since in the semi-classical formalism of these works
the particle coordinates depend parametrically on  time. Complex
topological phases are considered in, e.g., \cite {GarrisonW, MiniaturaSBB},
 which can arise from a non-Hermitean Hamiltonian.
The so-called Regge-poles are located in the complex region of momentum space.
(A brief review well suited for molecular physicists is in \cite {OzimbaM}).
 The plane of complex-valued interactions is the subject of \cite {Heuss}.

In addition, it can occasionally be useful to regard some physical
 parameter appearing in the theory as a complex quantity and the
 wave function to possess analytic properties with regard to them.
 This formal procedure might even include fundamental constants like $e,h$, etc.

\section {Analytic Theory of Complex Component-\\ Amplitudes}

\subsection {Modulus and Phase}

With the time-dependent {\SE} written as 
\beq
i\frac {\partial \Psi(x,t)}{\partial t}  = H(x,t)\Psi(x,t)
\label {TDSE}
\enq
(in which $t$ is time, $x$ denotes all particle coordinates, $H(x,t)$ is a real
 Hamiltonian and $\hbar =1$), the presence of $i$ in the equation causes
 the solution $\Psi (x,t)$  to be complex-valued.
 Writing $\Psi (x,t) $ in a logarithmic form and separating as
\beq
ln\Psi(x,t)=ln(|\Psi(x,t)|)  + i arg (\Psi(x,t))
\label {logPsi}
\enq
we have in  the first term the modulus $|\Psi (x,t)|$ and in the
 second term $arg$, the "phase" . It is the latter that expresses the
 signed or complex valued nature of the wave-function.
In this section we shall investigate what, if any, interrelations exist between
moduli and phases? Are they independent quantities or, more likely
 since they derive from a single \er {TDSE}, are they interconnected?
 The result will be of the form of  "reciprocal" relations, shown in
\er {RRim} and \er {RRre} below.
 Some approximate and heuristic connections between phases and moduli
have been known before (\cite{Mehra} Vol.5, Part 2, Sec.IV.5); \cite{Schiff}
- \cite {SternAI}; we shall return to these in  subsection \ref{OPMR}.

\subsection {Origin of Reciprocal Relations}

 Contrary to what appears at a first sight, the integral relations in \er {RRim}
 and \er {RRre} are not based on causality. However, they can be related
to another principle \cite {Khalfin}.
 This approach of expressing a general principle by mathematical formulae
can be traced to von Neumann \cite {Neumann} and leads in the present instance
to an "equation of restriction", to be  derived below. According to von Neumann
 complete description of physical systems must contain:
\begin{enumerate}
  \item A set of quantitative characterizations (energy, positions, velocities, charges,
  etc.).
  \item A set of "properties of states" (causality, restrictions on the spectra
 of self energies, existence or absence of certain isolated energy bands,
 etc.).
\end{enumerate}
As has been shown previously \cite{Perelman69}, both sets can be
 described by eigenvalue equations, but for the set (2) it is more direct to
 work with projectors $Pr$ taking the values 1 or 0.
        Let us consider a class of functions $f(x)$, describing the
 state of the system or a process, such that (for reasons rooted in physics)
 $f(x)$ should vanish for $ x \not\in D$ (i.e., for $ supp \ f(x) = D$,
 where $D$
 can be an arbitrary domain and  $x$ represents a set of variables).
  If $Pr_D (x)$ is the projector onto the domain $D$, which equals $1$ for
 $x\in D$ and $0$ for $x \not\in D$, then all functions
 having this state-property  obey an "equation of restriction" \cite
 {Perelman66}:
\beq
f(x) = Pr_D(x) f(x)
\label {projector}
\enq

The "equation of restriction"  can embody causality,
 lower-boundedness  of energies in the spectrum, positive wave-number in the
 outgoing wave (all these in non-relativistic physics) and interactions
 inside the light cone only, conditions of mass spectrality, etc.
 in relativistic physics. In the case of interest in this Chapter, the
 "equation of restriction" arises from the lower-boundedness of energies $(E)$, or
the requirement that (in non-relativistic physics) one must have $ E>0 $
(where we have arbitrarily chosen the energy lower bound as equal to zero).

 Applying to \er {projector} an integral transform
 (usually, a Fourier transform) $F_k$ , one derives by (integral) convolution,
 symbolized by $\bigotimes_k $, the expression
\ber
 f(k)& = & F_k [Pr_D (x)]{\bigotimes_k } f(k)
 \nonumber \\
    & = & \int F_{k-k'}[Pr_D (x)] f(k') dk'
\label {convolution}
\enr

 For functions of a single variable (e.g., energy, momentum  or time)
 the projector $ Pr_D (x)$  is simply $ \Theta(x)$, the Heaviside
step-function, or a combination thereof. When also replacing $x,k$ by the
variables $E,t$, the Fourier transform in \er{convolution} is given by
\beq
 F_t [\Theta(E)] =   \delta_+ (t)
                \equiv {1\over 2}[\delta (t) - \frac {i}{\pi}P({1\over t})]
\label {deltaplus}
\enq
where $P$ designates the principal part of an integral. Upon substituting into
\er {convolution} (with $k$ replaced by $t$) one obtains after a slight simplification

\beq
f(t)= \frac {i}{\pi} P\int_{-\infty}^{\infty} \frac{1}{t'-t}  f(t') dt'
\label {cauchy}
\enq
Real and imaginary parts of this yield the basic equations for the functions
appearing in \er{RRim} and
\er{RRre}, below. (The choice of the upper sign in these equations will be
 justified in a later subsection for the ground state component in several
 physical situations. In some other circumstances, such as for excited states in certain
systems, the lower sign can be appropriate.)

 \subsubsection {A general wave packet}

We can state the form of the conjugate relationship  in  a setting
 more general than $\Psi (x,t)$, which is just a particular, the coordinate
 representation of the evolving state.
For this purpose, we write the state function in a more general way, through
\beq
|\Psi(t)>= \sum_n \phi_n (t) |n>
\label {Psisum}
\enq
where $|n>$ represent some time-independent orthonormal set and $\phi_n (t)$
 are the  corresponding amplitudes. We shall write generically $\phi (t)$
 for any of  the  "component-amplitudes" $\phi_n (t)$ and derive from it,
  in \er {chidef} below, a new function $\chi_{\pm}(t)$ that  retains all the
 {\it fine-structured} time variation in $ ln \Psi (t)$ and is free of the
 large scale variation in the  latter. We then derive in several physically
 important  cases, but not in all, reciprocal relations between the modulus and
 phase of $\chi(t)$
 taking  the form
\beq
\frac{1}{\pi} P \int_{-\infty}^{\infty}dt' [ln|\chi(t')|]/(t-t') = \pm arg \chi (t)
\label {RRim}
\enq
and
\beq
\frac{1}{\pi} P \int_{-\infty}^{\infty}dt' [arg\chi(t')]/(t-t') = \mp ln|\chi (t)|
\label{RRre}
\enq
 The sign alternatives depend on the location of the zeros (or singularities)
 of $\chi(t)$.
 The above  conjugate, or reciprocal, relations are the main results in
this section. When \er{RRim} and \er{RRre} hold, $ln|\chi(t)|$ and $arg\chi(t)$
 are "Hilbert transforms" \cite {Titchmarsh1, Caratheodory}.

Later in this section, we shall specify the analytic properties of the functions
involved and obtain exact formulae similar to \er{RRim} and \er{RRre}, but
 less simple and harder to apply to observational data of, say, moduli.

 In subsection \ref{SC} we give conditions under which
 \er{RRim} and \er{RRre} are exactly or approximately valid.
 Noteworthy among these is the nearly adiabatic (slowly
 evolving) case, which relates to the Berry phase \cite {Berry84}.

\subsection {Other Phase-Modulus Relations}
\label{OPMR}

As a prelude to the derivation of our results, we note here some of the
 relations between phases and moduli that have been known previously.  The
 following is a list (presumably not exhaustive) of these relations. Some of
 them are standard text-book material.

\subsubsection {(a) The equation of continuity}

This was first found by  Schr\"{o}dinger in 1926 starting with \er {TDSE},
 which he called  the "eigentliche Wellengleichung".
 [Paradoxically,  this got translated  to "real wave equation" \cite {Mehra}.]
  In the form
\beq 2m\frac{\partial \ln|\Psi(x,t)|}{\partial t} + 2 \vec \nabla
\ln|\Psi(x,t)|.\vec \nabla arg[\Psi(x,t)]
 + \vec \nabla \cdot  \vec \nabla arg[\Psi(x,t)]=0
\enq
 (where $m$ is the particle mass), it is clearly a (differential) relation
 between the modulus and the phase. As such, it does not show up any
 discontinuity in the phase \cite {NugentPG}, whereas \er {RRim} and \er
{RRre} do that. We further note, that the above form depends on the
 Hamiltonian and looks completely different for, e.g, a Dirac electron.

\subsubsection {(b) The WKB formula}

In the classical region of space, where the potential is less than the
energy, the standard formula leads to an approximate relation between phase and
 modulus in the form of the following path integral (\cite{Schiff}, Section 28)
\beq
arg \ \Psi(x) = \pm C \int_0^{x(t)} | \Psi(x)|^{-2} dx
\enq
where  C is a normalization constant. This and the following example do not
 arise from the time dependent {\SE}; nevertheless, time
 enters naturally in a semi-classical interpretation \cite {Kay}.

\subsubsection {(c)  Extended systems}
Extending  some previous heuristic proposal \cite {SelloniCCP, AncilottiT}
 the phase in the polarized state
of a one-dimensional solid of macroscopic length $L$ was expressed
in \cite {Resta} as: \beq arg \ \Psi(x) = Im \ln\int_0^L
e^{\frac{2 \pi i x}{L}} | \Psi(x)|^2 dx \label{ext} \enq It has
been noted \cite{Resta}, that the phase in  \er{ext} is
gauge-independent. Based on
 the above mentioned heuristic conjecture (but fully justified, to our mind, in the
 light of our rigorous results), Resta  noted that "Within a finite
system two alternative descriptions [in terms of the squared modulus of the wave
function, or in terms of its phase] are equivalent" \cite{Resta94}.

\subsubsection {(d) Loss of phase in a quantum measurement}

In a self-consistent  analysis of the interaction between an observed system
 and the apparatus (or environment), Stern {\it et al} \cite {SternAI} proposed
 both a
 phase-modulus relationship [\cite {SternAI}, \eq (3.10)] and a deep lying
 interpretation.
 According to the latter, the decay of correlation between states in a
 superposition can be seen, equivalently, as the effect of either the
 environment upon the system or the back reaction of the system on
 its environment. The reciprocal relations refer to the wave-function
 of the (microscopic) system and not to its surroundings, thus
 there is only a change of correlation not a decay.  Still it seems
 legitimate to speculate that the dual representation of the change
 that we have found (namely, through the phase or through the modulus) might
 be an expression of the reciprocal effect of the coupling between the
 system (represented by its  states) and its environment (acting through
 the potential).

\subsection {The Cauchy-Integral Method for the Amplitudes}
\label{TCIM}

Since the amplitude $\phi(t)$ arises from integration of \er{TDSE}, it can be
 assumed to be uniquely given. We can further
 assume that $\phi(t)$ has no zeros on the real t-axis, except at those special
 points, where this is demanded by symmetry. The reason
 for this is that, in general, $\phi(t)=0$ requires the solution of two
 equations, for the real and the imaginary parts of $\phi(t)$ and this cannot be achieved
 with a single variable: a real $t$. (Arguing from a more physical angle, if there
 is a zero somewhere on a the real $t$ axis, then a small change in some parameter in the
 Hamiltonian, will shift this zero to a complex $t$. However, this small
 change cannot change the physical content of the problem and thus we can
 just as well start with the case where the zeros are away from the real
 axis.) We can therefore perform the decomposition of $\ln \phi(t)$,
  following \cite{PaleyW,Davison}:
\beq
\ln \phi(t) = \ln\phi_{+}(t) + \ln\phi_{-}(t)
\label{phi+-}
\enq
where $ \ln\phi_{+}(t)$ is analytic in a portion of the
complex $t$-plane that
 includes the real axis (or, as stipulated in \cite {PaleyW},
 "including  a strip of finite width about
 the real axis" ) and a large semicircular region above it and $ \ln\phi_{-}(t)$
 is analytic in the corresponding portion below and including the real axis.
 By defining new functions $\chi_{\pm} (t)$ we separate off those parts of
 $ \ln\phi_{\pm}(t)$ that do not vanish on the respective semicircles, in the form:
\beq
\ln \phi_{\pm} (t) = P_{\pm} (t) + \ln\chi_{\pm}(t)
\label{chidef}
\enq
where $\ln\chi_{+}(t)$  and $\ln\chi_{-}(t)$  are respectively analytic in the upper and
 lower half of the complex $t$-plane and vanish in their respective half
 planes for large $|t|$. The choices for suitable $P_{\pm} (t)$ are not unique,
 and only the end result for $\ln \phi_{\pm} (t)$ is. In the interim stage we apply
 to the functions $\ln\chi_{\pm}(t)$ Cauchy's theorem  with a contour $C$ that
 consists of an infinite semicircle in the upper $(+)$, or lower $(-)$ half
 of the complex $t'$-plane traversed clockwise $(+)$ or anti-clockwise $(-)$ and
 a line along the real $t'$ axis from  $-\infty$ to $+\infty$ in which the point $t'=t$ is
avoided with a small semicircle. We obtain:
\beq
\oint_C \frac{ \ln\chi_{\pm}(t')}{(t'-t)} dt' =  \pm 2 \pi i \ln\chi_{\pm}(t)
\qquad \mbox{or zero}
\label{HTchi}
\enq
depending on whether the small semicircle is outside or inside the half-plane
 of analyticity and the sign $\pm$ is taken to be consistently throughout.
Further, writing the logarithms as:
\beq
\ln \chi_{\pm}(t) = \ln |\chi_{\pm}(t)|  + i arg \ \chi_{\pm}(t)
\label {logchi}
\enq
and separating real and imaginary parts of the functions in \er{HTchi} we derive
 the following relations between the amplitude moduli and phases in the wave-function:
 \beq
 (\frac{1}{\pi}) P \int_{-\infty}^{\infty}
 \frac{[\log|\chi_{-} (t')| - \log|\chi_{+} (t')|]}{(t'-t)} d t'
 =  arg \ \chi_{+}(t) + arg \ \chi_{-}(t) = arg \ \chi (t)
\label {HTlogmodchi}
\enq
and
\beq
 (\frac{1}{\pi}) P \int_{-\infty}^{\infty}
 \frac{[ arg \ \chi_{+}(t') - arg \ \chi_{-}(t')]}{(t'-t)} d t'
 = \log|\chi_{-} (t)| + \log|\chi_{+} (t)|= \log|\chi(t)|
 \label {HTargchi}
\enq

\subsection {Simplified Cases}
\label{SC}

We shall now concentrate on several cases where relations \er{HTlogmodchi} and
\er{HTargchi} simplify. The most favorable case is where $\ln\phi (t)$ is
analytic in one
 half-plane, (say) in the lower half, so that $\ln\phi_{+} (t)=0$. Then one obtains
 reciprocal relations between observable amplitude moduli and phases as in
\er{RRim} and \er{RRre}, with the upper sign holding. Solutions of the
 \SE \ are expected to be regular in the lower half of the complex $t$-plane (which
corresponds to positive temperatures), but singularities of $\ln\phi (t)$
 can still arise from zeros of $\phi (t)$. We turn now to the location of
 these zeros.

\subsubsection {A. The near-adiabatic limit}

We wish to prove that as the adiabatic limit is approached the zeros of
 the component amplitude for the "time dependent ground state " (TDGS, to be
 presently explained) are such that for an overwhelming number of zeros
$ t_{r}, \quad Im \ t_{r}> 0$ and for a fewer number of other zeros
$ |Im \ t_s | \ll \frac{1}{\Delta E} \ll \frac{2 \pi}{\omega}$,
 where ($\Delta E$ is the characteristic spacing of the eigen-energies of the Hamiltonian ,
 and $\frac{2 \pi}{\omega}$ is the time scale (e.g., period) for the temporal variation of
 the Hamiltonian. TDGS is that solution of the \SE \ (\ref{TDSE}) that is initially
 in the ground state of $H(x,0)$, the Hamiltonian at $t=0$. It is known that
 in the extreme adiabatic (infinitesimally slow) limit a system not crossing
 degeneracies stays in the ground state ("the adiabatic principle"). We shall
 work in the nearly adiabatic limit, where the principle is approximately,
but not precisely true.

Expanding $\Psi(x,t)$ in the  eigenstates $|n>$ of $H(x,0)$, we have
\beq
\Psi(x,t)= \sum_n C_n (t) <x|n>
\enq
and we  assume (for simplicity's sake) that the expansion can be halted
 after a finite number (say, $N+1$) of terms, or that the coefficients
 decrease in a sufficiently fast manner (which will not be discussed here).
 Expressing the matrix of the Hamiltonian  $H$ as $G h_{nm}(t)$ where $h_{nm}(t)$ is of
 the order of unity and $G$ positive, we obtain (with the dot denoting
 time-differentiation)
\beq
\dot C_n (t)= -i G \sum_m h_{nm} (t) C_m (t)
\enq
The adiabatic limit is characterized by:
\beq
|\dot h_{nm} (t)| \ll |G|
\enq
We shall find that in the TDGS [i.e. $\Psi_g (x,t)$], the coefficient
 $C_g (t)$ of $<x|g>$ has the form:
\beq
C_g (t) = B_{gg} (t) e^{-i G \varphi_g} + \sum_{m}  B_{gm} (t) e^{-i G \varphi_m}
\label{CGDEF}
\enq
Here  $\varphi_m = \varphi_m (t)$ are time-integrals of the eigenvalues $e_m (t)$
of the matrix $h_{nm} (t)$:
\beq
 \varphi_m (t) = \int_0^t e_m (t') dt'
\enq
In the sum the value $m=g$ is excluded and (as will soon be apparent)
$\frac{B_{gm}}{B_{gg}}$ is small of the order of
\beq
\frac{|\dot h_{nm} (t)|}{G}
\label{smfac}
\enq
To find the roots of $C_g (t) = 0$ we divide \er{CGDEF}
by the first term shown and
 transfer the unity to the left hand side to obtain an equation of the
 form:
 \beq
1= c_1 (t) e^{-iG \delta e_1 t} +c_2 (t) e^{-iG \delta e_2 t} + \ldots
\qquad \mbox{to N terms}
\label{rooteq}
\enq
where $ \delta e_1 t$, etc. represent  the differences $ \varphi_m- \varphi_g$
and are necessarily
 positive and increasing with $t$, for non-crossing eigenvalues of $h_{nm} (t)$.
 (They are written in the form shown to make clear their monotonically
 increasing character and are exact, by the mean value theorem, with $\delta e_1$,
 etc. being some positive function of $t$.)  $c_1 (t)$, etc. are small near the
 adiabatic limit, where $G$ is large. It is clear that \er{rooteq} has solutions
 only at points where $Im t> 0$. That the number of (complex) roots of \er{rooteq}
 is very large in the adiabatic limit, even if \er{rooteq} has only a few number
 of terms,  can be seen upon writing $e^{ i t |h_{nm}|} = z$ and regarding \er{rooteq} as a
 polynomial equation in $z^{-1}$. Then the number of solutions increases with $G$.
 Moreover, these solutions can be expected to recur periodically provided
 the $\delta e$'s approach to being commensurate.

It remains to investigate the zeros of $C_g (t)$ arising from having divided
 out by\\ $B_{gg} (t) e^{-i G \varphi_g}$.
 The position and number of these zeros  depend
 only weakly on $G$, but depends markedly on the form that the time dependent
 Hamiltonian $H(x,t)$ has. It can be shown that (again due to the smallness of
 $c_1, c_2, \ldots )$ these zeros are near the real axis. If the Hamiltonian can be
 represented  by a small number of sinusoidal terms, then the number of
 fundamental roots will be small. However, in the $t$-plane these will recur
 with a period characteristic of the periodicity of the Hamiltonian. These
 are relatively long periods compared to the recurrence period of the roots of
 the previous kind, which is characteristically shorter by  a factor of
 \beq
\frac{|\dot h_{nm} (t)|}{G}
\enq
This establishes our assertion that the former roots are
 overwhelmingly more numerous  than those of the latter kind.
Before embarking on a formal proof, let us illustrate the theorem with
 respect to a representative, though specific example. We consider the time
 development of a doublet  subject to a {\SE} whose
 Hamiltonian  in a doublet representation is \cite {MooreStedman,EYB1}
\beq
H(t)= G/2\left( \begin{array}{cc}
  -cos(\omega t) &  sin(\omega t) \\
  sin(\omega t)  &  cos(\omega t)
\end{array} \right)
\label{h}
\enq
Here $\omega $ is the angular frequency of an external disturbance.
The eigenvalues of \er{h} are $\frac{G}{2}$ and $-\frac{G}{2}$. If $G>0$, then in the
ground state the amplitude of $|g>$ (=the vector  {\tiny $ \left(\begin{array}{c}
 1 \\ 0  \end{array} \right) $}  in  \er{h}) is
\beq
C_g =  \cos (Kt) \cos ( \omega t/2) + (\omega/2K) \sin( K t) \sin (\omega t/2)
         +i(G/2K) \sin(K t) \cos( \omega t/2)
\label{CGspec}
\enq
with
\beq
K =0.5 \sqrt{G^2 + \omega^2} \approx G/2, \qquad \mbox{since} \qquad G/\omega \gg 1
\enq
Thus the amplitude in \er{CGspec} becomes:
\ber
C_g (t) &\approx& \exp ( i K t) \cos ( \omega t/2) + (\omega/2K) \sin( K t) \sin (\omega t/2)
\nonumber \\
 &\approx& \exp ( i G t/2)[ \cos ( \omega t/2) - i (\omega/2G)  \sin (\omega
 t/2)] + i \exp ( - i G t/2) (\omega/2G)  \sin (\omega t/2)
 \nonumber \\ & &
 \label{CGaprox}
\enr
This is precisely of the form \er{CGDEF}, with the second term being smaller
than  the first by the small factor shown in \er{smfac}.
  Equating (\ref{CGaprox}) to zero and dividing by the
 first  term, we recover the form in \er{rooteq}, whose right hand side consists
 now of just one term. For an integer value of $G/\omega=M$ (say) which is large
 and $\exp (-i \omega t) = Z$, the resulting equation in $Z$ has about $M$ roots with
 $|Z|>1$ (or, what is the same, $Im \ t >0$). As noted above, further roots of $C_g
 (t)$ will arise from the neighborhood of $\cos (\omega t/2) = 0$, or $Z=-1$.
 [The upper state of the doublet states has the opposite properties, namely about $M$ roots
 with  $Im \ t <0$. We have treated this case (in collaboration with M. Baer)
in a previous work \cite {EYB1}.]

A formal derivation of the location of the zeros of $C_g (t)$ for a general
 adiabatic Hamiltonian can be given, following proofs of the adiabatic
 principle (e.g., \cite{BornF}-\cite {Messiah}). The last source,
\cite {Messiah} derives an evolution
 operator $U$, which is written there , with some slight notational change,
 in the form
 \beq
U(t) = A(t) \Phi(t) W(t)
\label{APHW}
\enq
[\eq XVII.86 in the reference source]. Here $A(t)$ is a unitary transformation
 [\eq XVII.70] that "takes any set of basis vectors of $H(x,0)$ into a set of
 basis vectors of $H(x,t)$ in a continuous manner" and is independent of $G$.
 In the previous worked example its components are of the form $\cos( \omega t/2)$ and
 $sin( \omega t/2)$ \cite {Messiah}. The next factor in \er{APHW} is diagonal [XVII.68]
 and consists of terms of the form:
 \beq
\Phi(t)= \exp (-i G \varphi_m ) \delta_{n m}
\enq
Finally, the unitary transformation W(t) was shown to have a near-diagonal
 form (\cite {Messiah}, Eq. XVII.97)
\beq
W(t)= \delta_{n m} + ( \frac{|\dot h (t)|}{G}) \delta W_{nm}
\enq
The $gg$-component of the evolution matrix $U$ is just $C_g$ and is, upon
 collecting the foregoing,
\beq
C_g (t) = \sum_m A_{gm} (t) \exp (-i G \varphi_m)[\delta_{mg} +
( \frac{|\dot h (t)|}{G}) \delta W_{mg} ]
\enq
This can be rewritten as
\beq
C_g (t) = A_{gg}(t) [1+ ( \frac{|\dot h (t)|}{G}) \delta W_{gg}]
 \exp(-i G \varphi_g)
 + ( \frac{|\dot h (t)|}{G}){\sum_m}' A_{gm} (t) \exp (-i G \varphi_m)\delta
 W_{mg}
\enq
with the summation excluding $g$. This is again of the form of \er{CGDEF},
 establishing the generality of the location of the  eigenvalues for the
 nearly adiabatic case.

\subsubsection {B. Cyclic wave-functions}

This is a particularly interesting case, for two reasons. First,
 time-periodic potentials such that arise from external periodic forces,
 frequently give rise to cyclically varying states. (According to the authors of
\cite {LiuHL}: "The universal existence of the cyclic evolution is guaranteed
 for any quantum system".) The second reason is that the Fourier
 expansion of the cyclic state spares us the consideration of the convergence
 of the infinite-range integrals in \er{RRim} and \er{RRre}; instead, we need to consider
 the convergence of the (discrete) coefficients of the expansion.
In this section we show that in a broad class of cyclic functions one
 half of the complex $t$-plane is either free of amplitude-zeros, or has
 zeros whose contributions can be approximately neglected. As already noted
 above, in such cases, the reciprocal relations connect observable phases
 and moduli (exactly or approximately). The essential step is that a
 function $\phi (t)$ cyclic in time with period $2 \pi$ can be written as a sine-cosine
 series. We assume that the series terminates at the $N$'th trigonometric function,
with $N$ finite.
 We can write the series as a polynomial in $z$, where $z = \exp (it)$,
 in the form:
 \ber
 \phi(t) &=& \sum_{m=0}^{2N} c_m z^{m-N} \\
  &=& z^{-N} c_0 \chi (t)  \nonumber \\
 &=& z^{-N} c_0 \sum_{m=0}^{2N} \frac{c_m}{c_0} z^{m}
\enr
If $\phi(t)$ is a wave function amplitude arising from a Hamiltonian that is
 time-inversion-invariant then we can choose $\phi(-t) = \phi^{*} (t)$ for real $t$,
 where the star denotes the complex conjugate. Then the coefficients
 $c_m$ are all real. Next, factorize in products as:
 \beq
 \chi(t) = \prod_{k=1}^{2N}(1-z/z_{k})
\enq
where $z_k$ are the (complex) zeros of $\chi(t)$ or $\phi (t)$, $2N$ in number.
 Then the decomposition shown in \er{chidef}, namely $\ln\chi (t)= \ln\chi_{+} (t)+
  \ln\chi_{-} (t)$, will be achieved with:
 \beq
 \ln\chi_{+} (t) = \sum_{k=1}^{R} \ln(1-z/z_{k+}),
 \qquad |z_{k+}| \geq 1,
 \label{logchi+}
 \enq
 \beq
 \ln\chi_{-} (t) = \sum_{k=R+1}^{2N} \ln(1-z/z_{k-}),
 \qquad |z_{k-}| < 1,
 \label{logchi-}
 \enq
provided that $R$ of the roots are on or outside the unit circle in the
 $z$-plane and $2N-R$ roots are inside the unit circle. The results in \er{HTlogmodchi}
 and \er{HTargchi} for the phases and amplitudes can now be applied directly. But
 it is more enlightening to obtain the coefficients in the complex
 Fourier-series for the phases and amplitudes. This is easily done for \er{logchi+},
 since for each term in the sum:
 \beq
|z/z_{k+}| = |\exp(it)/z_{k+}| < 1,
\label{zovzk+}
\enq
and the series expansion of each logarithm converges. (When, in
\er{zovzk+} equality reigns, which is the case when the roots are upon the unit circle,
the convergence is "in the mean" \cite {Zigmund}.) Then the  n'th Fourier coefficient
 is simply the coefficients of the  term $\exp(i n t)$ in the expansion, namely,
 $-(1/n) (1/z_{k+})^n$.

The corresponding  series-expansion of $\ln\chi_{-} (t)$ in \er{logchi-} is not legitimate,
since now in every term:
\beq
|z/z_{k-}| = |\exp(it)/z_{k-}| > 1,
\label{zovzk-}
\enq
Therefore we rewrite:
\ber
\ln\chi_{-} (t) = - \sum_{k=R+1}^{2N} \ln(-z_{k-}) + (2N-R)it
+ \sum_{k=R+1}^{2N} \ln(1- z_{k-}/z).
\enr
Each logarithm in the last term can now be expanded and the $(-n)$'th
 Fourier coefficient arising from each logarithm is $-(1/n)(z_{k-})^n$. To this
 must be added the $n=0$ Fourier coefficient coming from the first,
 $t$-independent term and that arising from the expansion of second term as
 a periodic function, namely
 \beq
i t = -2 i \sum_n (-1)^n \sin (n t)/n
\enq
For the Fourier coefficients of the modulus and the phase we note that,
 because of the time-inversion invariance of the amplitude, the former is
 even in $t$ and the latter is odd. Therefore the former is representable as a
 cosine series and the latter as a sine series. Formally:
 \beq
\ln(\chi)= \ln|(\chi)| + i arg (\chi)
          = \sum_n A_n \cos (n t)+ i \sum_n B_n \sin (n t)
\enq
When expressed in terms of the zeros of $\chi$, the sine-cos coefficients of the
log modulus and of the phase are respectively:
\beq
A_0 = - \sum_{k=R+1}^{2 N} \ln|z_{k-}|
\label{A0}
\enq
[This is written in terms of $|z_{k-}|$,  the moduli of the roots $z_{k-}$, since the roots
 are either real or come in mutually complex conjugate pairs. In any case,
 this constant term can be absorbed in the polynomial $P(t)$ in \er{chidef}.]
\beq
A_n = \left[\sum_{k=1}^{R} 1/(z_{k+})^n + \sum_{k=R+1}^{2N} (z_{k-})^n \right]/n
\label{An}
\enq
 \beq
B_n = \left[\sum_{k=1}^{R} 1/(z_{k+})^n - \sum_{k=R+1}^{2N}
[(z_{k-})^n - 2 (-1)^n] \right]/n
\label{Bn}
\enq
Equations (\ref{A0})-(\ref{Bn}) are the central results of this subsection.
Though somewhat complicated, they are easy to interpret, especially  in the limiting cases
 (a) -(d), to follow. In the general case, the equations show that the
 Fourier coefficients are given in terms of the amplitude zeros.
(a) When there are no amplitude zeros in one of the half planes, then only
 one of the sums in \er{An} or  \er{Bn} is non-zero ($R$ is either $0$ or $2N$).
 Consequently, the Fourier coefficients of the log modulus and of the phase
 are the same (up to a sign) and the two quantities are logically
 interconnected as functions of time. The connection is identical with that
  exhibited in \eqs (\ref{RRim}) and (\ref{RRre}).
In the two-state problem formulated by \er{h}, the solution (\ref{CGspec}) is cyclic
 provided  $K/\omega$ is an integer. A "Mathematica" output of the zeros of \er{CGspec}
 for $K/\omega=8$ gives the following results: None of the zeros is located
 in the lower half plane, $7$ pairs and an odd one is in the upper half plane
 proper, a pair of zeros is on the real $t$-axis. The reciprocal integral
 relations in \er{RRim} and \er{RRre} are verified numerically, as seen in figure
 \ref{fig: rec}.
\begin{figure}
\vspace{8cm}
\includegraphics{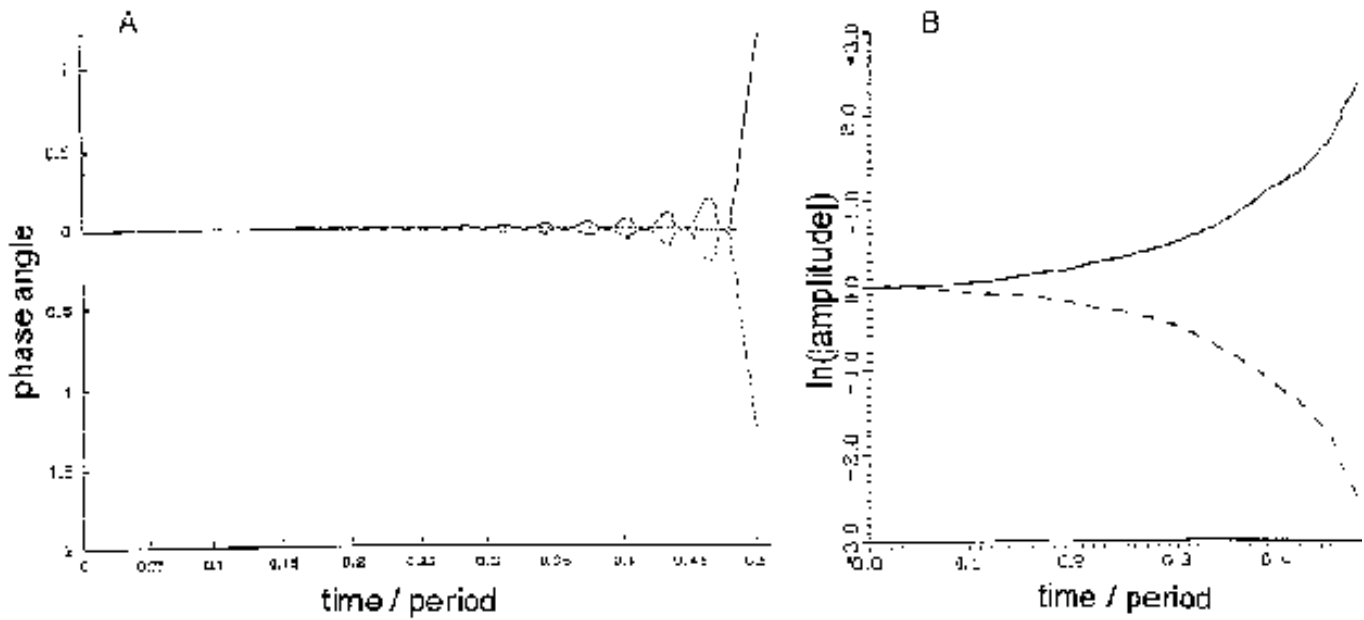}
\caption {Numerical test of the reciprocal relations in \er{RRim} and \er{RRre} for
$C_g$ shown in \er{CGspec}. The values computed directly from \er{CGspec} are
plotted upwards
and the values from the integral downwards (by broken lines) for $K/\omega=16$.
 The two curves are clearly  identical.
(a) $\ln|C_g (t)|$ against ($t$ /period).
The modulus is an even function of $t$.
(b) $\arg C_g (t)$ against ($t$/period). The phase is odd in $t$.}
\label {fig: rec}
\end{figure}
 (The equality between the Fourier coefficients $A_n$ and $B_n$ was verified
 independently.)
(b) It is a characteristic of the above two state problem (with general
     values of $K/\omega$), and of other problems of similar type that there is one
 or more roots at or near $z_{k+} =-1$ ($t= \pm \pi$; the generality of the occurrence
 of these roots goes back  to a classic paper on conical intersection
 \cite{NeumannW}.)
 By inspection of the second sum in \er{Bn} we find that, if all the roots
located in the upper half plane are of this type, then $A_n = B_n$ up to small
 quantities of the order of $(z_{k+}+1)$.
 Then again \eqs (\ref{RRim}) and (\ref{RRre}) can be employed.
(c) As a corollary to the previous observation (and an important one in
 view of the stipulation in subsection \ref{TCIM}, that the wave function has no real
 zeros) a small shift in the location of a zero originally at $t= \pm \pi$ into the
 complex plane either just above or just below this location, will only have
 a small consequence on the Fourier coefficients. Therefore, zeros of this
 type do not violate the assumptions of the theory.
(d) If either $|z_{k+}| \gg 1$ or  $|z_{k-}| \ll 1$, it is clear from \er{An} and
\er{Bn} that the contribution of such roots is small. This circumstance is important for
 the following reason: Suppose that the model is changed slightly by adding
 to the potential a small term, e.g., adding $\epsilon \cos 2 \omega t$ to a diagonal matrix
 element in \er{h}, where $\epsilon$ is small. (In \cite {ZwanzigerG}  terms of this type were
 used to describe the nonlinear part of a Jahn-Teller effect.) Necessarily,
 this term will introduce new zeros in the amplitude. It can be shown that
 this addition will add new roots of the order  $|z_{k+}| \approx 1/\epsilon$ or
$|z_{k-}| \approx \epsilon$. The effects of these are asymptotically negligible. In other
 words, the formulae (\ref{An}) and (\ref{Bn}) are stable with respect to small
variations in the model. [A similar result is known as Rouche's  theorem
 about the stability of the number of zeros in a finite domain
 (\cite {Titchmarsh2} Section  3.42).]

\subsubsection {C. Wave packets}

A time varying wave function is also obtained with a time-independent
 Hamiltonian by placing the system initially into a superposition of energy
 eigenstates ($|n>$), or forming a wave-packet. Frequently, a coordinate
 representation is used for the wave function which then may be written as
\beq
\Psi (x,t) =\sum_m a_m (t)  \exp (-i E_m t) <x|m>
\label{wavpac}
\enq
where $<x|m>$ are solutions of the time independent {\SE}, with
 eigen-energies $E_m$ that are taken as non-degenerate and increasing with $m$.
 In this coordinate-representation, the "component-amplitudes" in the
 Introduction are just fancy words for $\Psi (x,t)$ at fixed $x$ (so that the
 discrete state label $n$ that we have used in \er {Psisum} is equivalent to the
continuous variable $x$)  and ($\phi_n (t)$ is simply $<x(t)|\Psi(x,t)>$. The results in
 the earlier section are applicable to the present situation. Thus, to
 test \er{RRim} or \er{RRre}, one would look for any fixed position $x$ in space at
 the moduli (or state populations) as a function of time, as with repeated
 state-probing set ups. In turn, by some repeated interference experiments
 at the same point $x$, one would establish the phase and then compare
 the results with those predicted by the equations. (Of course, the same
 equations can also be used to predict one quantity, provided the time
 history of the second is known.)

As in previous sections, the zeros of $\Psi(x,t)$ in the complex $t$-plane
 at fixed $x$ are of interest. This appears a hopeless task, but the situation
is not that bleak. Thus, let us consider a wave packet initially localized
 in the ground state in the sense that in \er {wavpac}, for some given $x$,
\beq
\sum_{m>0} |a_m <x|m>|^2 < | a_0 <x|0>|^2.
\label{con}
\enq
Then we expect that for such value(s) of the coordinate, the $t$- zeros of the
 wave-packet will be located in upper $t$-half plane only. The reason for this
 is similar to the reasoning that led to the theorem about the location of
 zeros in the near-adiabatic case. (Paragraph A, above.)
Actually, empirical investigation of wave-packets appearing in the
 literature indicates that the expectation holds in a broader range
 of cases, even when the condition (\ref{con}) is not satisfied. It should
 be mentioned that much of the wave packet work is numerical and it
 is not easy to theorize about it. (A review describing certain aspects
 of wave-packets is found in \cite {Kosloff}.)

We now present some examples of
 studied wave-packets for which the
 reciprocal relations hold (exactly or approximately), but have not been
 noted.

(a) Free-particle in one dimension.

 The Hamiltonian consists only of the kinetic energy of the particle having
 mass $m$ (\cite {Tomonaga}, \cite {Schiff} Section 28).
 The (unnormalized) energy eigenstates
 labeled by the momentum index $k$ are:
 \beq
\psi_k (x) = \exp (i k x)
\enq
with corresponding energies $E_k = k^2 / 2m$. Initially the wave packet is centered
 on $x=0$ and has mean momentum $K$. As shown in \cite {Tomonaga}, the coefficients $a_k$
 appearing in \er{wavpac} are
\beq
a_k = \exp [ -(k-K)^2 \Delta^2]
\enq
where $\Delta (>0)$ is the root mean square width in the initial wave packet.
 The expanding wave packet can be written as:
\beq
\ln\Psi (x,t) = -1/2 \ln[\Delta +(it/2m \Delta)] -
\frac{[x^2-4 i \Delta^2 K (x-K t/2m)]}{[4 \Delta^2+2 i t/m]} \ + \ constant
\label{logpsiwavpac}
\enq
which is clearly analytic in the lower half $t$-plane. (The singularity arises
 because free electron wave functions are not normalizable.) We can therefore
 identify this function with $\ln\phi (t) = \ln\phi_{-} (t)$ in \er{phi+-},
 and put $\ln\phi_{+} (t) =0$.
 As a numerical test  we have inserted \er{logpsiwavpac} in \er{RRim} and
 \er{RRre}, integrated numerically and found (for $K=0$) precise agreement.

(b) "Frozen Gaussian Approximation".

Semi-analytical and semi-classical wave packets suitable for calculating
 evolution on an excited state multi-dimensional potential energy surface
 were proposed in pioneering studies by Heller \cite {Heller81}. In this  method (called
 the Frozen Gaussian Approximation) the last two factors in the summand of
 \er{wavpac} were replaced by  time dependent gaussians. The time dependence arose
 through  having time varying average energies, momenta and  positions.
 Specifically, each coefficient $a_m$ in  \er{wavpac} was followed by a function
 $g(x,t)$ of the form:
  \ber
 \ln g(x,t) = &-& m \omega (x-<x>_t)^2 / 2 + i <p>_t (x-<x>_t)
 \nonumber \\
 &+& i \int_0^t (<p>_t^2 /m -<g|H|g>_t) dt
\label{logg}
\enr
where $\omega$ is an energy characteristic of the upper potential surface, the
 angular brackets are the average position and momenta of the classical
 trajectory and the Dirac bracket of the Hamiltonian $H$ is to be evaluated
 for each component $g$ separately.

 For a set of gaussians it is rather difficult to establish the analytic
 behavior of \er{logg}, or of \er{wavpac}, in the $t$-plane. However, with a single
 gaussian (in one spatial dimension) and a harmonic potential surface one has
classically:
\beq
<x>_t =x_0 \cos \omega t,
\label{xt}
\enq
\beq
<p>_t = (m)\frac{d<x>_t}{dt} = -(\omega x_0 m ) \sin \omega t,
\label{pt}
\enq
\beq
<g|H|g>_t = \omega/2
\label{gHg}
\enq
Substituting these expressions into \er{logg} one can see after some algebra
 that $\ln g(x,t)$ can be identified  with $\ln\chi_{-} (t) + P(t)$ shown in subsection
 \ref{TCIM}. Moreover,  $\ln\chi_{+} (t) = 0$.
  It can be verified, numerically or algebraically,
 that the log-modulus and phase of $\ln\chi_{-} (t)$ obey the reciprocal relations
(\ref{RRim}) and (\ref{RRre}).
In more realistic cases (i.e., with several gaussians) \eqs (\ref{xt})- (\ref{gHg}) do not
 hold. It still may be true that the analytical properties of the wave packet
 remain valid and so do relations (\ref{RRim}) and (\ref{RRre}). If so, then these
 can be thought of as providing numerical checks on the accuracy of
approximate wave-packets.

(c) Expanding waves.

As a further application we turn to the expanding potential problem
(\cite {BerryK}-\cite {Grosche}), where we shall work from the amplitude-modulus to the phase.
The time dependent potential is of the form:
\beq
V(x,t)= \zeta^{-2} (t) V(x/\zeta(t))
\label{Vexpan}
\enq
Here $\zeta^2 (t)= c + t^2$, which differs from the more general case considered in
\cite {BerryK}-\cite {Grosche}, by putting their time scale factor $a=1$ and making the potential
 real and regular for real $t$, as well as time-inversion invariant. Then $c$ is positive
 and, in \cite {BerryK, DodonovMN} $b=0$. The Hamiltonian is singular at
 $t = \pm i \sqrt{c}$, away from the real axis.
As first shown in  \cite {BerryK}, the generic form of the solution of the time
dependent  {\SE}  is the same for a wide range of potentials.
 We shall consider the ground state for a harmonic potential $V(x)= 1/2 m \omega_o^2 x^2$.
The log (amplitude-modulus) of the ground state wave function (in the
 coordinate representation) is according to \cite {BerryK} for real $t$
\beq
\ln|\phi (x,t)| = -(1/4) \ln(c+t^2) - 1/2 [ m \omega x^2/(c + t^2)],
\label{logexwf}
\enq
where $ \omega^2  = \omega_o^2 + c$.
Processing the expression in \er{logexwf} as in \er{phi+-}, we can arbitrarily decompose
 $\phi (x,t)$ into factors that are analytic above and below the real $t$-axis.
 Thus, let us suppose that in \er{logexwf} a fraction $f_1$ of the first term and
 a fraction $f_2$ of the second term is analytic in the upper half and,
 correspondingly, fractions $(1-f_1)$ and $(1-f_2)$ are analytic in the lower half.
 Explicitly, for complex $t$
\ber
 \ln|\phi (x,t)| &=& Re \ \{ - 1/2[ f_1 \ln(\sqrt{c}-it) +
 (1-f_1) \log(\sqrt{c}+it)]
\nonumber \\
 &-& 1/2 (m \omega x^2/\sqrt{c}) [f_2/(\sqrt{c}-it)
\nonumber \\
 &+&
 (1-f_2)/(\sqrt{c}+it)]\}.
\enr
Next, for the log term (which normalizes the wave function), we have to
 choose, as in \er{chidef}, suitable functions $P_{\pm}(t)$ that will
 "correct" the behavior of that term along the large semi-circles.
 Among the multiplicity of choices
 the following are the most rewarding (since they completely cancel the
 log term):
\ber
 P_{+} (t) &=& -f_1 (1/2) \ln(\sqrt{c}-it)
 \nonumber \\
 &=& -(f_1/4)[\ln(t^2+c) -2i \arctan (t/\sqrt{c})],
 \nonumber
\enr
\ber
P_{-} (t) &=& -(1-f_1) (1/2) \log(\sqrt{c} + i t)
 \nonumber \\
 &=& -(1/4)(1-f_1)[\ln(t^2+c)+ 2i \arctan (t /\sqrt{c})].
 \nonumber
\enr
The right hand side of \er{HTlogmodchi} comes from the second term of \er{logexwf} alone
 and is
\beq
\arg \chi (x,t) = (1-2 f_2) [m \omega x^2/(t^2+c)](t/4\sqrt{c}).
\enq
To complete the phase of the wave function, $\arg \phi (x,t)$, we have to
 reinstate the terms $P_{\pm} (t)$ that
 were removed in \er{chidef} so as to get $\chi_{\pm} (x,t)$. The result
 is:
\beq
\arg \phi (x,t) = -(1-2f_1)(1/2) \arctan (t/\sqrt{c}) +
 (1-2 f_2)[m \omega x^2/(t^2+c)](t/4 \sqrt{c})
\enq
This establishes the functional form of the phase for real (physical)
 times. The phase of the solution given in  \cite {BerryK, DodonovMN} has
 indeed this functional form. The fractions $f_1$ and $f_2$ cannot be determined from our
 equations (\ref{logchi}) and (\ref{HTlogmodchi}).
 However, by comparing with the wave functions
 in \cite {BerryK, DodonovMN}, we get for them the following values.
 \beq
 1-2 f_1 = \omega /\sqrt{c},   \qquad  1 - 2 f_2 = 4 \sqrt{c} /\omega
 \enq
In the excited states for the same potential, the log modulus contains
 higher order terms in $x$ ($x^3$, $x^4$, etc.) with coefficients that depend on
 time. Each term can again be decomposed (arbitrarily) into parts analytic
 in the $t$-half-planes, but from elementary inspection of the solutions in
 \cite {BerryK, DodonovMN} it turns out that every term except the lowest
 (shown in \er{Vexpan}) splits up
 equally (i.e. the $f$'s are just $1/2$) and there is no contribution to the
 phases from these terms. Potentials other than the harmonic can be treated
 in essentially identical ways.

\subsection {Consequences}
 The following theoretical consequences of  the reciprocal relations can be noted:

(i) They unfold a connection between parts of time-dependent wave functions
 that arises from the structure of the defining \er{TDSE} and some simple
 properties of the Hamiltonian.

(ii) The connection holds separately for the coefficient of each state
 component in the wave function and is not a property of the total wave
 function (as is, e.g., the "dynamical" phase \cite{Berry84}).

(iii) The relations pertain to the fine, small-scale time-variations in the
 phase and the log modulus, not to their large scale changes.

(iv) One can define a phase that is given as an integral over the log of
 the amplitude modulus and is therefore an observable and is gauge-invariant.
This phase (which is unique, at least in the cases for which \er{RRim} holds)
 differs from other phases, those that are, e.g. a constant, the dynamic
 phase or a gauge-transformation induced phase, by its satisfying the
 analyticity requirements laid out in subsection \ref{OPMR}.

(v) Experimentally, phases can be obtained by measurements of
 occupation probabilities of states using \er{RRim}. (We have
 calculationally verified this for the case treated in \cite {AharonovKPR}.)

(vi) Conversely, the implication of \er{RRre} is that a geometrical phase
 appearing on the left-hand-side entails a corresponding geometric
   probability change, as shown on the right-hand-side. Geometrical decay
probabilities have been predicted in \cite {Berry90} and experimentally
 tested in \cite {ZwanzigerRC}.

(vii) An important ingredient in the analysis  has been the
positions  of zeros of $\Psi(x,t)$ in the complex t-plane
 for a fixed x.  Within quantum mechanics the zeros have not  been
 given much attention, but they have been studied in a mathematical context
 \cite {Titchmarsh2} and in some classical wave-phenomena (\cite {Shvartsman}
 and references cited therein).
 Their relevance to our study is evident since at its zeros the phase of
 $\Psi(x,t)$ lacks definition. Future theoretical work shall focus on a systematic
 description of the  location of zeros.
  Further, practically oriented work will seek out computed or experimentally
 acquired time dependent wave functions for tests or application of
 the present results.

(viii) Finally, and probably most importantly, the relations show that changes
(of a nontrivial type) in the phase
 imply necessarily a change in the occupation number of the state components
 and vice versa. This means that for time-reversal-invariant situations,
 there is (at least) one partner state with which the phase-varying state
communicates.

\section { Non-Linearities That Lead to Multiple Degeneracies}

       In previous sections of this review we have treated  molecules and other
localized systems in which a linear electron-nuclear coupling resulted in  a
single degeneracy, or conical intersection ({\it ci}) of the electronic potential
energy surfaces. A notable, symmetry-caused example of this is the linear
$E \otimes \epsilon$ Jahn-Teller effect (a pair of degenerate electronic states, that can
happen under trigonal or higher symmetry, which is coupled to two
energetically degenerate displacement modes)
\cite {JahnT, Englman72,BersukerPolinger}.
Still, some time ago nonlinear coupling was also considered
within the $E \otimes \epsilon$ case in \cite {O'Brien, Ham71} and subsequently in \cite
{ZwanzigerG}. Such coupling can result in a more complex situation, in which
there is a quadruplet of {\it ci}'s, such that one {\it ci} is situated at the origin of
the mode coordinates (as before) and three further {\it ci}'s are located farther
outside in the plane, at points that possess trigonal symmetry.

 As of late, non-linear coupling has become of increased interest, partly
 through evidence for
a weak linear coupling in the metallic cluster $Na_3$ \cite {KoppelMeiswinkel, Yarkony99a}
(computations of vibrational levels in a related molecule $Li_3$ were performed
 in \cite
{VarandasYX, VarandasX}), and partly by attempts to computationally locate
{\it ci}'s in the potential energy landscape with a view to estimate their effect
on inter-surface, non-adiabatic transitions \cite { Yarkony99b}. The method
used in the last reference was based on the acquisition of the geometric phase by the
total function as a {\it ci} is circled \cite { HerzbergL, Longuet, Stone,
Berry84}. Independently, the authors of \cite {KoizumiBersuker} found theoretically a
causal connection between the number of {\it ci} effectively circled (one or
four) and the important question of the nature of the ground state. They
showed that, contrary to what had been widely thought before, the ground
state may be either a vibronic doublet or a singlet, depending on the distance
(which is a function of the parameters in the vibronic Hamiltonian) of these
trigonal {\it ci}'s from the centre.  (A similar instance of "quantum phase
transition" was noted for a threefold degenerate system in \cite {KoizumiBBP} and,
earlier, for an icosahedral system, in \cite {MoateODBLP}).

In a different field, location and characteristics of {\it ci}'s on diabatic
potential surfaces have been recognized as essential for the evaluation of
dynamic parameters, like non-adiabatic coupling terms, needed for the
dynamic and static properties of some molecules (\cite {SaxeLY}  -\cite
{WaschewskyKMKB}, \cite {BaerLAAB}). More recently, pairs of {\it ci}'s  have
been studied \cite {MebelBL2000, MebelBL2001} in greater detail. These
studies arose originally in connection with a {\it ci} between the $1^2 A' $ and
$2^2 A'$  states found earlier in computed potential energy surfaces for $C_2 H$
in $C_s$ symmetry \cite {ThummelPPB}. Similar {\it ci}'s appear between the
potential surfaces of the two lowest excited states $^1A_2$ and $^1B_2 $ in
$H_2 S$  or of  $^2 B_2$ and $^2A_1$  in $Al-H_2$ within $ C_ {2v} $ symmetry \cite
{Yarkoni98}. A further, closely spaced pair of {\it ci}'s has also been found
between the $ 3^2 A' $  and  $ 4^2 A' $   states of the molecule $C_2 H$. Here the
separation between the twins varies with the assumed $C-C$ separation, and
they can be brought into coincidence at some separation \cite {MebelBL2001}.

        In this section we investigate the phase changes that
characterize the double and trigonal (or cubic) {\it ci}'s. We shall find
 that the Berry phases upon circling around the {\it ci}'s can
take the values of $0 $ or $2N\pi $ (where $N$ is an integer). It can be
shown that the different values of $N$  can be made experimentally
observable (through probing the state populations after inducing changes in
the amplitudes of the components), in a way that is not marred by the fast
oscillating dynamic phase.
    Apart from the results regarding the integer $N$ in the Berry phase, the
difference between our approach to the phase changes and those in some
previous works, especially in \cite {Yarkoni98, Yarkony99b}, needs to be
noted. While these consider the topological phase belonging to the total wave
packet, we continue in the spirit of the previous section 3 and treat the open
phase belonging to a single component of the wave packet. (For the
topological, full-cycle phase the two are equivalent, but not for the open
phase, that is present at interim stages.) Explicitly, we write the (in general)
time ($t$)-dependent molecular wave function $\Psi (t)$ as a superposition of
(diabatic) electronic states $\chi_k$ as
\beq
\Psi (t) =  \sum_k a_k (t) \chi_k
\label {Psi}
\enq
 where the amplitudes $a_k$ are functions of the nuclear
coordinates. In the previous section we have developed and used
the reciprocal relations between the phases ($arg \ a_k$) and the
("observable") moduli ($|a_k|$).

We also describe a "tracing" method to obtain the phases after a full cycling.
We shall further consider wave-functions whose phases at the completion of
cycling differ by integer multiples of $2\pi $ (a situation that will be written, for
brevity, as "$2N \pi $ "). Some time ago, these wave functions have been
shown to be completely equivalent, since only the phase factor (viz., $
e^{iPhase}$ ) is  observable \cite {WY}; however, this is true only for a set of
measurements that are all made at instances where the phase difference is
$2N\pi $. We point out simple, necessary connections between having a
certain $2N\pi $ situation and observations made prior to the achievement of
that situation. The phase that is of interest in this review is the Berry phase of
the wave-function \cite {Berry84}, not its total phase, though this distinction
will not be re-stated.

\subsection {Conical Intersection Pairs}
\label{Conical Intersection Pairs}

    We treat this case first, since it is simpler than the trigonal case. The
molecular displacements are denoted by $x$ and $y$ (with suitable choice of
their origins and of scaling). Then, without loss of generality we can denote
the positions of the {\it ci} pairs in Cartesian coordinates by

\beq
x=\pm 1, \qquad y=0
\label {xyroots}
\enq
or, in polar coordinates, where $x=q \cos \phi$, $y= q \sin \phi $, by
\beq
q=1 , \qquad \phi =0, \pi
\label{qroots}
\enq
To obtain potential surfaces for two electronic states which will be degenerate
at these points, we write a Hamiltonian as a 2x2 matrix in a diabatic
representation in the following form:
\ber
H(x,y) &=& K \left(
\begin{array}{cc}
-(x^2-1) & y f(x)\\
 y f(x) &  (x^2-1)
\end{array} \right)
\nonumber \\
 &=& K  \left(
\begin{array}{cc}
  -(q^2 \cos^2 \phi-1)  & q \sin \phi f(q \cos \phi) \\
 q \sin \phi f(q \cos \phi) &  (q^2 \cos^2 \phi-1)
\end{array} \right)
=  H(q, \phi)
\label {pairHamiltonian}
\enr
whose two eigenvalues are
\ber
E_{\pm} (x,y)  &=& \pm K \sqrt{(x^2 - 1)^2  + [y f(x)]^2 }
\nonumber \\
              &=& \pm K \sqrt{(q^2 \cos^2 \phi - 1)^2 +[q \sin \phi f(q \cos \phi)]^2}
\nonumber \\
             &=& E_{\pm}(q,\phi)
\label {eigenv}
\enr
For $K$ a (positive) constant and $f(x)$ a function which is non zero at
$x=\pm 1$, the Hamiltonian in \er {pairHamiltonian} can be taken as a model
that yields the postulated {\it ci} pairs, since the two eigenvalues coincide just
at the points given by \er {xyroots} or \er {qroots}. There may be more general
models that give the same two {\it ci}'s. (Note, however, that if $f(x)$ had a
zero at $x=\pm 1$, the degeneracy of energies would not be conical.)
    We now make the above model more specialized and show that
different values of the Berry phase can be obtained for different choices of $
f(x)$. For definiteness we consider specific molecular situations, but these are
just instances of wider categories. (The notation of Herzberg \cite{Herzberg} is used.)

\subsubsection {$1A_1 $  and $2A_2$ states in $C_{2v}$ symmetry
(Exemplified by $ ^1 A_1 ^{(1)} $ and   $ ^1 A_1 ^{ (2)} $ in bent $HCH$.)}

If the $x$ coordinate represents a mode displacement that transforms as
$B_1$ (e.g., an asymmetric stretch of $CH$) and $y$ transforms as $A_1$  (a
flapping motion of the $H_2$, this coordinate being the same as $y$ in figure 169
of \cite {Herzberg}), then $f(x)$  in \er{pairHamiltonian} can be taken as a
constant.  Without loss of generality we put for this case $f(x) = 1$ and find
that  cycling adiabatically counterclockwise around the ${\it ci}$ that is at
$(-1,0)$
induces (in the component that is unity at $\phi=0$) a topological phase of $\pi$, and
that around $(1,0)$ yields $-\pi$. Cycling either fully inside or outside $q=1$ (the latter
case encircling both {\it ci}'s), gives zero phase. We now describe a
"continuous (phase-) tracing method" that obtains in an unambiguous way the
phase of a {\it real} wave function. The alternative, "adiabatic cycling" method
of the previous section gave the same phase change in terms of  the evolution
of the {\it complex} solution of the time dependent {\SE} in the extremely slow
(adiabatic) limit. Other methods will be briefly referred to.

\subsection {Continuous Tracing of the Component Phase}
\label{Continuous Tracing of the Component Phase}

 In this method one notes that
real-valued solutions of the time independent Hamiltonian of a $2 \times 2$  matrix
form can be written in terms of an $ \theta (\phi,q)$, which is twice the "mixing
angle", such that the electronic component which is "initially" $1$ is
$\cos[\theta (\phi,q)/2]$, while that which is initially $0$ is
$\sin [\theta (\phi,q)/2]$. For the second matrix form in \er{pairHamiltonian} (in which, for
simplicity $f(x)=1$), we get
\beq
\theta (\phi,q)= \arctan \frac {q \sin \phi}{q^2 \cos^2 \phi \ - 1}
\label {theta}
\enq
One can trace the continuous evolution of $\theta$  (or of $\frac{\theta}{2}$) as
$\phi$  describes the circle $q=constant$. This will yield the topological phase
(as well as intermediate, open-path phase during the circling). We illustrate
this in the next two figures for the case $q>1$
(encircling the ${\it ci}$'s).
 \begin{figure}
\vspace{6cm}
\includegraphics{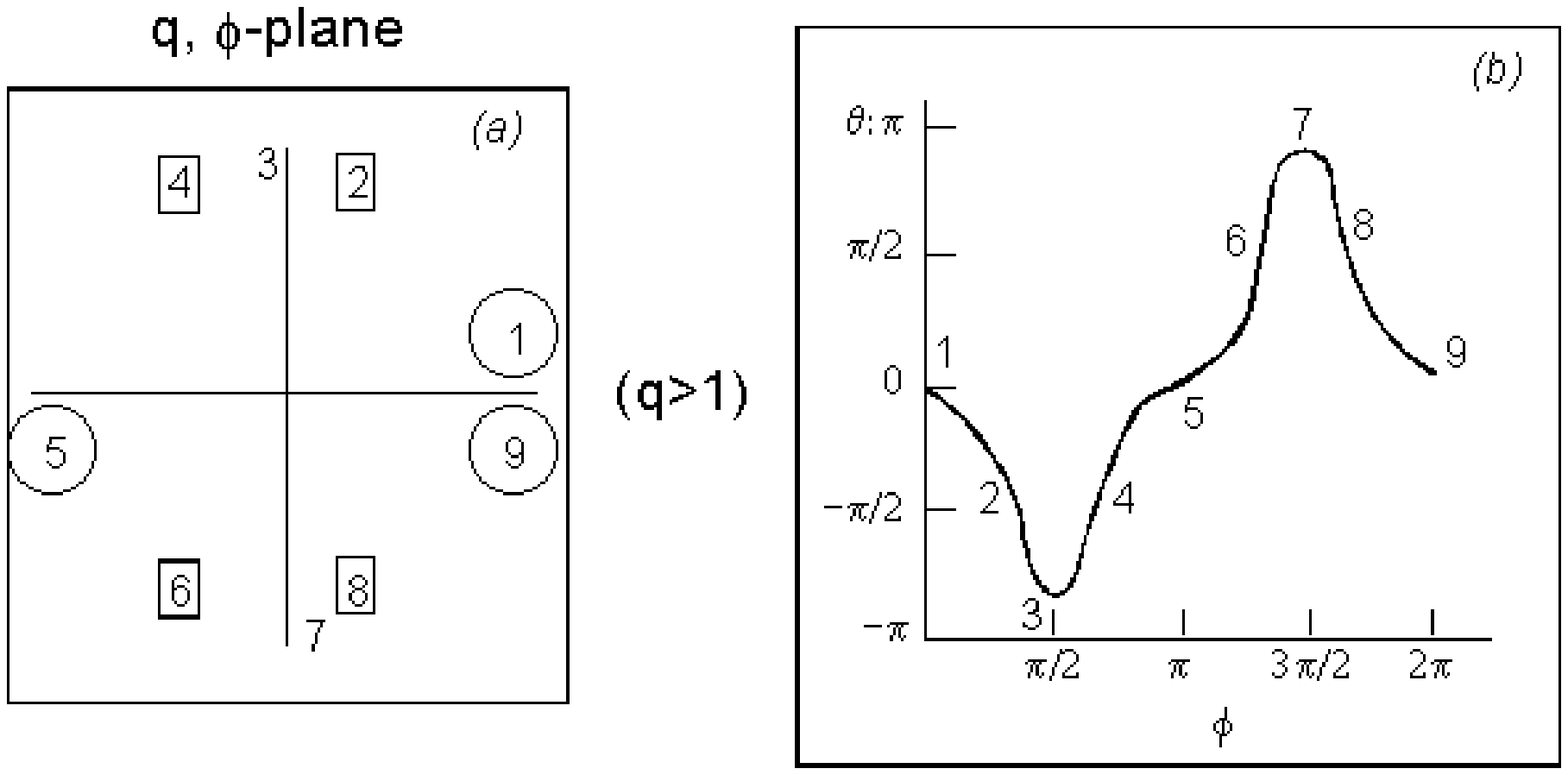}
\caption {Phase-tracing for the case of $1A_1$ and $2A_2$ states in $C_{2v}$
symmetry:
(a) The left-hand side shows the labels for the significant stages during the
circling in the $(q, \phi)$-plane. In this and the following figures, numbers
in circles represent the positions of zeros in the argument of the arctan in
the expression of the angle, (\er{theta}), numbers in squares are poles and
free numbers are other significant stages in the circling.
(b) The angle $\theta $  in \er{theta} as function of the circling angle. The
numbers correspond to those on the adjacent part (a) of the figure.
 (Note: The angle $\theta $ is defined as twice the transformation or mixing
angle.) The circling is with $q>1$, namely outside the {\it ci} pair.}
\label {fig:4.1}
\end{figure}
\begin{figure}
\vspace{14cm}
\includegraphics{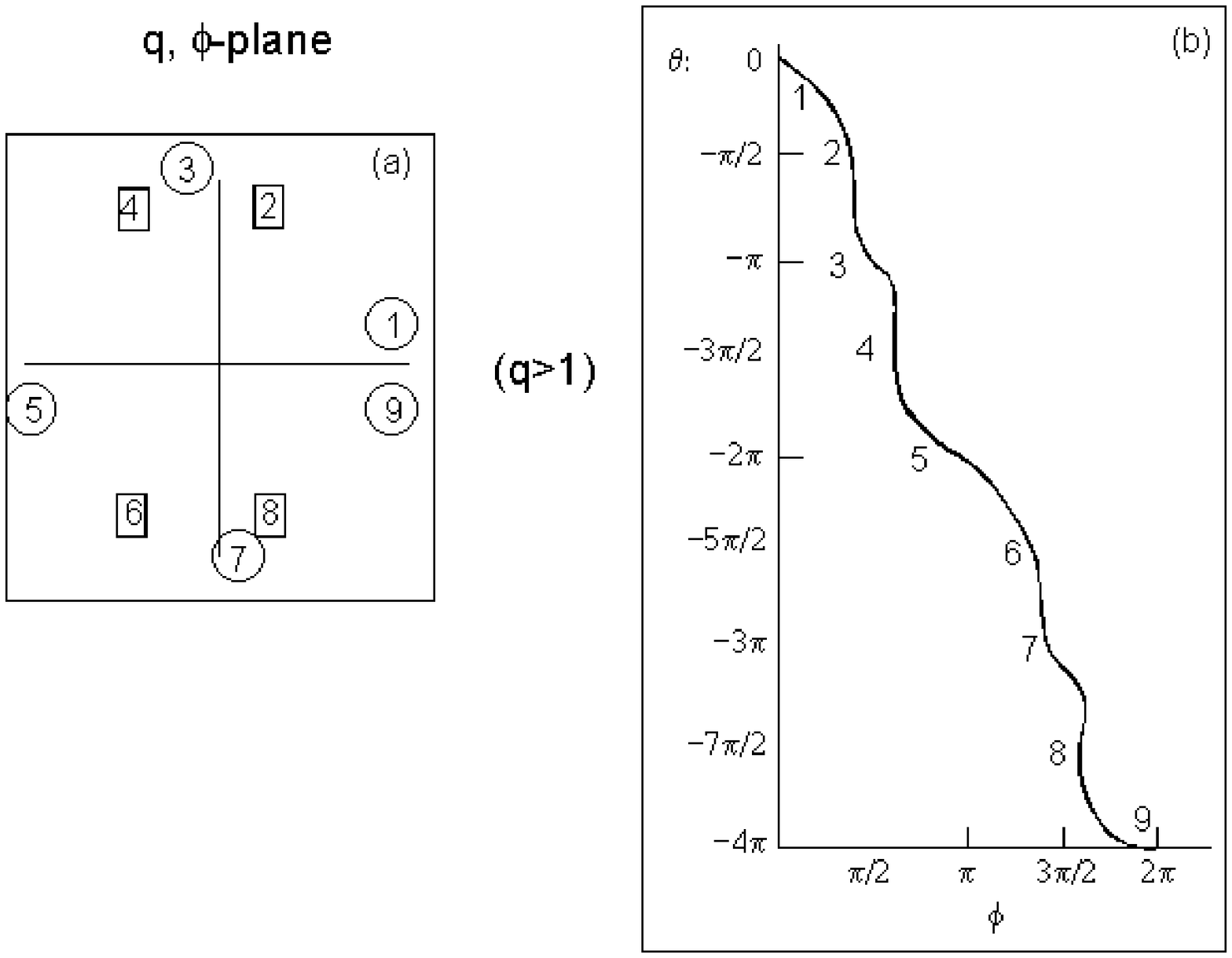}
\caption {Phase-tracing for circling outside the {\it ci} pair (for the model in
$A_1$ and $B_2$  states in $C_{2v}$  symmetry. The Berry phase (half the angle shown at
 the extremity of the figure) is here $-2\pi$.}
\label {fig:4.2}
\end{figure}
\begin{figure}
\vspace{7cm}
\includegraphics{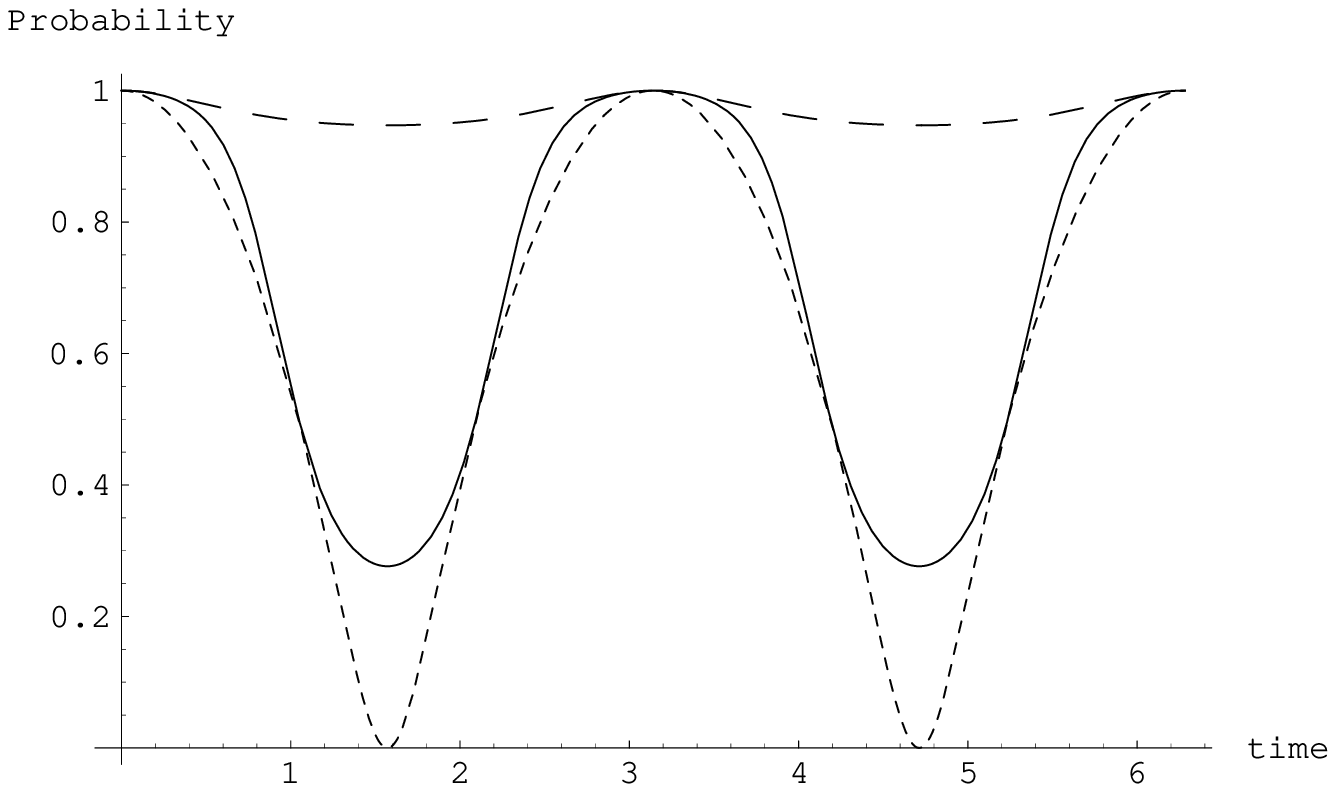}
\caption {Probabilities in different models during adiabatic circling around {\it
ci}'s. The square moduli of component amplitudes as function of time are seen
to be different for different models. Long-dashed lines: the model in
$1A_1 $  and $2A_2$ states in $C_{2v}$ symmetry for circling inside the ci's.
Full lines: the model in $1A_1 $  and $2A_2$ states in $C_{2v}$ symmetry
for circling outside the ci's. Broken lines: the model in
$A_1$ and $B_2$  states in $C_{2v}$  symmetry (with the "$xy$" off-diagonal matrix element)
 for circling outside the {\it ci}'s.
In the latter model, circling inside the {\it ci}'s gives probabilities that would
be indistinguishable from unity on the figure (and are not shown).}
\label {fig:4.3}
\end{figure}

In the sub-figure (\ref{fig:4.1}.a) several important stages in the circling are labeled with
Arabic numerals. In the adjacent sub-figure (\ref{fig:4.1}.b) the values of $ \theta
(\phi,q)$ are plotted as $\phi$  increases continuously. The labeled points in
the two sub-figures correspond to each other. (The notation is that points
which represent zeros of $ \tan \theta$ are marked with numbers surrounded
by small circles, those which represent poles are marked by numbers placed
inside squares, other points of interests that are neither zero nor poles are
labeled by free numbers.)  The zero value of the topological phase
($\frac{\theta }{2}$) arises from the fact that at the point $2$  (at which ${\it \phi}$
=$ \frac{\pi}{2} $), $\theta$ retraces its values, rather than goes on to decrease.

\subsubsection {$A_1$ and $B_2$  states in $C_{2v}$  symmetry
(Exemplified by $^2 A_1$  and $^2 B_2$  in $AlH_2$  \cite{ChabanGY})}

Symmetry considerations forbid any non-zero off-diagonal matrix
elements in \er{pairHamiltonian} when $f(x)$  is even in $x$, but they can be
non-zero if $f(x)$  is odd, e.g., $f(x)=x$. (We note  that $x$  itself transforms
as $B_2$ \cite {Herzberg}.)
Figure \ref{fig:4.2} shows the outcome for the phase by the continuous phase
tracing method for cycling outside the {\it ci}'s ($q > 1$). The difference
between the present case and the previous one (in which $f(x)$ was an even
function of $x$) is that now, in the second half of circling in the  $q,\ \phi$ -
plane, the wave-function component angle $\theta$  does not retrace its path,
but goes on decreasing.
(It is interesting to remark here on an analogy between the present
results and the well known results of contour integration in the complex $z$ -
plane. An integration of $(z^2 - 1)^{-1} $ over a path that encircles the two
poles of the function gives zero result, but the same path integration of $ z
(z^2 - 1)^{-1} $, gives $2\pi i$. However, the analogy does not work fully.
Thus, a simple multiplication of the integrand by a positive constant alters the
residues, but not the phase.)

However, the resulting Berry phase of $-2 \pi $ depends on (i) having
reached the adiabatic limit and (ii) circling well away from the {\it ci}'s; i.e., it is
necessary that the circling shall be done with a value of $q$ that is either
much smaller or much larger than $1$. A contrary case not satisfying these
conditions, e.g., when either $q < 3$ or $ K < 60$ , would give a Berry phase
of about $4\pi $, $6\pi $, $\ldots$, or a number  $N \approx 2,3,\ldots$ rather than $1$,
 as might have been expected. What is perhaps remarkable is that even in the not quite
adiabatic or not very large $q$ cases, $N$ (though plainly different from $1$) is still
close to being an integer. More study may be needed on this result, especially
in view of the possibility of observable consequences of the value of $N$. The cases  of "$1A_1 $  and $2A_2$ states in $C_{2v}$
symmetry" and of "$A_1$ and $B_2$ states in $C_{2v}$ symmetry" are, of course,
inequivalent, since they arise from different
Hamiltonians. Their non equivalence results not only in different topological
phases (zero and $2 \pi$), but in different state occupation probabilities.
These are defined as the probabilities of the systems being in one of the
states $\chi_k$, of which the superposition in \er{Psi} is made up. In Figure
\ref{fig:4.3} we show these probabilities as functions of time for systems that differ by
their having different functions $f(x)$ in the off-diagonal positions of the
Hamiltonian. The differences in the probabilities are apparent.

\subsubsection {Trigonal  Degeneracies}
\label{Trigonal Degeneracies}

The simplest way to write down the 2-by-2 Hamiltonian for two states such
that its eigenvalues coincide at trigonally symmetric points in $(x,y)$- or
$(q,\phi)$ -plane is to consider the matrices of vibrational-electronic coupling  of
the $E \otimes \epsilon$ Jahn-Teller problem in a diabatic electronic state
representation. These have been constructed by B. Halperin, and listed in
Appendix IV of \cite {Englman72}, up to the third order in $q$. The first order
or linear coupling in the displacement  coordinates is of the well known form
(shown by the first term  in the Hamiltonian presented below) and yields the
familiar {\it ci}  at the origin, $q=0$. When one adds to this the quadratic
coupling, designated $I(E)$  in  section IV.3 (A) of the above reference and
quoted  below, one obtains three further, trigonally situated {\it ci}, namely at
either $ \phi =0, \pm \frac{2 \pi}{3} $, or $\phi = \pi , \pm \frac{4\pi}{3}$,
depending on whether the signs of the linear and quadratic couplings are the
same or opposite. The distance of the trigonal {\it ci}'s from the origin varies
with the relative magnitudes of the couplings: the higher the strength of the
quadratic term, the nearer the trigonal {\it ci} are to the center. This was, of
course, the physical basis of \cite {KoizumiBersuker}, in which a ground vibronic
{\it singlet} state for strong quadratic coupling was found.
The resulting Hamiltonian is of the form (to be compared with the two
matrices  in \er {pairHamiltonian} above):
\ber
H(x,y) &=& K \left(
\begin{array}{cc}
  -(x -2\kappa (x^2-y^2)) & y+4\kappa x y   \\
  y+4\kappa x y & (x -2\kappa (x^2-y^2))
\end{array} \right)
\nonumber \\
&=& K \left(
\begin{array}{cc}
  -(q \cos \phi  -2\kappa q^2 \cos 2 \phi) &
  q \sin \phi + 2\kappa q^2 \sin 2\phi   \\
  q \sin \phi + 2\kappa q^2 \sin 2\phi & (q \cos \phi  -2\kappa q^2 \cos 2 \phi)
\end{array} \right)
\nonumber \\
&=&  H(q, \phi)
\label {trigHamiltonian}
\enr
where $\kappa $  represents the ratio of the strength of the quadratic coupling
to the linear one. The trigonal {\it ci} 's are at a distance $ q=(2 \kappa)^{-1}$,
with angular positions as described above. Employing now the continuous
phase tracing method introduced in subsection \ref{Conical Intersection Pairs},
one again obtains the graphs for the mixing angle. There are now
three cases to consider, namely
(i) for cycling that encloses all four {\it ci}'s , ($q > (2 \kappa)^{-1}$) the
resulting phase acquired being now $2 \pi$ (shown in Fig. 1.5)
\begin{figure}
\vspace{10cm}
\includegraphics{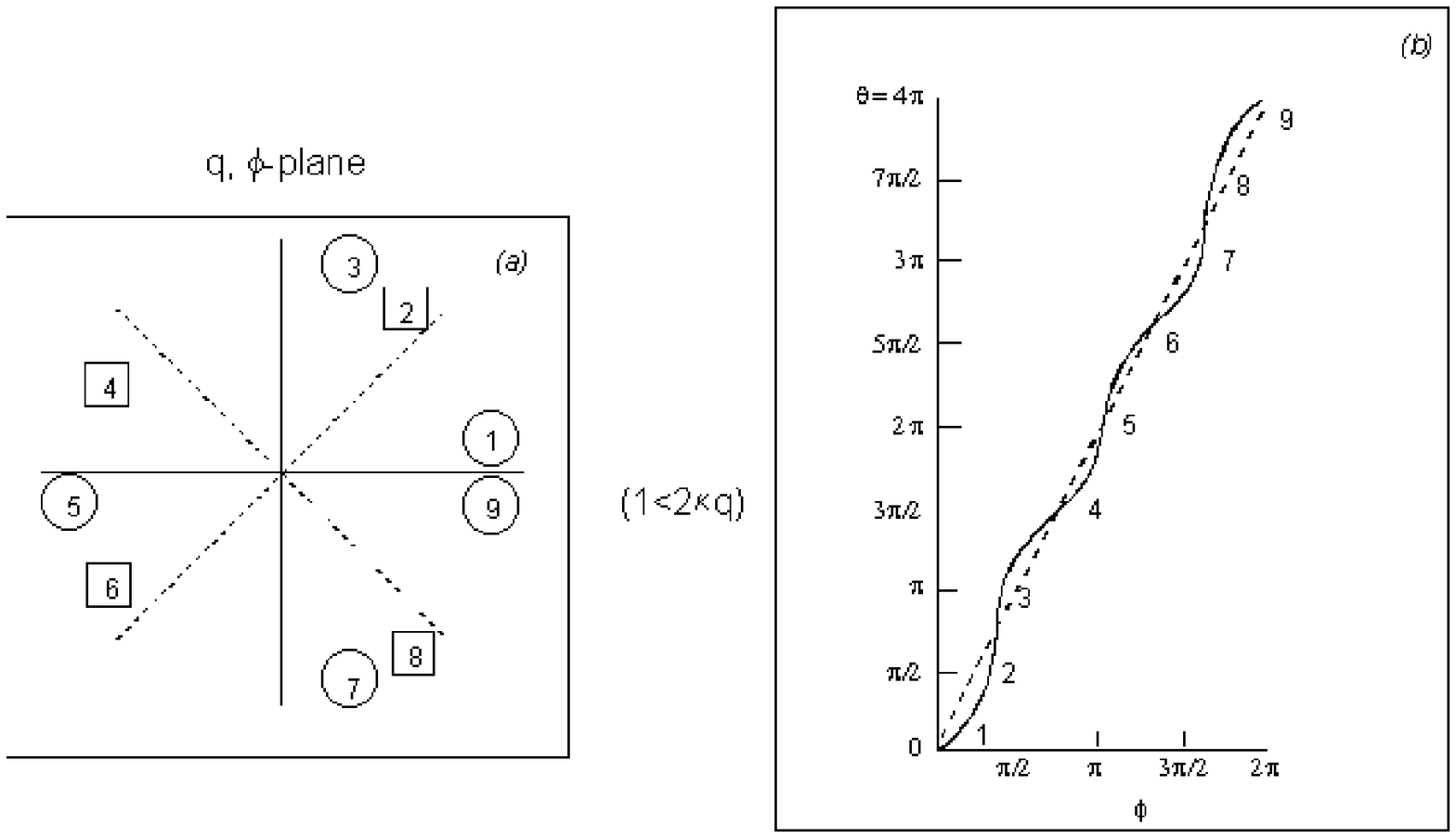}
\label {fig:4.4}
\caption {Phase-tracing for the case of trigonal degeneracies when
 the circle encompasses all four {\it ci}'s and the Berry phase is $2\pi$.}
\end{figure}
This is an even multiple of $\pi$, as expected for four {\it ci}'s \cite{HerzbergL},
but differs from $4 \pi$ (or from zero). Then
(ii) for intermediate radius cycling  ($ (2 \kappa)^{-1} > q > (4 \kappa)^{-1}$)
(which is shown in Fig. \ref{fig:4.5})
\begin{figure}
\vspace{7cm}
\includegraphics{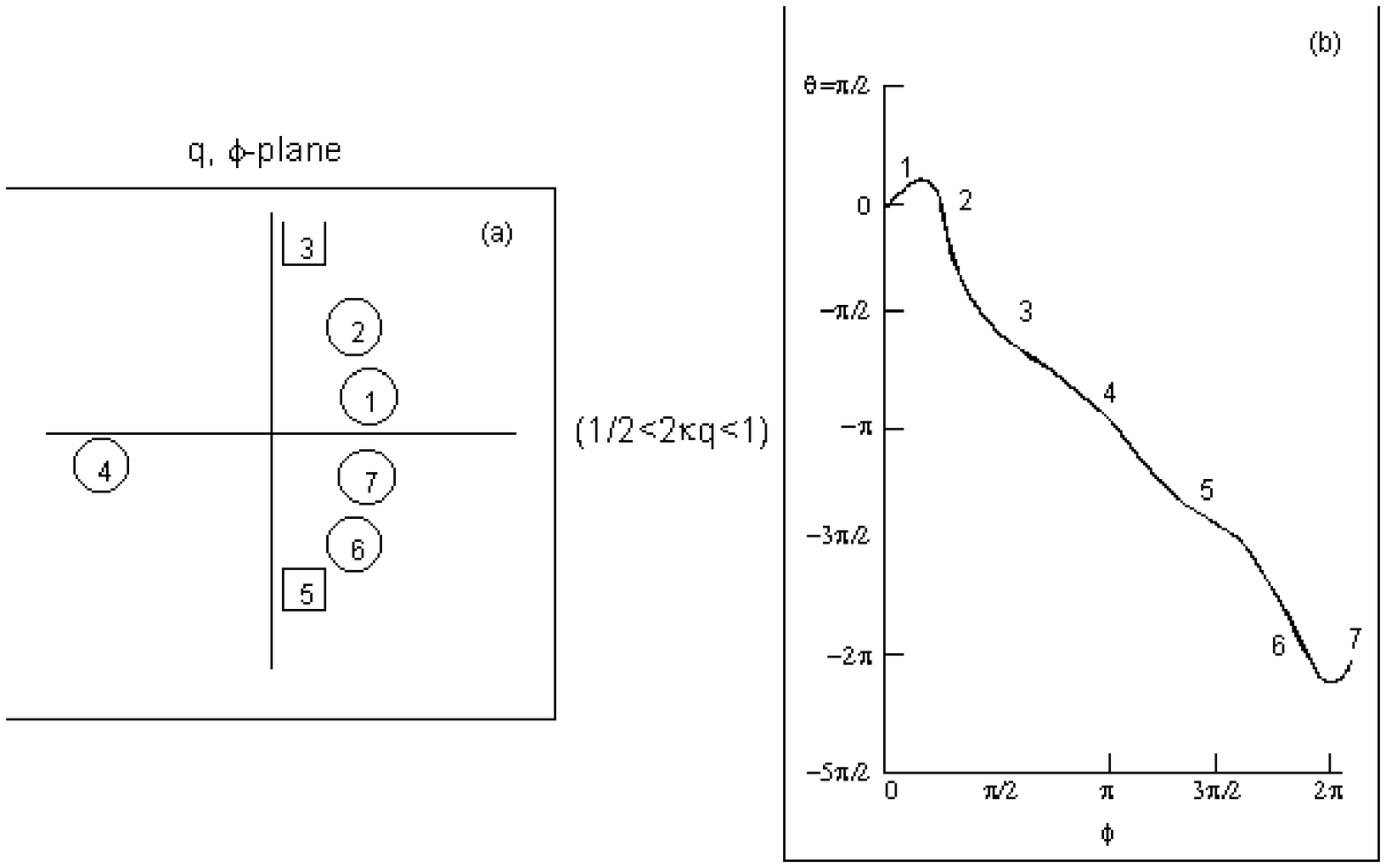}
\caption {Phase-tracing for the trigonal
degeneracies. The drawings (which are explained in the caption to
 figure \ref{fig:4.2}) are for intermediate-radius ($q$) circling.}
\label {fig:4.5}
\end{figure}
that terminates with a Berry phase of $-\pi$, and lastly
(iii) for small radius cycling $(q< (2 \kappa)^{-1})$. The last case has
also the Berry phase of $ \pi $ , but differs from the intermediate case
(ii), in that the initial increase is absent.

It might be asked what happens when one adds further couplings beyond the
quadratic one? In the next higher order one finds a scalar cubic term of the
form:
\beq
q^3 cos 3 \pi \bf I
\label {cubic}
\enq
where $\bf I $ is the unit matrix. This  gives rise to three trigonally aligned
degeneracies (\cite {Englman72}, Appendix  IV). However, these are parabolic
(touching) degeneracies, not conical intersections, and do not cause changes
of sign in the wave function upon circling round them. Higher order terms (not
listed in that Appendix) can give rise to additional {\it ci }'s of trigonal
symmetry, but the strength of these terms is expected to be less and therefore
the resulting {\it ci}'s  will be farther outside, where are without importance
for low lying states. Still, their presence is of interest for
revealing the connection between the Berry angle and the number of {\it ci}'s
circled and we shall presently obtain nonlinear coupling terms to an
arbitrarily high power of $q$.

\subsection {The Adiabatic to Diabatic Transformation (ADT)}

    Several years ago Baer proposed the use of a matrix $A$, that
transforms the adiabatic electronic set to a diabatic one \cite {Baer75}. (For
a special two-fold set this was discussed in \cite {HobeyM, McLachlan}.)
Computations performed with the diabatic set are much simpler than those
with the adiabatic set. Subject to certain conditions, $A$ is
 the solution of a set of first order partial differential equations.
 $A$ is unitary and has the form of a "path-ordered" phase factor, in
which the phase can be {\it formally} written as
\beq
\int_{\bf{R_0}}^{\bf R} \bf f^{IJ} (\bf R) \cdot d{\bf R}
\label {phangle}
\enq
Here the integrand is the off-diagonal gradient matrix element between adiabatic
electronic states,
\beq
\bf f^{IJ} (\bf R)= <I| \vec \nabla |J>
\label {momentum}
\enq
($\vec \nabla$ is the derivative with respect to $ \bf R$.)
We stress that in this formalism, $I$ and $J$ denote the complete adiabatic
electronic state, and not a component thereof. $|I>$  and $|J>$ contain
the nuclear coordinates, designated by ${\bf R}$, as parameters.
The above line integral was used and elaborated in calculations of nuclear
dynamics on potential surfaces by several authors (\cite{Yarkony99b,Yarkoni98},
(\cite {LepetitK} -\cite{CharutzBB}).
(For an extended
discussion of this and related matters the reviews of Sidis \cite{Sidis} and
Pacher et al. \cite {PCK} are especially informative.)

The possibility of a non-gradient component in the integrand introduces some
difficulty and an alternative formulation has been proposed
\cite{Yarkoni98,Yarkony99b}. (At positions that are close to a {\it ci}, the
alternative approximates well to the angle shown above in \er {phangle}.)

 The ADT, computed for $\it ci$ pairs in \cite {MebelBLin2} and denoted
(in their figures 1-3) by $\gamma (\phi |q)$, is related to the  "open path
phase" defined below in \er{gammak}, but is identical with it (and with Berry's
angle) only at $\phi = 2\pi$. At this value the computed results of \cite {MebelBLin2} are
in agreement with those that were derived with the model Hamiltonian in
\er{pairHamiltonian}. However, in some of the cases, when the coupling terms
became zero, the sign that the phase $\gamma (\phi |q)$ acquires might
become ambiguous  (e.g., whether it is even or odd under reflection about the
line $\phi = \pi $). In the above analytic models the signs are given
unambiguously.

Since up to date summaries about the practical implementation of the line
integral have been given recently (in \cite{MebelBL2001, BE1}; as also in  the
chapter by Baer in the present volume), and the method was applied also to a
pair of {\it ci}'s \cite {MebelBL2001}, we do not elaborate here on the form of
the phase associated with one or more {\it ci}'s, as obtained through this
method .

\subsection { Direct Integration}

The open path phase \cite {Pati, JainPati} associated with a component-
amplitude can be obtained as the imaginary part of an integral:
\beq
\gamma_k (t) = Im \ \int_0^t dt' \frac{\partial a_k (t')}{\partial t'}/ a_k (t')
\label {gammak}
\enq
where, as before at several places in this chapter, $a_k(t)$ is the amplitude of the
$k$-component in the solution of the time - dependent {\SE} in the near-
adiabatic limit. The (complex) amplitude in the integrand is (in general)
 non-vanishing (unlike the real wave-function amplitude in the strictly adiabatic
solution) and thus the integral is non-divergent. However, in practice, even fairly
close to the adiabatic limit, the convergence is very slow, due to oscillations in
the amplitude, noted in the previous section and in \cite {EYB1}-\cite {EYB3}.
For this reason, the formula in \er{gammak} is not a convenient one to use.
Still, using the formula for increasing values of the adiabaticity
parameter (i.e. increasing $K$  in \er{pairHamiltonian} to $K>10^2$), we have
evaluated the topological phase for the case with trigonal symmetry  and have
found it to converge close to the value $2N \pi$, with N=1 (and not 0).
Because of the diffiulties in its practical implementation, we shall
 not further consider the direct integration method.

\subsection { Higher Order Coupling in some Jahn-Teller and Renner-Teller
Effects}

A systematic derivation of forms of coupling that is super-linear in the
nuclear motion amplitude was given, partly based on Racah coefficients, in
\cite {Englman72}, Appendix IV, but these went up only as far as the third
order in the amplitudes $q$. As will shortly be made apparent, there is some
theoretical need to obtain higher order terms. For the $E \otimes \epsilon $
 Jahn-Teller case, the form of coupling to arbitrary powers was given in \cite
{ThompsonM}. Here we give a different and arguably simpler derivation using
the vector-coupling formalism of Appendix IV in \cite {Englman72}, the
complex representation form given in \cite {Griffith62}, and a mathematical
induction type of argument.

\subsubsection {Complex Representation}

The mode coordinates, transforming respectively as the $-1$ and $1$ components
of the $E$ (doubly degenerate) modes, have the form:
\beq
(g_{-1},g_{1})= -(i/\sqrt{2})(g_{\theta} ,g_{\epsilon})
\left(
\begin{array}{cc}
 -1 & 1 \\
 \ i & i
\end{array}
\right)=(i/\sqrt{2})( q e^{-i \phi}, - q e^{i \phi} )
\label {complexq}
\enq
where the extreme right-hand member recalls the "modulus" $q$ and the "phase"
 $\phi$ used in the
real representation. The vector coupling -coefficients for e.g. octahedral
symmetry,  can be
obtained from Table A.20 in the Appendix 2 of \cite {Griffith64}, upon
performing the transformation shown (\ref{complexq}) on the real representation.
In the following Table we show the Clebsch-Gordan coefficients defined in
\er {combination} below.

\begin{table}
  \centering
  \begin{tabular}{cccccccc}

  A: & B: &  & C: &$A_1$&$A_2$ &  E  &  \\
  E  & E  &  & -- & -- & -- & -- & -- \\
  a: & b: &  & c: &$a_1$ &$a_2$ & 1 & 1 \\
     &    &U:& -- & -- & -- & -- & -- \\
   1 &  1 &  &  & 0 & 0 & -i & 0 \\
   1 & -1 &  &  & $1/\sqrt{2}$ & $-i/\sqrt{2}$ & 0 & 0 \\
  -1 &  1 &  &  & $1/\sqrt{2}$ & $ i/\sqrt{2}$ & 0 & 0 \\
  -1 & -1 &  &  & 0 & 0 & 0 & 1 \\
\end{tabular}
\caption{The coupling- coefficients $U(A B C | a b c)$ for the complex form
of a doubly degenerate representation, following G.F. Koster {\it et al} "Properties
of the thirty-two point groups" (MIT Press, Cambridge, Mass., 1963), pp. 8, 52.}
\label{group}
\end{table}

To obtain non-linear coupling terms, we consider two linearly independent, not
identical  $E$-modes, namely
\ber
& & (g_{-1}, g_1) \\
& & (G_{-1}, G_1)
\label {gG}
\enr
and construct bilinear expressions from these. The combination that
transforms as components of an $E$-mode is given by
\beq
\chi(E|x)= \sum_{a,b} U(E E E|a b x) g_a G_b
\label{combination}
\enq
The above Table for U immediately shows that there are only two bilinear
combinations of $g$ and $G$, namely those $ a=1$, $b=1$ and $a=-1$, $b=-
1$. These lead to the quadratic terms belonging to the components $-1$ and
$1$:  $ \chi (E|-1)= g_1 G_1$ and   $ \chi (E|1)=g_{-1} G_{-1}$

Eq. A IV.4 of \cite {Englman72} tells us which ket-bra operator
$|d><e|$ is multiplied by the above combinations or, equivalently, where in the
$2 \times 2$
electronic- nuclear coupling matrix each of these terms sit. Here again we
adopt for the electronic kets a complex representation, analogous to that
shown in \er {complexq}. To use the vector coupling coefficients for these, we
recall that in the complex representation the bra's transform as the
corresponding  "{\it minus-label} ket"'s (cf. Eq. (2.34) in \cite {Griffith62}).
Using again the vector coupling coefficients, we see that $ \chi (E|-1)$ is the
factor that multiplies $|1><-1| $ (and $ \chi (E|1)$ is the factor that
multiplies $|-1><1|$). In the usual matrix notation [in which rows and
columns are taken in the order $(-1,1)$] this means that in the upper right
corner one has (for linear coupling ) $g_{-1}$ and $G_{-1}$, and (for quadratic
coupling),  $g_1 G_1$, and similarly in the lower left corner $g_1$, $G_1$ ,
as well as $g_{-1} G_{-1}$. Both linear and quadratic terms will be multiplied
by different constants, whose values depend on the physical situation and
cannot be given by symmetry considerations, except that the electronic-
nuclear interaction must be Hermitean and invariant under the symmetry
operations of the group.
The same construction can be employed to derive bilinear terms on the
diagonal part of the coupling matrix. Using again the $U$-coefficients in the
Table, one obtains the forms (not normalized)
\beq
(g_1 G_{-1}+ g_{-1} G_1)(|1><1| +|-1><-1|)
\label {A1diag}
\enq
where each factor belongs to the $A_1$ representation and
\beq
(g_1 G_{-1} - g_{-1} G_1)(|1><1| -|-1><-1|)
\label{A2diag}
\enq
where each factor belongs to the $A_2$ representation.

\subsubsection {Squaring of Off-diagonal Elements}

The method shown affords easy generalization to higher order coupling
in the important case where a single mode is engaged, i.e., $G_{\pm 1} =
 g_{\pm 1} = \pm \frac{1}{i \sqrt{2}} q e^{\pm i \phi}$.
Then the two off-diagonal terms derived above are, after physics based
constant coefficients have been affixed, in the upper right corner
\beq
(qe^{-i \phi} -2\kappa q^2 e^{2 i \phi}) |1> <-1|
 \label {offdiag}
\enq
with another, Hermitean conjugate expression on the other (lower left)
 off-diagonal position. These were previously given in a similar form in, e.g.,
\cite{ZwanzigerG, Englman72}. The $ A_1$  term in \er {A1diag} only
renormalizes the vibrational frequency. The $A_2$ term vanishes (for terms
up to second order in $q^2$).
Proceeding in the same way to get further terms by cross multiplying
the second order expression in  \er{offdiag}, and continuing the procedure, we
obtain in the upper right corner the following terms correct up to the fourth
order in $q$:
\beq
 q^3 e^{-i \phi}, q^4 e^{2i \phi}, q^4 e^{-4i \phi}
\label {newterms}
\enq
The first and second terms contain phase factors identical to those previously
met with in \er {offdiag}. The last term has the "new" phase factor
$e^{-4i\phi}$
(Though the power of $q$ in the second term is different from that in
 \er{offdiag}, this term enters with a physics-based coefficient that is independent
of $\kappa $ in \er{offdiag}, and can be taken for the present illustration as
zero. The full expression is shown below in \er{offdiag2} and the implications
of higher powers of $q$ are discussed thereafter.) Then a new off-diagonal
matrix element enlarged with the third term only,  multiplied by a (new)
coefficient $\lambda$, is
\beq
 (q e^{-i \phi} - 2\kappa q^2 e^{2 i \phi} + \lambda q^4 e^{-4 i \phi}) =
q e^{-i \phi} (1- 2\kappa q e^{3 i \phi}+ \lambda q^3 e^{-3 i \phi})
\label {offdiag1}
\enq
There is going to be an $A_1$ (scalar) term of the form, well known in the
literature (e.g., \cite{Englman72}), $q^3 \cos 3\phi$, and an $A_2$ (pseudo-
scalar) term of the form $q^3 \sin 3 \phi$.  We may once again suppose that the
coefficients of all these terms are independent (i.e., their physical origins are
different) and that we may discuss terms in diagonal and off-diagonal
positions separately.
 Let us consider the off-diagonal term, as given on the right hand member
 of \er{offdiag1}.
The vanishing of the first factor gives the traditional conical intersection ({\it ci})
at the origin. The zeros of the second factor give additional {\it ci}'s. These are
all trigonally positioned, due to the phase factors $e^{\pm 3i \phi}$, which
induce trigonal symmetry. The maximum number of trigonal {\it ci}'s (to this,
fourth order approximation in q) is clearly $3 \times 3=9$. Thus, to give a numerical
example in which $\kappa=0.15$, $\lambda= 0.003$, we obtain the following nine trigonal roots
of \er{offdiag1}
\ber
 &q=3.95, &\phi=0, \frac{2\pi }{3} , \frac{4\pi }{3}\nonumber\\
 &q=7.42, &\phi=0, \frac{2\pi}{3} , \frac{4\pi}{3}  \nonumber \\
 &q=11.37, &\phi=\pi , \frac{\pi}{3}, \frac{5\pi}{3}
\label {trigroots}
\enr
(Clearly the pseudo-scalar term vanishes at these points; so the {\it ci}
character at the roots is maintained, no matter whether there are or are not
$A_2$ terms. Also the vanishing of $A_2$ terms will not lead to new {\it ci}'s.)
On the other hand, by circling over a large radius path $q \rightarrow \infty$, so that all
{\it ci}'s are enclosed, the dominant term in \er {offdiag1} is the last one and
the acquired  Berry- phase is $- 4(2\pi )/2= - 4 \pi$.

To see that this phase has no relation to the number of {\it ci}'s
encircled (if this statement is not already obvious), we note that this last
result is true no matter what are the values of the coefficients $\kappa $ ,
$\lambda$, etc., provided only that the latter is non-zero. In contrast, the
number of {\it ci}'s depends on their values; e.g. for some values of the
parameters the vanishing of the off-diagonal matrix elements occurs for
complex values of $q$, and these do not represent physical {\it ci}'s. The
model used  in  \cite {Yarkony99a} represents a special case, in which it was
possible to derive a relation between the number of {\it ci}'s and the Berry-
phase acquired upon circling about them. We are concerned with more
general situations. For these it is not warranted, for example, to count up the
total number of {\it ci}'s by circling with a large radius.

\subsubsection {General Off-diagonal Coupling}

The construction given above to obtain off-diagonal non-linear
couplings up to order $q^4$ can be generalized to arbitrary order. Only the
final result is given. This gives for the off diagonal term in the upper right
corner:
\beq
K q e^{-i \phi} [1 + q^{-2} \sum_{m=1,\ldots}q^{3m} Q_{m+} e^{3 m i \phi}
+  \sum_{m=1,\ldots}q^{3m} Q_{m-} e^{-3 m i \phi}]
\label{offdiag2}
\enq
where $Q_{m+}$ and $Q_{m-}$ are polynomials in $q^2$  with coefficients
that depend on the physical system and whose leading terms will be $q^0$ .
When transformed back
to the real representation, by applying the inverse of the transformation in \er
{complexq}, one regains the expressions of \cite {ThompsonM}.
   Normally, for stable physical systems, it is expected that, with increasing
$m$ , $Q_{m+}$   and $Q_{m-}$  will numerically decrease and so will, in
each polynomial, the coefficients of successively higher powers $ q^2$. If we
assume only a finite number of summands in the above sums and that the
highest power of $q$  in \er {offdiag2} has the phase factor
$e^{3Mi\phi}$ (where $M$ is a positive or negative integer), then the path  along a very large
circle will add a topological phase of $(3M -1)\pi $. In general, $3|M|$  is
different (either smaller or larger) than the number of {\it ci}'s enclosed by the
large contour, though it equals the number of {\it ci}'s for the case $ M=+1$
treated in \cite {Yarkony99a}. When there are two or more different phase
factors with the same highest power of $q$, then the amount of topological
phase is not simply given, but can be determined, using the continuous
phase-tracing method described above in subsection
\ref{Continuous Tracing of the Component Phase}.

\subsubsection {Nonlinear Diagonal Elements}

Their forms are:
\ber
&A_1:& \sum_{m=0,\ldots} q^{3m} D_{1,m} (q) \cos 3 m \phi
\\
&A_2:& \sum_{m=0,\ldots} q^{3m} D_{2,m} (q) \sin 3 m \phi
\label{diag2}
\enr
where $D_{1,m}$  and $D_{2,m}$ are polynomials in $q^2$ with coefficients
that again depend on the physical system and whose leading terms is $q^0$.
The scalar term evidently does not produce a {\it ci}. The zeros of the $A_2$ term
(which is applicable for systems not invariant under time reversal)
by themselves do not add to or subtract from the {\it ci}'s.

\subsubsection {Generalized Renner-Teller Coupling}

The foregoing formulae in \eqs (\ref{offdiag2}, \ref{diag2}) can be immediately applied to
two physically interesting situations (not treated in \cite {ThompsonM}, but
very recently considered for a special model in \cite {BevilacquaMP}). The first
is the vibronic interaction in a system having inversion symmetry between a
doubly degenerate electronic state and an {\it odd} vibrational mode. The
second situation is the more common one of Renner-Teller coupling (e.g., a
linear molecule whose doubly degenerate orbital is coupled to a bending-type
distortion) \cite {BersukerPolinger}, formally identical to the previous. To write out the
coupling to any order, one simply removes in the previous formulae all terms
having odd powers of $q$. In the {\it real} representation, the coupling matrix
correct to the fourth harmonics in the angular coordinate has the following
form:
\beq
\left(
\begin{array}{cc}
  R_1  q^2 \cos 2\phi + R_2 q^4 \cos 4 \phi + \ldots &
   R_1  q^2 \sin 2 \phi - R_2 q^4 \sin 4 \phi + \ldots \\
 R_1  q^2 \sin 2 \phi - R_2 q^4 \sin 4 \phi + \ldots &
  -(R_1 q^2 \cos 2\phi + R_2 q^4 \cos 4 \phi + \ldots)
\end{array}
\right)
\label{RennerT}
\enq
where, as in the instances of \eqs (\ref{offdiag2},\ref{diag2}) above, $R_1$  and $R_2$
are polynomials in $q^2$ with coefficients that again depend on the physical
system and whose leading terms are of order $q^0$.

\subsubsection {Interpretation}

The key of constructing vibronic coupling terms for doubly degenerate states
and modes to an arbitrary order is the use of  a complex representation. The
formal essence of the method is that in the complex representation
$V(EEE|xyz)$ is nonzero only for a single $z$. (In Table \ref{group} there is only one
entry in a row. Figuratively speaking: All coupled "coaches" travel to a unique
"train –station" and all trains in that station consist of coupled coaches.
Moreover, this goes also for the coupling of coupled train, and so on.) From
our result we conclude that the Berry phase around more than one conical
intersection is not uniquely given by the number of conical intersections
enclosed, but is model dependent. This has consequences for experimental
tracing of the phase, as well as for computations of line integrals with the
purpose of obtaining non-adiabatic surface jumping in chemical
rearrangement processes (e.g., in \cite {DixonHYHLY} -\cite {BillingK}, \cite
{BaerCKB, CharutzBB}) and as discussed in section
\ref{Some Aspects of Phases in Molecules}.

\subsection {Experimental Phase Probing}

Experimental observation of topological phases is difficult, for one  reason
(among others) that the dynamic-phase part (which we have subtracted off in
our formalism, but is present in any real situation) oscillates in general
 much faster than the topological phase
and tends to dominate the amplitude behavior (\cite {BitterD}-
\cite{FuentesBV}). Several researches have addressed this difficulty, in
particular, by neutron-interferometric methods, which also can yield the
 open-path phase \cite {WaghRFI}, though only under restricted
 conditions \cite{Sjoquist}.

The continuous tracing method and other methods for cycling reviewed in this
section can be used in several very different areas. An example is a
mesoscopic system composed of quantum dots that is connected to several
capacitors. For this  a network of singularities was described in the parameter
space of the gate voltages \cite {PothierLUED}. It has been suggested that the
outcome of circling around these singularities, through a phased alteration of
the charges on the capacitors, is formally similar to that of circling around {\it
ci}'s \cite{Levine}. Although the physical effects are different (i.e., the
acquisition of a $\pi$- phase by the wave function has the effect of
transferring a single electronic charge), the results of circling obtained in this
section can be associated with quantized charges passing between quantum
dots. Some related topics, for which the results of this section can be used or
extended are: Phase behavior in a different type of multiple ci's, located in a
single point but common to several states. This  was studied in
\cite{ManolopoulosC} and for an electronic quartet state in \cite {BEV,BE2,YE1}. A
further future extension of the theory is to try to correlate the topological
phase with a general (representation-independent) property of the system (or
of the Hamiltonian).

The phases studied in the present work are those of material, Schr\"{o}dinger
waves, rather than of electromagnetic, light waves. Recently, it has been
shown that it is possible to freeze coherent information (=phases) from light
into material degrees of freedom and {\it vice versa} \cite {LiuDBH,
PhillipsFMWL}. This development extends the relevance of this section to
light, too. Among fields of application not directly addressed in their recent
work, let us quote from the authors of \cite{LiuDBH} quantum information transfer
\cite{DivicenzoT} and Bose-Einstein condensates.

\section {Molecular Yang-Mills Fields}

\subsection {A Nuclear Lagrangean}

One starts with the Hamiltonian for a molecule
$H({\bf r},{\bf R})$ written out in terms of the electronic coordinates $(\bf
r )$ and the nuclear displacement coordinates  ($ \bf R$, this
being a vector whose dimensionality is 3 times the number of
nuclei) and containing the interaction potential $V(\br,\bR)$. Then, following the Born-Oppenheimer scheme one can write
the combined wave-function $\Psi (\bf r ,\bf R) $ as a sum of an infinite
number of terms
\beq
\Psi ({\bf r} ,{\bf R}) =
\sum_k \zeta _k ({\bf r} ,{\bf R}) \chi_k ({\bf R})
\label {Psimol}
\enq
Here the first factor, $ \zeta_k ({\bf r} ,{\bf R})$
 in the sum is one of the solutions of the electronic
Born-Oppenheimer equation and its partner in the sum,
$\chi_k({\bf R})$ is the solution of the following equation for the nuclear
motion, with total eigenvalue $E_{k}$ :
\beq
\{-\frac{1}{2M} \partial_b \partial^b \delta^k_m  + V_m^k ({\bf R}) -
\frac{1}{M} \tau^k_{bm}({\bf R}) \partial^b
+ \frac{1}{2M} \tau^k_{bn}({\bf R})\tau^{b \ n}_m({\bf R}) \}\chi^m ({\bf R}) = E_{k} \chi^k ({\bf R})
\label{nuclearSE}
\enq
The symbol $M$ represents the masses of the nuclei in the molecule, which
for simplicity are taken to be equal.  $\delta ^k_m$ is the Kronecker delta.
The tensor notation is used in this section and the summation convention is
 assumed for all repeated indices not placed in parentheses.
In \er {nuclearSE} appears the "non-adiabatic coupling term" (NACT)
$\tau^k_{bm}$  (this being a matrix in the electronic Hilbert space, whose
components are denoted by labels $k,m$, and a "vector" with respect to the
$b$-component of the nuclear coordinate ${\bf R}$). It is given by an integral
over the electron coordinates:
\beq
\tau^k_{bm}({\bf R}) = \int d {\bf r} \zeta_k(\br,\bR) \partial_b \zeta_m(\br,\bR) :=
<k|\partial_b|m>= - <m|\partial_b|k>
\label {tau1}
\enq
The effective potential matrix for nuclear motion, which is a diagonal matrix for
the adiabatic electronic set, is given by:
\beq
V^k_m (\bR) =<k|V(\br,\bR)|m>
\label {potential1}
\enq
In the algebraic, group theoretical treatments of non-Abelian systems
(\cite{YM}, \cite{Jackiw} - \cite{ Weinberg}, \cite{MoodySW1}-\cite{Zygelman2}) the
NACT is usually written in a decomposed form as
\beq
\tau^k_{bm}({\bf R}) = d^{(b)}(\bR) (t_b)^k_m
\label {tau2}
\enq
where $ t_b$  is one of the set of constant (non-commuting) matrices (the
"generators") that define the Lie-group of the system. So far, with the
summation in \er{Psimol} over $k$ running over the full electronic Hilbert space
spanned by  $\zeta _k ({\bf r} ,{\bf R})$, the Hamiltonian treatment is exact.
We shall shortly see
that the truncation of the summation in \er{Psimol} (which in practice is
almost inevitable) has far-reaching effects in the Yang-Mills theory.
 Before that, we turn to an equivalent description, standardly used in
field theories but which has not been in use for the Born-Oppenheimer
treatment of molecules, namely  to write down a "nuclear" Lagrangean density
$\mathcal{L}_M$  for the vector $\bf {\psi} (\bf R)$  whose (transposed) row vector
form is
\beq
{\bf \psi}^T = (\chi_1, \chi_2, \chi_3,\ldots, \chi_N)*
\label {PsiT}
\enq
(The mixed, $\psi$ - $\chi$ notation here has historic causes.) The {\SE} is obtained
from the nuclear Lagrangean by functionally deriving the latter with respect to
$\bf {\psi} $. To get the {\it exact} form of the {\SE}, we must let $N$ in the
previous equation to be equal to the dimension of the electronic Hilbert space
(namely, infinite), but we shall soon come to study approximations in which
$N$ is finite and even small, e.g., $2$ or $3$. The appropriate nuclear Lagrangean density
is for an arbitrary electronic states:
\ber
{\cal L}_M (\psi, \partial_a \psi ) &=&
(2M)^{-1}(\partial^a \psi)^k (\partial_a \psi)_k -
M^{-1} \psi^k \tau_{ka}^{\ \ m} (\partial^a \psi )_m
\nonumber \\
&-& (2M)^{-1}(\psi)^k \tau_{kb}^{\ \ m} \tau_m^{\ bn} (\psi)_n
 - \psi^k V^m_k \psi_m
\label {NucLagrangean}
\enr
The non-Abelian nature of the formalism is apparent from the presence of
 non-diagonal matrices $\tau$ and $V$. $V$ can be diagonalized, leading to
adiabatic energy surfaces and states, but not simultaneously with the $(\tau
\partial )$ term.
Requiring now only {\it global} gauge invariance of the Lagrangean, we
obtain the usual phase-gauge theories \cite{MeadTruhlar79, Mead92},
incorporating a vector potential. However, requiring invariance under a {\it local}
gauge transformation we obtain the extension of the vector potential to a
Yang-Mills field \cite {YM, Jackiw}. [Actually, the local gauge invariance is not
a "luxury" because, if the Lagrangean is invariant under global
(constant) transformation, then it is also invariant under a gauge
transformation with general position dependent parameters (\cite {Weinberg},
section 15.2). A remark on nomenclature:"field" and "fields" are used
interchangeably.] Before obtaining the equation for the field, we return for a
moment to the (simpler) Abelian case.

\subsection {Pure {\it versus} Tensorial Gauge Fields }

To start, it is useful to put the previous result in a more elementary
setting, familiar in the context of electromagnetic force between charged
particles, say electrons. Thus, we  recapitulate as follows:

In an Abelian theory (for which $\Psi (\br ,\bR)$ in \er {Psimol} is a
scalar rather than a vector function, $N=1$), the introduction of a gauge field
$g(\bR)$  means pre-multiplication of the wave function $\chi (\bR) $ by
$\exp (i g \bR)$, where $g(\bR) $ is a scalar. This allows the definition of a
"gauge"-vector potential, in natural units:
\beq
A_a  = \partial_a g
\label {vecpot}
\enq
and if we define a field intensity tensor, as in electromagnetism,
by:
\beq
 F_{bc} = \partial_ b A_c   -  \partial_c A_b
\label {emfield}
\enq
 we find that $F_{bc}$ is zero, excluding singularities of $A_a$. Therefore a vector
potential arising from a gauge transformation $g$ does not give a true field
(since it can be transformed away by another gauge $-g$). Conversely, a
vector potential $ A_a$ for which $F_{bc}$ in \er{emfield} is not zero, gives a true
field and cannot be transformed away by a choice of gauge.

In a non-Abelian theory  (where the Hamiltonian contains non-commuting
matrices and the solutions are vector or spinor functions , with
$N$ in \er {Psimol} greater than $1$ ) we also start with a vector potential
$A_b$. ( In the manner of \er{tau2} this can be decomposed into components
$A^a _b$ , in which he superscript  $^a$ labels the matrices in the theory).
We next define the field intensity tensor  through a "covariant curl" by
\beq
F^a_{bc} = \partial_b A^a_c   -  \partial_c A^a_b  + C^a_{de} A^d_b A^e_c
\label {Ymfield}
\enq
Here $C^a_{bc}$ are the structure constants for the Lie-group defined by the set of
the non-commuting matrices $t_a$ appearing in \er{tau2} and which also
appear both in the Lagrangean and in the {\SE}. We further define the
"covariant derivative" by
\beq
(D_a \psi)_k = (\partial_a  \psi)_k - i{A_a ^b} (t_b)_k ^m { \psi_m}
\label {covder}
\enq
and write the field equations for $A$ and $F$ as
\beq
\partial_a F_b^{\ ac} = F_d^{\ cf} C^d_{be} A^e_f +
i \frac{\delta {\cal L}_M (\psi, D \psi)}{\delta D_c \psi_k} (t_b)^m_n \psi_k
\label {f.e.}
\enq
If the vector potential components $A^a_b$ have the property that the
derived field intensity, the Yang-Mills field in \er{Ymfield},  is non zero,
 then the vector potential cannot be transformed away by a gauge phase
$g (\bf R)$ through pre-multiplication of the wavefunction $\chi^m (\bR)$ by the
(unitary) factor
 $\exp ( ig (\bR))$. There is no $g(\bR)$ that will do this. Conversely, if there is
a $g (\bf R)$, one obtains a vector potential-matrix $A_a$  whose $km$
components satisfy
\beq
A_{am} ^k  =(\exp (g ({\bf R}))^{-1})_n ^k  \partial_a  [\exp (i g(\bR))]^n _m
\label{Aakm}
\enq
Thus, the existence of a (matrix-type) phase $g$ represents the "pure-gauge
case" and  the non-existence of $g$ the non-pure, Yang-Mills field case,
which cannot be transformed away by a gauge.

\subsection {The "Curl Condition"}

We now return to the nuclear Born-Oppenheimer  \er{nuclearSE} in the molecular
context.
Consider the derivative coupling term in it, having the form:
\beq
M^{-1}\tau^k_{\ bm} (\bR) \partial^b \chi^m (\bR)
\label {derivcoup}
\enq
Suppose that we want this to be transformed away by a pure gauge factor
having the form
\beq
[\exp (ig(\bR))]^k_m  =[G(\bR)]^k_m
\label {transmatr1}
\enq
where g and G are matrices. That is, we require:
\beq
\tau^k_{\ bm} (\bR) = [G(\bR)^{-1}]^k_s \partial_b [G(\bR)]^s_m
\enq
for all $b$, or
\beq
[G(\bR)]^s_k \tau^k_{\ bm}(\bR) = \partial_b [G(\bR)]^s_m
\label{G.tau}
\enq
The consistency condition for this set of equations to possess a (unique)
solution is that the field intensity tensor defined in \er {Ymfield} is zero
\cite{Baer75}. This is  also known as the "curl condition"
 and is written in an abbreviated form as
\beq
curl \  \tau = -\tau \times \tau
\label {curlcondition}
\enq
Under circumstances that this condition holds, an Adiabatic Diabatic Transformation
(ADT) matrix A exists, such that the adiabatic electronic set can be transformed
to a  diabatic one. Working with this diabatic set, at least in some part
of the nuclear coordinate space, was the objective aimed at in \cite {Baer75}.

Starting from a completely different angle, namely the nuclear Lagrangean and
the requirement of local gauge invariance, we have shown in the previous subsection
that if the very same curl condition is satisfied, there is a pure gauge field.
 If it
is not satisfied, then the field is not a gauge field, but something more
complicated, namely the Yang-Mills field. The set of equations that
give the pure gauge $g$ is identical to that which yields the ADT matrix $A$
which  was introduced in \cite {Baer75}. The
equivalence between a (pure) gauge phase factor and the ADT matrix does
not seem to have been made in the literature before, though the conditionality
of a pure gauge on the satisfaction of the "curl-relation" was common
knowledge. (Indeed, they are regarded as tautologously the same.) The
reason for the omission may have been that, possibly, the ADT matrix was not
thought to have the respectability of the pure gauge. (From a naive, superficial
angle it is not evident, why one and the same condition should guarantee the
elimination of the cross term in the molecular {\SE}, which is a
non-relativistic, second order differential equation, and the possibility of a pure
gauge for a hadron field, which obeys entirely different equations: e.g.,
relativistic, first order ones.)

\subsection{The Untruncated Hilbert Space}

We now recall the remarkable result of \cite {Baer75} that if the adiabatic
electronic set in \er{Psimol} is complete ($N=\infty$), then the curl condition is
satisfied and  the Yang-Mills field is zero, except at points of singularity of the
vector potential. (An algebraic proof can be found in \cite {Baer75}, Appendix 1. An
alternative derivation, as well as an extension, is given below.)
Suppose now that we have a (pure) gauge $g (\bR)$, that satisfies the
following two conditions:

(1) the electronic set (represented in the following by Greek indexes) is {\it
complete}, and

(2) the vector potential-matrix $A$ present in the Hamiltonian (or
in the Lagrangean) arises from a dynamic coupling: meaning, that
it has the form
\beq
A_{a  \beta}^{\ \alpha} \propto
<\alpha|\partial_a| \beta>
\label {A2}
 \enq
 Then, two things (that
are actually interdependent) happen: (I) The field intensity $F=0$
(II) There exists a unique gauge $ g(\bR)$ and, since $F=0$,  any
apparent field in the Hamiltonian can be transformed away by
introducing a new gauge. If, however, the above condition (1) does
not hold, i.e., the electronic Hilbert space is truncated, then
$F$ is in general not zero within the truncated set, In this
event, the fields $A$ and $F$ cannot be nullified by a new gauge
and the resulting Yang-Mills field is true and irremovable.

[Attention is directed to a previous discussion of what happens
when the electronic basis is extended to the complete Hilbert space, \cite
{MoodySW2} p. 60; especially eqs. (2.17)-(2.18). It is shown there that
in that event the full symmetry of the invariance group is regained (in
 effect, through
the cancellation of the transformation matrix operating on the electronic and
on the nuclear functions spaces). From this result it is only a short step to
conclude that the Yang-Mills field coming from electron-nuclear coupling must
be zero for a full set. However, this conclusion is not drawn in the article, nor
is the vanishing of the Yang-Mills field shown explicitly.]

As was already noted in \cite {Berry84},
the primary effect of the Yang-Mills field is to induce transitions ($\zeta _m \to
\zeta_k$) between the nuclear states (and, perhaps, to cause finite life-times).
As already remarked, it is not easy to calculate the probabilities of transitions
due to the derivative coupling between the zero order nuclear states (if for no
other reason, then because these are not all mutually orthogonal). Efforts made in
this direction are successful only under special circumstances, e.g. the
Perturbed Stationary State method \cite {BatesMcCarroll,MottMassey} for slow atomic
collisions. This  difficulty is avoided when one follows Yang and Mills to derive a
mediating tensorial force that provide an alternative form of the interaction
between the zero order states and,  also, if one
introduces the ADT matrix to eliminate the derivative couplings.

\subsection{An Alternative Derivation}

The vanishing of the Yang-Mills field intensity tensor can be shown to follow
from the gauge transformation properties of the potential and the field. It is well known
(e.g., \cite {Jackiw} section II) that under a unitary transformation described by
the matrix
 \beq
U=U(\bR)
\label {unitary}
\enq
(which induces a rotation in the nuclear function space) the vector potential
transforms as:
\beq
A_a= A_a (\bR) \rightarrow U^{-1} A_a U + U^{-1} \partial_a U
\label{Atrans}
\enq
whereas the field intensity transforms covariantly, homogeneously
as:
\beq
F_{ab} \rightarrow U^{-1} F_{ab} U
\label{Ftrans}
\enq
If now there exists a representation in which $A_a$ is zero, then in this
representation $F_{ab}$ is also zero (by \er{Ymfield}). Now, in a $U$-transformed
representation (which can be chosen to be completely general), one finds that
\beq
A_a \rightarrow  U^{-1} \partial_a U
\enq
since the first term in \er{Atrans} is zero, but not the second. Thus $A_a$ is not
zero. However, the transformed $F_{ab}$ has no such inhomogeneous term (see
\er{Ftrans}) and therefore, in the transformed representation $F_{ab}=0$ and this is true in
all representations, i.e., generally true.
The crucial assumption was that  there is a representation in which $A_a$ are all
zero, and this holds in any diabatic representation (where the electronic
functions $\zeta_k(\br,\bR)$ are independent of $\bR$).
Then also derivative matrices, defined
in \er{tau1}, are zero and so are the potentials $A_a$ depending linearly on the
derivative matrices. On the other hand, the possibility of a diabatic set is
rigorously true only for a full electronic set. The existence of such a set is thus
a (sufficient) condition for the vanishing of the Yang-Mills field intensity tensor
$F_{ab}$.

\subsection {General Implications}

The foregoing indicate that there are three alternative ways to represent the
combined field in the degrees of freedom written as $\br,\bR$.
\begin{enumerate}
  \item By starting with a Lagrangean having the full symmetry, including that
under local gauge transformations, and solving for $\Psi(\br,\bR)$ (this being
a solution of the corresponding {\SE} in the variables $\br,\bR$).
 The solutions can then
be expanded as in \er{Psimol}, utilizing the full electronic set (the first factor on the
sum \er{Psimol}), or, for that matter, employing any other full electronic set.
  \item Projecting the nuclear solutions $\chi_k (\bR)$ on the Hilbert space of the electronic
states $\zeta_k(\br,\bR)$ and working in the projected Hilbert space of the nuclear
coordinates $\bR$. The equation of motion (the nuclear \SE)
is shown in \er{nuclearSE} and the Lagrangean in \er{NucLagrangean}.
In either expression, the terms
with $\tau^k_{bm}$ represent couplings between the nuclear wave functions
$\chi_k (\bR)$
and $\chi_m (\bR)$, i.e. (virtual) transitions (or admixtures) between the nuclear
states. (These may represent
transitions also for the electronic states, which would get expressed in finite
electronic lifetimes.) The expression for the transition matrix is not elementary,
 since the coupling terms are of a derivative type.

    Now the Lagrangean associated with the nuclear motion is not invariant under a
 local gauge transformation. For this to be the case the Lagrangean needs to
include also an "interaction field". This field can be represented either as a
vector field (actually a four-vector, familiar from electromagnetism), or as a
tensorial, Yang-Mills type field. Whatever the form of the field, there are
always two parts to it. First, the field induced by the nuclear motion itself,
secondly an "externally induced field", actually produced by some other
particles $\br',\bR'$ , which are not part of the original formalism. (At our
convenience, we could include these and then these would be part of the
extended coordinates $\br,\bR$. The procedure would then result in the appearance
of a potential interaction, but not having the "field".) At a first glance, the field
(whether induced internally or externally) is expected be a Yang-Mills type
tensorial field since the system is non-Abelian, but here we meet a surprise.
When the couplings  $\tau^k_{bm}$ are of the derivative form shown in \er{tau1}
{\it and} when a complete set is taken for the electronic states $\zeta_k(\br,\bR)$,
then the Yang-Mills field
intensity tensor $F_a^{\ bc}$ induced by the $\br,\bR$ system vanishes and the induced field
is a  "pure gauge field". Just as the induced 4-vector potential in the Abelian
case can be transformed away by a choice of gauge, so also can the  $\tau^k_{bm}$
interaction terms. (See our previous proof. This shows that the vanishing of
the Yang-Mills tensor is the condition for the possibility to transform away the
interaction term.) This serves as a reminder that with choice of a  full
electronic set, the solutions $\Psi(\br,\bR)$ are exact and there is no residual
interaction between different $\Psi(\br,\bR)$'s. Such interaction can, of course, be
externally induced by an "external" Yang-Mills field intensity tensor$F_a^{\ bc}$ . This
is rooted, as before, in $\br',\bR'$, and it  could be got rid off by including these in
the Hamiltonian.
\item There is finally the case that the electronic set $\zeta_k(\br,\bR)$
is not a complete set.
Then, neither $\Psi(\br,\bR)$ in \er{Psimol}, nor the nuclear \er{nuclearSE} is exact.
Moreover,
the truncated Lagrangean in \er{NucLagrangean} is not exact either and this shows up by its
not possessing a full symmetry (namely, lacking invariance under local gauge
transformation). We can (and should) remedy this by introducing a Yang-Mills
field, which is not now a pure gauge field. This means that the internally
induced Yang-Mills field cannot be transformed away by a (local) gauge
transformation and that it brings in (through the back door, so to speak) the
effect of the excluded electronic states on the nuclear states, these being now
dynamically coupled between themselves.
\end{enumerate}

At this stage, it would be too ambitious to extrapolate the implications of the above
molecular theory for to elementary particles and forces but, by analogy with
the fully worked out molecular model and disregarding any complications due
to the fully relativistic covariance, one might argue that particle states are also
eigenstates of some operators ("hamiltonians") and constitute full sets.
Interactions between different particles (leptons, muons, etc.) exist and when
these interactions (in their minimal form) are incorporated in the formalism,
one gets exact eigenstates (and at this stage, as yet, {\it no} interaction fields ). It is
only when one truncates the particle state-manifolds to finite subsets, which
may have some internal symmetry (as the $SU(2)$ multiplets: "neutron, proton",
or "electron, neutrino"), that one finds that one has to pay some price for the
approximation involved in this truncation. Namely, the Lagrangean loses its
original gauge-invariance, which is the formal reflection of the fact that the
original interaction field is not fully accounted for in the truncated
representation. To remedy both the formal deficiency and the neglect of part
of the interaction, one has to introduce some new forces (electromagnetic, or
Yang-Mills types and possibly others). These do both jobs.

        Moreover, if the molecular analogy is further extended, these residual
forces play a further role, in addition to the two already mentioned (namely,
restoring formal invariance and reinstating the missed interaction). They bring
in extra degrees of freedoms (e.g., photons), which act on the particles (but,
supposedly, not between themselves). (In a vernacular locution, the tail that
was wagged by the dog, can also wag the dog.) In the consistent scheme that
we describe here, these extra degrees of freedom are illusory in that the
residual forces are only convenient expressions of the presence of some
other particles, and would be eliminated by including these other particles in a
broader scheme.
        Evidently, the above description steers clear of field theory and is not
relativistic (covariant). These, as well as other shortcomings that need to be
supplied, require us to stop our speculations at this stage. 

\subsection {An Extended (Sufficiency) Criterion for the Vanishing of the
Tensorial Field}

We define the field intensity tensor $F_{bc}$ as a function of a so far
undetermined vector operator $X=X_b$ and of the partial derivatives
$\partial_b$:
\beq
F_{bc \ mn}(\bX) = \partial_b \bX_{c \ mn}   -
\partial_c \bX _{b \ mn} - [\bX_{b \ mk} \bX_{c \ kn} - \bX_{c \ mk} \bX_{b \ kn}]
\label{Fbcmn}
\enq
 (The summation convention for double indices,
e.g. $k$ in the above, is assumed, as before. However, we no longer make
 distinction between covariant and contravariant sets.)
We set ourselves the task to find anti-hermitean operators $\bX_b$  such that
\beq
F_{bc \ mn}(\bX) = 0
\label{FbcmnXb}
\enq
The matrix elements are given by:
\beq
\bX_{b \ km}  :=<m|\bX_b|n> := \int d \br \zeta_m (\br,\bR) \bX_b \zeta_n (\br,\bR)
\label{Xbkm}
\enq
i.e., the brackets represent integration over the electron coordinate $\br$. The
$\zeta_m (\br,\bR)$ are a real orthonormal set for any fixed $\bR$. By anti-hermiticity of the
derivative operator $\partial_b$, we have already noted that
\beq
<m|\partial_b|n>= -<n|\partial_b|m>
\label{mpbn}
\enq
Now (with $\partial_b$ designating a differential that operates to the right until it
encounters a closing bracket symbol) one finds that
\beq
\partial_b <m|\bX_c|n> = <(\partial_b m)|\bX_c|n> +<m|\partial_b (\bX_c|n)>
\label {totalpartial}
\enq
and, further, that
\ber
& &\partial_b \bX_{c \ mn}   - \partial_c \bX_{b \ mn}
  \nonumber \\
&=& <(\partial_b m)|\bX_c|n> - <(\partial_c m)|\bX_b|n> +
 <m|\bX_c (\partial_b n)> - <n|\bX_b (\partial_c m)>
\nonumber \\
&+& Commut.
\label{pbXcmn}
\enr
where we designate:
\beq
Commut. \equiv <m|[\partial_b \bX_c - \partial_c \bX_b]|n>
\label{mpbXcomu}
\enq
Subtracting in the first four terms from the $\bX$'s the derivatives and adding to
compensate, we have for \er {pbXcmn}
\ber
&=& <(\partial_b m)|(\bX_c -\partial_c)|n> - <(\partial_c m)|(\bX_b - \partial_b)|n> +
 <m|(\bX_c - \partial_c )|(\partial_b n)>=
\nonumber \\
 &-& <n|(\bX_b - \partial_b)|(\partial_c m)> + <(\partial_b m)|(\partial_c n)>
 - <(\partial_c m)|(\partial_b n)>
\nonumber \\
&+& Commut.
\label{pbX-pcn}
\enr
We have ignored a term $<m|(\partial_c \partial_b - \partial_b \partial_c)|n>$,
which is zero by the commutativity of derivatives.
The crucial step is now, as in  \cite {Baer75}
and in other later derivations, the evaluation of the fifth and sixth terms by
insertion of $|k><k|$ (which is the unity operator, when $k$ is summed over a
complete set)
\ber
& &<(\partial_b m)|(\partial_c n)> -<(\partial_c m)|(\partial_b n)>
\nonumber \\
&=& <(\partial_b m)|k><k|\partial_c|n> - <(\partial_c m)|k><k|\partial_b|n>
\nonumber \\
&=& - < m|\partial_b|k><k|\partial_c|n> + < m|\partial_c|k><k|\partial_b|n>
\label{pbmpcn}
\enr
where \er{mpbn} has been used. We replace any derivative  $\partial$ by  $\partial - \bX$ and
compensate, so as  to get for \er{pbmpcn} the expression
\ber
&=& < m|\partial_b-\bX_c|k><k|\partial_c - \bX_b|n> -
 < m|\partial_c-\bX_c|k><k|\partial_b-\bX_b|n>
\nonumber \\
&+& < m|\bX_b|k><k|\partial_c-\bX_c|n> -
< m|\bX_c|k><k|\partial_b -\bX_b|n>
\nonumber \\
&+& < m|\partial_b- \bX_b|k><k|\bX_c|n> - < m|\partial_c-\bX_c|k><k|\bX_b|n>
\nonumber \\
&+& < m|\bX_b|k><k|\bX_c|n> - < m|\bX_c|k><k|\bX_b|n>
\label{mpb-Xbk}
\enr
We now put
\beq
\bX_b  = \partial_b + f_b(\bR)
\label{fbdef}
\enq
where
the function $f_b(\bR)$ is a c-number (not an operator) and can be
taken outside brackets (where the integration variable is $\br$). Then
we find that first three lines in \er{mpb-Xbk} cancel and so do the four
matrix elements in \er{pbX-pcn} (involving $\partial_c -\bX_b$). The surviving
contributions to the right hand side of \er{pbXcmn} are, first, the last line of
\er{mpb-Xbk}, which is nothing else than the square brackets in the
expression \er{Fbcmn} for the field intensity tensor and, secondly, the "Commut."
term in \er{pbXcmn}, defined in the line following \er{pbXcmn}. For this term to
vanish for all values of $n,m$ , we require that:
\beq
\partial_b \bX_c - \partial_c \bX_b = 0
\enq
or, in view of \er{fbdef}, that the function $f_b(\bR)$ be the gradient of a scalar
$G(\bR)$
\beq
f_b(\bR)= \partial_b G(\bR)
\enq

In conclusion, we have shown that the non-Abelian gauge-field intensity
tensor $F_{bc} (\bX)$ shown in \er{Fbcmn} vanishes when:
\begin{enumerate}
  \item the electronic set is complete, and
  \item the $\bX$ operator has the form $\bX_b  = \partial_b + \partial_b
  G(\bR)$
\end{enumerate}

It will be recognized that this generalizes the result proved by Baer in
\cite {Baer75}. Though that work did establish the validity of the curl-condition for the
derivative operator as long as some 25 years ago and the validity  is nearly
trivial for the second term taken separately, the same result is not self-evident
for the combination of the two terms, due to the non-linearity of $F(\bX)$.
 An important special case is when $G(\bR)=\bR^2/2$. Then
\beq
\bX_b= \partial_b+\bR_b
\enq
and the last expression is recognized as a multiple of the creation operator
$a_b^+$. This result paves the way for second-quantized or field theoretic
treatments.
    An additional extension is to the time derivative operator, appropriate
when the electronic states are time dependent. This extension is elementary
(though this has not been noted before), since the key relation that leads to the
vanishing of the field intensity, $F_{bc}=0$, is \er{mpbn}, and this  also holds when the
subscript $b$ stands for the time variable. What makes this result of special
interest is the way that it provides an extension of the results to relativistic
theories, in particular to a combination of Hamiltonians that (for the electron) is
the Dirac Hamiltonian and (for the nuclear coordinates) is the Schr\"{o}dinger
Hamiltonian.

\subsection {Observability of Molecular States in a Hamiltonian Formalism}

We now describe the relation between a purely formal calculational device,
like a  gauge transformation that merely admixes the basis states, and
observable effects.

Let us start, for simplicity,  with a Hamiltonian $H(\br,\bR)$ for
two types of particles. The particles can have similar or very different masses,
but for clarity of exposition we continue to refer to the two types of particle as
electrons ($\br$) and nuclei ($\bR$). As before, we posit solutions of the time
independent {\SE} that have the form shown in \er{Psimol} but, for
completeness, we attach an energy label $e$ to each solution:
\beq
\Psi^e (\br,\bR)  = \sum_k \zeta_k (\br,\bR) \chi^e_k (\bR) \qquad (e=0,1\ldots)
\label{Psie}
\enq
The electronic factor in the sum $\zeta_k (\br,\bR)$ arises from the familiar
 Born-Openheimer electronic Hamiltonian defined for a fixed $\bR$. Since this
Hamiltonian is independent of the nuclear set $\chi^e_k (\bR)$, it does not carry the
$e$-label. As is well known, with each $k$ there is associated a potential surface
$V_k (\bR)$ (the eigen-energies of the electronic Hamiltonian). Therefore by holding
the nuclear positions fixed for a sufficiently long time and choosing an
excitation wave-length appropriate to $V_k (\bR)$, it is possible to excite into any
of the mutually orthogonal  electronic states, $\zeta_k (\br,\bR)$. The dependence of
these functions on both of their variables can therefore be experimentally
obtained.
Turning now to the nuclear equation, \er{nuclearSE} above, when the derivative
terms are excluded,  this equation yields the nuclear set $\chi^e_k (\bR)$ with a set of
(constant) eigen-energies $E^e_k$ for any given diagonal $V_k$.
The set $\chi^e_k (\bR)$
is orthogonal for different $e$'s and the same $k$, but not orthogonal for different
$k$'s and the same $e$ (say, the lowest energy $e=0$) or different $e$'s.
Returning to \er{Psie}, it is clear that we can select any stationary
eigen-state $\Psi^e(\br,\bR)$ of the combined system by exciting with the proper
wavelength for a sufficiently long time (in this case, of course, without
constraint on $\bR$). Thus, the dependence of any of these superpositions on the
two variables $\br,\bR$ can also be ascertained and $\Psi^e(\br,\bR)$ thereby operationally
obtained. By computing the projections
\beq
<\zeta_k (\br,\bR)|\Psi^e(\br,\bR)>
\label{ovlap}
\enq
(in which both factors have been experimentally determined) we obtain the
nuclear cofactors $\chi^e_k (\bR)$. [See again \er{Psie}.]
Actually, one could have written, instead of \er{Psie}, a different
superposition, sometimes called the "crude Born-Oppenheimer" wave-function:
\beq
\Psi^e (\br,\bR)  = \sum_k \zeta_k (\br,\bR_0) \chi^e_k (\bR|\bR_0)
 \qquad (e=0,1\ldots)
\label{Psie2}
\enq
in which the electronic state refers to a fixed nuclear position $\bR_0$ rather than to
all values of the nuclear coordinate. This electronic state can be operationally
obtained in a manner similar to, but actually more simple, than that which has
already been proposed to obtain $\zeta_k (\br,\bR)$ in \er{Psie}, namely,  by exciting at a
wavelength corresponding to $V_k(\bR_0)$ and probing the $\br$-dependence
of $\zeta_k (\br,\bR_0)$.
Determining $\Psi^e (\br,\bR)$ as before and forming the projection
$<\zeta_k (\br,\bR_0)|\Psi^e(\br,\bR)>$ we
again obtain (gedanken-experimentally) the nuclear factors $\chi^e_k (\bR|\bR_0)$. While
this procedure is legitimate (and even simpler than the previous), it suffers
from the more problematic convergence of the superposition (\ref{Psie2}) in
comparison to (\ref{Psie}).
One could next try \er{Psie} with a truncated superposition, say involving
only $N$ summand terms (in practice $N=2$ or $3$ are common), rather than an
infinite number of terms. The electronic functions $\zeta_k (\br,\bR)$
 ($k=1,\ldots,N$) can be
determined as before, and so can be the associated nuclear factors
 $\chi^e_k (\bR)$,
but here one risks to come upon inconsistencies, when from the
observationally obtained full wave function $\Psi^e(\br,\bR)$ one computes the overlaps
$<\zeta_k (\br,\bR)|\Psi^e(\br,\bR)>$ for any  $k$ above $N$.
Then the truncated sum on the right hand
side vanishes, while the computed overlap on the left hand side will in general
be non-zero. In a sense it may be said that it is this inconsistency that the
introduction of the Yang-Mills field tries to resolve.
The resulting eigen state $\Psi^e(\br,\bR)$ is an "entangled state", in the
terminology of measurement theory \cite {Neumann}. While there appears to
be no problem in principle to extract by experiment any $\zeta_k (\br,\bR)$ (as already
indicated), the question arises whether one can put the nuclear part into any
particular $k$-state $\chi^e_k (\bR)$. This does not appear possible for the form in
\er{Psie}
and the source of the difficulty may again be the presence of derivatives in the
nuclear equation.
Can one select some observable nuclear set? It turns out that the set
$\phi^e_h(\bR)$ in the transformed eigen state
\beq
\Psi^e (\br,\bR)  = \sum_{kh} \zeta_k (\br,\bR)
[G(\bR)^{-1}]_{kh} \phi^e_h (\bR) \qquad (e=0,1\ldots)
\label{Psie3}
\enq
is observable. The matrix $G(\bR)$ is the gauge factor introduced in \er{transmatr1}. The
product $\zeta_k (\br,\bR)G(\bR)^{-1}$ is independent of $\bR$.
[Recall that $G(\bR)$ is identical with
the ADT matrix $A$]. Then  $\phi^e_h (\bR)$ can be selected by exciting an $e$-state such
as in \er{ovlap} and then selecting one of the $\br$ states. The coefficient of the
selection will be (apart from a phase factor) the nuclear state $\phi^e_h (\bR)$.

However, this procedure depends on the existence of the matrix $G(\bR)$,
(or of any pure gauge), which predicates the expansion in \er{Psimol} for a full
electronic set. Operationally, this means the pre-selection of a full electronic
set in \er{Psie2}. When the pre-selection is only to a partial, truncated electronic
set, then the relaxation to the truncated nuclear set in \er{ovlap} will not be
complete. Instead, the now truncated set nuclear in \er{ovlap} will be subject to a
Yang-Mills force $F$. It is not our concern to fully describe the dynamics of the
truncated set under a Yang-Mills field, except to say (as we have already
done above) that it is the expression of the residual interaction of the
electronic system on the nuclear motion.

\subsection {An Interpretation}

 As shown in \er{tau1}, the gauge field $A^b_c$ is simply related
 to the non-adiabatic coupling elements $\tau^k_{\ bm}$. For an infinite
 set of electronic adiabatic states ($N=\infty$ in \er {Psimol}), $F_{bc}=0$.
This important results seems to have been first established by
\cite {Baer75} and was later rederived by others. [In the original formulation
of \cite {Baer75} only the
contracted form of the field $A^b_c$ (appearing in the definition of $F$)
\beq
A^k_{\ cm} = A^b_c (t_b)^k_m
\enq
enters. This has the form
\beq
A^k_{\ cm} =\tau^k_{\ cm}(\bR)
\enq
If the intermediate summations are over a complete set,
 then
\beq
(F_{bc})^k_m = F^a_{\ bc}(t_a)^k_m = 0]
\enq
This result extends the original theorem \cite{Baer75} and is true due to the
linear independence of the t-matrices \cite {Jackiw}.]
The meaning of the vanishing of $F$ is that, if $\chi_k (\bR)$ is the partner of the
electronic states spanning the whole Hilbert space, there is no indirect
coupling (via a gauge-field) between the nuclear states; the only physical
coupling being that between the electronic and nuclear coordinates, which is
given by the potential energy part of $H(\br,\bR)$.
    When the electronic $N$-set is only part of the Hilbert space (e.g., $N$ is
finite), then the underlying electron-nuclei coupling gets expressed by an
additional, residual coupling between the nuclear states. Then $F^a_{\ bc}\neq 0$ and the
Lagrangean has to be enlarged to incorporate these new forces.

[We further make the following tentative conjecture (probably valid only under
restricted circumstances, e.g., minimal coupling between degrees of
freedom): In Quantum Field Theories, too, the Yang Mills residual fields,
$A$ and $F$, arise because the particle states are truncated (e.g., the
proton-neutron multiplet is an isotopic doublet, without consideration of excited
states). Then, it is within the truncated set that the residual fields reinstate the
neglected part of the interaction. If all states were considered, then
eigenstates of the form shown in \er{Psimol} would be exact and there would be no
need for the residual interaction negotiated by $A$ and $F$.]

\section {Lagrangeans in Phase-Modulus Formalism}

\subsection {Background to the Non-Relativistic and Relativistic
Cases}

 The aim of this section is to show how the modulus-phase
 formulation,
 which is the keytone of our Chapter, leads very directly to the
 equation of continuity and to the Hamilton-Jacobi equation.
 These equations have formed the basic building blocks in Bohm's
 formulation of non-relativistic Quantum Mechanics \cite{Bohm}. We begin with
the non-relativistic case, for which the simplicity of the derivation has mainly
 pedagogical merits, but then we go over to the relativistic case that involves
new results, especially regarding the topological phase. Our conclusions (presented
in the penultimate subsection) are that, for a broad range of commonly encountered
 situations, the relativistic treatment will not affect the presence or absence
 of the Berry-phase that arises from the Schr\"{o}dinger equation.

The earliest appearance of the non-relativistic \ce is due to
Schr\"{o}din-ger himself \cite{Schrodinger,Mehra}, obtained from
his time-dependent wave-equation.
 A relativistic \ce (appropriate to a scalar field and formulated in terms of
the field amplitudes) was found by Gordon  \cite{Gordon1}. The \ce
for an electron in the relativistic Dirac theory
\cite{Dirac1927,Greiner2} has the well known form
\cite{Bogoliubov}: \beq {\partial_\nu J^\nu} = 0 \label{div} \enq
where the four-current $J^{\nu}$ is given by \beq J^{\nu} = {\bp
\gamma^\nu \psi} \label{J} \enq (The symbols in this equation are
defined below).
 It was shown by Gordon \cite{Gordon2}, and further discussed by Pauli
\cite{Pauli} that, by a handsome trick on the four-current, this
can be broken up into two parts $J^{\nu} = J_{(0)}^{\nu} + J_{(1)}^{\nu}$
(each divergence-free), representing
respectively a conductivity current (``Leitungsstrom''):
\beq
J_{(0)}^{\nu} = -\frac{i}{2mc} \{ [(\hbar \partial^\nu - i \frac{e}{c} A^\nu) \bp] \psi -
\bp [(\hbar \partial^\nu + i \frac{e}{c} A^\nu) \psi] \}
\label{J0}
\enq
 and a polarization current \cite{Gordon2}:
\beq
J_{(1)}^{\nu} = -\frac{i \hbar}{2mc} \partial_\mu (\bp \gamma^\mu \gamma^\nu \psi),
\qquad \nu \ne \mu
\label{J1}
\enq
 Again, the summation convention is used, unless we state otherwise.
As will appear below, the same strategy can be used upon the
Dirac Lagrangean density to obtain the \ce and \hje in the modulus-phase
representation.

Throughout, the space coordinates and other vectorial quantities are
 written either in vector form $\vec x$, or with Latin indices $x_k$
 $(k=1,2,3)$, the time ($t$) coordinate is $x_0 =ct$. A four vector will
 have Greek-lettered indices, such as $x_\nu$ $(\nu =0,1,2,3)$ or the partial
derivatives $\partial_\nu$. $m$ is the electronic mass, $e$ the charge.

\subsection {Non-Relativistic Electron}

The phase-modulus formalism for  non-relativistic electrons was
discussed at length by Holland \cite{Holland}, as follows:

The Lagrangean density ${\cal L}$ for the non-relativistic
electron is written as
\beq
 {\cal L}= -\frac{\hbar^2}{2m} \vec \nabla \psi^* \cdot \vec \nabla \psi -
 e V \psi^* \psi -
\frac{i e \hbar}{2 c m} \vec A \cdot (\psi^* \vec \nabla{\psi} -
\vec \nabla{\psi^* } \psi )+
 {1 \over 2} i \hbar (\psi^* \dot{\psi} - \dot{\psi^* } \psi )
 \label{nrLag}
 \enq
Here dots over symbols designate time derivatives. If now the modulus $a$
 and phase $\phi$ are introduced through
 \beq
\psi = a e^{i\phi } \label{nrpsi}
\label{psi1}
\enq
 the Lagrangean density takes the form
\beq
{\cal L}= -\frac{\hbar^2}{2m} [(\vec \nabla a)^{2}
+a^{2}(\vec \nabla \phi)^{2}] -e a^{2} V +
 \frac{e \hbar}{c m} a^2 \vec \nabla \phi \cdot \vec A  -
\hbar a^{2} {\partial \phi \over \partial t}
\label{nrLag2}
\enq
The variational derivative of this with respect to $\phi$ yields
the \ce
 \beq
 {\delta {\cal L} \over \delta \phi} = 0 \rightarrow
 { \partial \rho \over \partial t} + \vec \nabla \cdot (\rho \vec v) = 0
 \label{nrec}
 \enq
 in which the charge density is defined as: $\rho= e a^2$ and the velocity is
 $\vec v = \frac{1}{m}(\hbar \vec{\nabla} \phi - \frac{e}{c}\vec A)$.

 Variationally deriving with respect to $a$ leads to the \hje
 \beq
 {\delta  {\cal L} \over\delta a} = 0 \rightarrow
 {\partial S \over \partial t} +
  \frac{1}{2m}(\vec \nabla S - \frac{e}{c} \vec A)^2 + e V =
 {\hbar^2 \nabla^2 a \over 2 m a}+ \frac{e^2 \vec A^2}{2 m c^2}
 \label{nrhje}
 \enq
 in which the action is defined as: $S = \hbar \phi$. The right hand side of the
above equation
 contains the "quantum correction" and the electromagnetic correction.
 These results are elementary, but their derivation illustrates the advantages
 of using the two variables, phase and modulus, to obtain equations of motion
that are
 substantially different from the familiar Schr\"{o}dinger equation and have
 straightforward physical interpretations \cite{Bohm}. The interpretation is,
 of course, connected to the modulus being a physical observable (by Born's
 interpretational postulate) and to the phase having a similar though somewhat
 more problematic status. (The "observability" of the phase has been
 discussed in the literature by various sources, e.g. in \cite{Mandel} and,
 in connection with a recent development, in \cite{EY1,EY2}. Some of
its aspects have been reviewed in section 2 of this Chapter.)

 Another possibility to represent the quantum mechanical Lagrangian density is
using the logarithm of the
 amplitude $\lambda = \ln a, \quad a=e^\lambda$. In that particular
representation the Lagrangean density takes the
 following symmetrical form:
 \beq
{\cal L} = e^{2\lambda} \{ -\frac{\hbar^2}{2 m}[(\vec \nabla \lambda)^{2} +
(\vec \nabla \phi)^{2}]
-\hbar {\partial \phi \over \partial t}-e V +  \frac{e \hbar}{c m} \vec \nabla
\phi \cdot \vec A \}
\label{symmlagr}
\enq

\subsection{Similarities Between Potential Fluid Dynamics and Quantum
 Mechanics}

In writing the Lagrangean density  of quantum mechanics in the modulus-phase
representation,
\er{nrLag2}, one notices a striking similarity between this Lagrangean density
 and that
of potential fluid dynamics (fluid dynamics without vorticity) as represented in
the work of Seliger and Whitham \cite{Seliger}. We recall briefly some parts of
their work which are relevant and then discuss the connections with quantum mechanics.
The connection between fluid dynamics and quantum mechanics
of an electron was already discussed by Madelung \cite{Madelung} and in Holland's book \cite{Holland}.
However, the discussion by Madelung refers to the equations only and does not
address the variational formalism which we discuss here.

If a flow satisfies the condition of zero vorticity, i.e. the velocity field
$\vec v$ is such that
$\vec \nabla \times \vec v = 0$, then there exists a function $\nu$ such that
$\vec v = \vec \nabla \nu$.
In that case one can describe the fluid mechanical system with the following
Lagrangean density:
\beq
{\cal L} = [- \frac{\partial \nu}{\partial t}
 - {1 \over 2} (\vec \nabla \nu)^2 - \varepsilon(\rho) -\Phi] \rho
\label{Asheractpc}
\enq
in which $\rho$ is the mass density, $\varepsilon$ is the specific internal
energy and $\Phi$ is some arbitrary
function representing the potential of an external force acting on the fluid.
Taking the variational
derivative with respect to $\nu$ and $\rho$, one obtains the following equations:
\ber
\frac {\partial \rho }{\partial t}  & + &
\vec \nabla \cdot (\rho \vec \nabla \nu) = 0
\label{ContinS2p}
 \\
\frac {\partial \nu}{\partial t} & = & -\frac{1}{2} (\vec \nabla \nu)^2 - h -
\Phi.
\label{nueq2p}
\enr
in which $h =  {\partial (\rho \varepsilon) \over \partial \rho}$ is the
specific enthalpy.
The first of those \eqs is the continuity equation, while the second is Bernoulli's
equation.

Going back to the quantum mechanical system described by \er{nrLag2}, we
introduce
the following variables $ \hat \nu = \frac{\hbar \phi}{m}, \quad  \hat \rho = m
a^2 $.
In terms of these new variables the Lagrangean density in \er{nrLag2} will take
the form:
\beq
{\cal L} = [- \frac{\partial \hat \nu}{\partial t}
 - {1 \over 2} (\vec \nabla \hat \nu)^2 -
  \frac{\hbar^2}{2 m^2} \frac{(\vec{\nabla}\sqrt{\hat \rho})^2}{\hat \rho}  -
\frac{e}{m} V] \hat \rho
\label{fluidquantum}
\enq
in which we assumed that no magnetic fields are present and thus $\vec A = 0$.
When compared with \er{Asheractpc} the following correspondence is noted:
\beq
 \hat \nu \Leftrightarrow \nu, \qquad  \hat \rho \Leftrightarrow \rho, \qquad
 \frac{\hbar^2}{2 m^2} \frac{(\vec{\nabla}\sqrt{\hat \rho})^2}{\hat \rho}
\Leftrightarrow \varepsilon,
\qquad   \frac{e}{m} V \Leftrightarrow \Phi.
\label{corres}
\enq
The quantum "internal energy"
$ \frac{\hbar^2}{2 m^2} \frac{(\vec{\nabla}\sqrt{\hat \rho})^2}{\hat \rho}$
depends also on the derivative of the density, unlike
the fluid case, in which internal energy
is a function of the mass density only. However, in both cases the
internal energy is a positive quantity.

Unlike classical systems in which the Lagrangean is quadratic in the time
derivatives of the degrees of freedom, the Lagrangeans of both quantum and fluid
dynamics are linear in the time derivatives of the degrees of freedom.

\subsection{Electrons in the Dirac Theory}

 (Henceforth, for simplicity, the units $c=1$,
$\hbar=1$ will be used, except at the end, when the results are
discussed.) The Lagrangean density for the particle is in the
presence of external forces
 \beq {\cal L} = {i\over
2}[\bar{\psi}\gamma ^\mu ({\partial_\mu} + ieA_\mu)\psi -
  ({\partial_\mu} - i e A_\mu) \bar{\psi}\gamma ^\mu \psi ]
 - m \bar{\psi} \psi
\label{Lagr}
\enq
Here $\psi$ is a four-component spinor, $A_\mu$ is a four-potential and the
$4 \times 4$ matrices $\gamma ^\mu $ are given  by
\beq
\gamma^0 = \left(\begin{array}{cc} I & 0 \\ 0 & -I \end{array}
 \right) \qquad
\gamma^k = \left (\begin{array}{cc} 0 & \sigma^{k} \\ -\sigma^{k} & 0
\end{array}
\right) \qquad (k=1,2,3)
\label{gamma}
\enq
where we have the $2 \times 2$ matrices
\ber
 I & = & \left( \begin{array}{cc} 1 & 0 \\ 0 & 1 \end{array} \right) \qquad
\sigma^{1} = \left( \begin{array}{cc} 0 & 1 \\ 1 & 0 \end{array} \right)
\nonumber \\
\sigma^{2} & = & \left( \begin{array}{cc} 0 & -i \\ i & 0 \end{array} \right)
\qquad
\sigma^{3} = \left( \begin{array}{cc} 1 & 0 \\ 0 & -1 \end{array} \right)
\label{sigma}
\enr
Following \cite{Gordon2} we substitute in the Lagrangean density, \eq (\ref{Lagr}), from
 the Dirac equations \cite{Bogoliubov}, namely, from
\beq
 \psi = \frac{i}{m} \gamma^\nu (\partial_\nu +ieA_\nu)\psi \qquad \bp =
 -\frac{i}{m} (\partial_\nu -ieA_\nu) \bp \gamma^\nu
\label{psi}
\enq
 and obtain
\beq {\cal L} = \frac{1}{m}(\partial_\nu -ieA_\nu ) \bar{\psi}
\gamma^\nu \gamma^\mu (\partial_\mu +ieA_\mu )\psi - m \bar{\psi}
\psi \label{Lagr2} \enq
 We thus obtain a Lagrangean density, which
 is equivalent to \er{Lagr} for all solutions of the Dirac equation,
 and has the structure of the non - relativistic Lagrangian density, \eq (\ref{nrLag2}).
Its variational derivations with respect to $\psi$ and $\bp$ lead to the solutions
shown in \eq (\ref{psi}), as well as to other solutions.

 The Lagrangean density  can be separated into two terms
\beq {\cal L}= {\cal L}^0 +{\cal L}^1 \label{Lagrsep} \enq
according to whether the summation symbols $\nu$ and $\mu$ in
(\ref{Lagr}) are equal or different. The form of  ${\cal L}^0$ is
\beq
{\cal L}^0 = \frac{1}{m}(\partial^\mu -ieA^\mu ) \bar{\psi}
(\partial_\mu +ieA_\mu )\psi + m \bar{\psi} \psi
\label{L0}
\enq
Contravariant $V^\mu $ and covariant $V_\nu $ four-vectors are
connected through the metric $ g^{\mu \nu} ={\rm diag} \
(1,-1,-1,-1)$ by \beq V^\mu =g^{\mu \nu} V_\nu \enq \label{contra}

The second term in \er{Lagrsep},  ${\cal L}^1$ will be shown to be smaller than
the first in the near non-relativistic limit.

Introducing the moduli $a_i$ and phases $\phi_i$ for the four spinor
 components $\psi_i$ ($i=1,2,3,4$), we note the following relations
(in which no summations over $i$ are implied):
\ber
\psi_i & = & a_i e^{i\phi_i} \nonumber \\
\bp_i & = & \gamma^0_{ii}a_i e^{-i\phi_i} \nonumber \\
\bp_i \psi_i & = & {a^2_i} \gamma^0_{ii}
\label{relations}
\enr

The Lagrangean density  \er{Lagr2} rewritten in terms of the phases and moduli takes
a form  that is much simpler (and shorter), than that which one would
obtain by substituting from \er{psi1} into \er{Lagr}. It is given by
\beq
{\cal L}^0 = \frac{1}{m} \sum_i \gamma^0_{ii} [\partial_\nu a_i \partial^\nu
 a_i +a_i^2 ((\partial_\nu \phi_i+e A_\nu) (\partial^\nu \phi_i+e A^\nu)
 -m^2)]
\label{Lagrmod2}
\enq
When one takes its variational derivative with respect to the
 phases $\phi_i$, one obtains the \ce in the form
\beq
{\delta {\cal L}^0 \over \delta \phi_i} =
 - {\delta {\cal L}^1 \over \delta \phi_i}
\label{ce1}
\enq
The  right hand side will be treated in a following subsection, where we
shall see that it is small in the nearly non-relativistic limit and
that it vanishes in the absence of an electromagnetic field.
 The left hand side can be evaluated to give
\beq
{\delta {\cal L}^0 \over \delta \phi_i} = -\frac{2}{m}
\partial_\nu [a_i ^2 (\partial^\nu \phi_i +eA^\nu )]
\equiv 2 \partial_\nu J^\nu_i
\qquad ({\rm no \ summation \ over} \ i )
\label{ce2}
\enq
The above defined currents are related to the conductivity current by
the relation:
\beq
J_{(0)}^\nu = \sum_i \gamma_{ii}^0 J^\nu_i
\label{currentsum}
\enq
Although the conservation of $ J^\nu_i $ separately is a stronger
result than the result
obtained in \cite{Pauli}, one should bear in mind that the present
result is only approximate.

The variational derivatives of ${\cal L}^0$ with respect
to the moduli $a_i$ give the following equations:
\beq
{\delta {\cal L}^0 \over \delta a_i}
=  -\frac{2}{m} [ \partial_\nu \partial^\nu a_i
-a_i ((\partial_\nu \phi_i+e A_\nu) (\partial^\nu \phi_i+e A^\nu)
-m^2)]
 \label{hje2}
 \enq
 The result of interest in the expressions shown in  \er{ce2}
 and  \er{hje2} is that, although one has obtained expressions that include
corrections to the non-relativistic case, given in \er{nrec} and \er {nrhje},  still both the
 \ces and the \hjes involve each spinor component separately.
 To the present approximation there is no mixing between the components.

\subsection{The Nearly Non-Relativistic Limit}

In order to write the previously obtained \eqs in the nearly non-relativistic
limit, we introduce phase differences $s_i$ that remain finite in
the limit $c \rightarrow \infty$. Then
\beq
\phi_i = \gamma^0_{ii}(- m x_0 +s_i),  \qquad \partial_0 \phi_i=
\gamma^0_{ii}(-m + \partial_0 s_i), \qquad \vec \nabla \phi_i=
\gamma^0_{ii} \vec \nabla s_i.
\label{deltaphi}
\enq
We reinstate the velocity of light $c$ in this and in the next subsection,
 in order to appreciate the order of magnitude of the various terms.
When contributions from ${\cal L}^1$ are neglected, the expression in \er{hje2}
equated to zero gives the following equations, in which
the large $(i=1,2)$ and small $(i=3,4)$ components are separated.
\ber
&& \partial_t s_i + \frac{1}{2m}(\vec\nabla
s_i-\frac{e}{c} \vec A)^2 + eA_0 =
 \frac{ \nabla^2 a_i}{2m a_i} +  \frac{e^2}{2 m c^2} \vec A^2
\nonumber\\
&& +\frac{1}{2 m c^2}[ -\frac{\partial^2_t a_i}{a_i}
+(\partial_t s_i)^2 + 2 e A_0 \partial_t s_i + e^2 A_0^2 - e^2
\vec A^2 ]
\nonumber\\
&&(i=1,2)  \label{hjesepa} \\
&& \partial_t s_i+\frac{1}{2m}(\vec\nabla s_i-\frac{(-e)}{c} \vec A)^2 + (-e)A_0 =
 \frac{ \nabla^2 a_i}{2m a_i} +  \frac{e^2}{2m c^2} \vec A^2
\nonumber\\
&&+\frac{1}{2 m c^2}[- \frac{\partial^2_t a_i}{a_i}
+(\partial_t s_i)^2 + 2(-e)A_0 \partial_t s_i + e^2A_0^2 - e^2
\vec A^2 ]
\nonumber\\
&&(i=3,4).
 \label{hjesepb}
 \enr
In the same manner we obtain the following \eqs from \er{ce2}
\ber
&& \partial_t \rho_i + \vec \nabla \cdot (\rho_i \vec v_i)=
\nonumber\\
&& \frac{1}{c^2}\partial_t [\rho_i (\frac{\partial_t s_i + e A_0}{m})]
\nonumber\\
&&(i=1,2), \qquad \rho_i = m a_i^2, \qquad \vec v_i = \frac{\vec \nabla s_i -
\frac{e}{c}\vec A}{m} \label{cesepa} \\
&& \partial_t \rho_i + \vec \nabla \cdot (\rho_i \vec v_i)=
\nonumber\\
&& \frac{1}{c^2} \partial_t [\rho_i (\frac{\partial_t s_i + (-e) A_0}{m})]
\nonumber\\
&&(i=3,4), \qquad \rho_i = m a_i^2, \qquad
\vec v_i = \frac{\vec \nabla s_i - \frac{(-e)}{c} \vec A}{m}
\label{cesepb}
 \enr
The first lines give the non-relativistic part of the \hje and the \ce shown in
\er{nrhje} and \er{nrec}, while the second lines contribute relativistic
corrections. All terms from ${\cal L}^0$ are of the non-mixing
type between components.
There are further relativistic terms, to which we now turn.

\subsection{The Lagrangean-Density Correction Term}

As noted above, ${\cal L}^1$ in \er{Lagrsep} arises from terms in which $\mu
\neq \nu$. The corresponding contribution to the four-current was evaluated in
\cite{Gordon2,Pauli} and was shown to yield the polarization current. Our result
is written in terms of the magnetic field $\vec H$ and the electric field
$\vec E$, as well as the spinor four-vector $\psi$ and the vectorial $2 \times 2$
sigma matrices given above in \er{sigma}.
\beq
{\cal L}^1 = -\frac{e}{mc} \bp (\vec H \cdot \vec \sigma ) \left(
 \begin{array}{cc }I & 0 \\ 0 & I \end{array}\right) \psi +\frac{ie}
{mc} \bp ( \vec E \cdot \vec \sigma ) \left( \begin{array}{cc} 0 & I \\
  I & 0 \end{array}\right) \psi
\label{L1}
\enq
These terms are analogous to those on p. 265 of \cite{Dirac2}. It will be
 noted that the symbol $c$ has been reinstated as in the previous subsection,
 so as to
facilitate the order of magnitude estimation in the nearly non-relativistic
limit. We now proceed based on \er{L1} as it stands, since the
transformation of \er{L1} to modulus and phase variables and functional
derivation gives rather involved expressions and will not be set out here.

To compare ${\cal L}^1$ with
${\cal L}^0$ we rewrite the latter in terms of the phase variables introduced
in \er{deltaphi}:
\ber
{\cal L}^0 &=& 2 \sum_i \gamma^0_{ii}\{ -\frac{1}{2m} [(\vec \nabla a_i)^{2}
+a_i^{2}(\vec \nabla s_i)^{2}] -(\gamma^0_{ii} e) a_i^{2} A^0 -
a_i^{2} {\partial s_i \over \partial t}\}
\nonumber \\
&+& \frac{2 e}{m c} \sum_i a_i^2 \vec \nabla s_i \cdot \vec A
+O(\frac{1}{c^2})
\label{L0b}
\enr
which contains terms independent of $c$ as well as terms of the order $O(\frac{1}{c})$
and $O(\frac{1}{c^2})$.

In \er{L1}, the first, magnetic-field term admixes different components of the
spinors both in the \ce and in the \hje. However, with the z-axis chosen as the
direction of $\vec H $, the magnetic-field  term does not contain phases and
does not mix component amplitudes. Therefore, there is no contribution from
this term in the \ces and no amplitude mixing  in the \hjes.
The second, electric-field term is non-diagonal between the large and
small spinor components, which fact reduces its magnitude by a further small
factor of $O(particle \ velocity/c)$. This term is therefore of the same
small order $O(\frac{1}{c^2})$, as those terms in the second line in \eq
(\ref{hjesepa}) and in \eq (\ref{cesepa}) that refer to the upper components.

We conclude that in the presence of electromagnetic fields the components
remain unmixed, correct to the order $O(\frac{1}{c})$.

\subsection {Topological Phase for Dirac Electrons}

  The topological (or Berry) phase \cite {Berry84, ShapereW, JainPati})
 has been discussed in previous sections of this Chapter.  The physical picture
for it is that when a periodic force, slowly
 (adiabatically) varying in time, is applied to the system then, upon a full
 periodic evolution, the phase of the wave function  may have a part
that is independent of the amplitude of the force. This part  exists in addition to
that part of the phase which depends on the amplitude of the force and which
 contributes  to the the usual, "dynamic" phase. We shall now discuss
 whether a relativistic electron can have a Berry phase when this is absent in
 the framework of the Schr\"{o}dinger equation, and vice versa.
 (We restrict the present discussion to the nearly non-relativistic limit,
 when particle velocities are much smaller than c.)

The following lemma is needed for our result. Consider a
 matrix Hamiltonian $h$ coupling two states, whose energy difference is
$2m$:
\beq
h = \left( \begin{array}{cc}
m+E_1 cos(\omega t+\alpha) & E_2 sin(\omega t) \\
  E_2 sin(\omega t)  & -m-E_1 cos(\omega t+\alpha)
\end{array} \right)
\label{h2}
\enq
The Hamiltonian contains two fields,  periodically
varying in time, whose intensities
 $E_1$ and $E_2$ are non-zero. $\omega$ is their angular frequency, and is
 (in appropriate energy units) assumed to be much smaller than the field
 strengths. This ensures the validity of the adiabatic approximation
 \cite{EY2}. $\alpha $ is an arbitrary angle. It is assumed that
 initially, at $t=0$, only the component with the
positive eigen-energy is present. Then after a full revolution
the initially excited component acquires or does not acquire a Berry phase
 (i.e., returns to its initial value with a changed or unchanged sign)
 depending on whether $|E_1|$  is greater or less than $m$ (=half the energy difference).

\noindent
Proof:
When the time dependent Schr\"{o}dinger equation is solved under adiabatic
 conditions, the upper, positive energy component has the coefficient:
the dynamic phase factor $\times C$, where
\beq
C=\cos[\frac{1}{2} \arctan (\frac{E_2 \sin(\omega t)}{m+E_1 \cos(\omega t+\alpha )})]
\label{coeff}
\enq

 Tracing the $\arctan$ over a full revolution by the method described
  in section 4 of this Chapter
 and noting the factor $\frac{1}{2}$ in \er{coeff} establishes our result.
(The case that $|E_1|=m$ needs more careful consideration, since it leads to a
 breakdown of the adiabatic theorem. However, this  case will be of
no consequence for the  results.)

We can now return to the Dirac equations, in which the time varying forces
enter through the four-potentials $(A_0 ,\vec A)$. [The "two states" in \er{h}
refer now to a large and to a small (positive- and negative-energy) component
in the solution of the Dirac \eq in the near nonrelativistic limit.]
 In the expressions (\ref{hjesepa}-\ref{hjesepb})
 obtained for the phases $s_i$ and arising from the Lagrangean
${\cal L}^0 $, there is no coupling between different components and therefore
the small relativistic correction terms will clearly not introduce or eliminate a
Berry phase. However, terms in this section supply the diagonal matrix
 elements in \er{h}. Turning now to the two terms in \er{L1}, the first, magnetic
field term again does not admix the large and  small
components, with the result that for either of these components
previous treatments based on the Schr\"{o}dinger or the Pauli equations
\cite{Greiner2,Holland} should suffice.
Indeed, this term was already discussed by Berry \cite{Berry84}. We thus need to
consider only the second, electric-field term which admixes the two types
 of components. These are the source of the off-diagonal matrix-elements in
\er{h}.
However, we have just shown that in order to introduce a new topological
 phase, one needs field strengths matching the electronic
 rest energies, namely electric fields of the order of $10^{14} Volt/cm $.
(For comparison we note that the electric field that binds an electron in a
hydrogen atom is four orders of magnitudes smaller than this. Higher fields
 can also be produced in the laboratory, but, in general,  are not of the
 type that can be used to guide the motion of a charged particle during a
 revolution.) As long as we exclude from our considerations such enormous
 fields, we need not contemplate relativistically
 induced topological phases. Possibly, there may be cases (e.g. many electron systems
or magnetic field effects)
 that are not fully covered by the model represented in \er{h}. Still,
 the latter model should serve as an
indicator for relativistic effects on the topological phase.

\subsection {What Have We Learned about Spinor Phases?}

  This part of our Chapter has shown
that the use of the two variables, moduli and phases, leads in a direct way
 to the derivation of the
continuity and Hamilton-Jacobi equations for both scalar and spinor wave functions.
 For the latter case we show that the differential equations for each spinor
component are (in the nearly non-relativistic limit) approximately decoupled.
 Because of this decoupling (mutual independence) it appears that the reciprocal
relations between phases and moduli derived in section 3 hold
to a good approximation for each spinor component separately, too.
 For velocities and electro-magnetic field strengths that are normally
 below the relativistic scale, the Berry phase obtained from the
 Schr\"{o}dinger \eq  (for scalar fields)
will not be altered by consideration of the Dirac equation.

\section {Conclusion}

This Chapter has treated a number of properties that arise from the presence
of degeneracy in the electronic part of the molecular wave fuction. The existence
of more than one electronic state in the superposition that
describes the molecular state demands attention to the phase relations
 between the
different electronic component-amplitudes. Looked at from a different angle, the phase
relations are the consequence of the complex
form of the molecular wave functions, which is grounded in the time dependent
{\SE}. Beside reviewing numerous theoretical and experimental works
 relating to the phase-properties of complex wave functions,
 the following general points have received emphasis in this Chapter:
 (1) Relative phases of components that make up, by the superposition principle,
     the wave function are observable.
 (2) The analytic behavior of the wave function in a complex parameter-plane
 is in several instances traceable to a physics-based "equation of restriction".
 (3) Phases and moduli in the superposition are connected through reciprocal
 integral relations.
 (4) Systematic treatment of zeros and singularities of component amplitudes
are feasible by a phase tracing method.
 (5) The molecular Yang-Mills field is conditioned by the finiteness of the
     basic Born-Oppenheimer set.
Detailed topics are noted in the Abstract.

\vspace{2cm}

\noindent
{\Large \bf Acknowledgments} \\

The authors are indebted to Prof. Michael
Baer for many years of exciting collaboration, to Dr. B. Halperin
for advice, to Prof. Mark Pere'lman for discussion and permission to
quote from his preprint ("Temporal Magnitudes and Functions of Scattering
Theory"), to
Prof. Shmuel Elitzur for suggesting the approach leading to
"Alternative Derivation" in section 5 and to Prof. Igal Talmi for
an inspiration \cite{Talmi}.


\begin{thebibliography} {99}
\bibitem {Jammer}           
M. Jammer, {\it The Conceptual Development of Quantum Mechanics} (McGraw-Hill,
New York, 1966)
\bibitem {Mehra}
J. Mehra and H. Rechenberg, {\it The Historical Development of Quantum Theory}
(Springer-Verlag, New York, 1987) Vol. 5,  Part 2
\bibitem {Cao}
T.Y. Cao, {\it Conceptual Development of 20th Century Field Theory} (University
Press, Cambridge, 1997)
\bibitem {Dirac1}
P.A.M. Dirac, Proc Roy. Soc. (London) A {\bf 112} 661 (1926)
\bibitem {Wigner1}
E.P. Wigner, {\it The Place of Consciousness in Modern Physics} in C. Muses and
A.M. Young (editors), {\it Consciousness and Reality} (Outerbridge and Lazzard,
New York, 1972); reprinted in E.P.Wigner, {\it Philosophical Reflections and Syntheses}
(Springer-Verlag, Berlin, 1997)
\bibitem {Steiner}
M. Steiner, {\it The Applicability of Mathematics as a Philosophical Problem}
(Harvard University Press, Cambridge Mass.,1998)
\bibitem {Dirac2}
 P.A.M. Dirac {\it The Principles of Quantum Mechanics}
(Clarendon Press, Oxford, 1958)
\bibitem {AB}
Y. Aharonov and D. Bohm, Phys. Rev. {\bf 115} 485 (1959)
\bibitem {Berry84}
M.V. Berry, Proc. Roy. Soc. (London) A {\bf 392} 45 (1984)
\bibitem {AA}
Y. Aharonov and J. Anandan, Phys. Rev. Lett. {\bf 58} 1593 {1987}
\bibitem {JainPati}
S.R. Jain and A.K. Pati, Phys. Rev. Lett. {\bf 80} 650 (1980)
\bibitem {ChengFung}
C.M. Cheng and P.C.W. Fung, J. Phys.A {\bf 22} 3493 (1989)
\bibitem {MooreStedman}
D.J. Moore and G.E. Stedman, J.Phys. A {\bf 23} 2049 (1990)
\bibitem {Pati}
 A.K. Pati, Phys. Rev. A {\bf 52} 2576 (1995)
\bibitem {Averbukh}
I.Sh. Averbukh and N.F. Perel'man, Soviet Phys. Uspekhi
 {\bf 34} 572 (1991)
\bibitem {Shapiro2}
 I. Sh. Averbukh, M. Shapiro, C. Leichtle and W.P. Schleich,
 Phys. Rev. A {\bf 59} 2163 (1999) 
\bibitem {Heiblum}
 R. Schuster, E. Buks, M. Heiblum, D. Mahalu, V. Umansky
 and H. Shtrikman, Nature {\bf 385} 417 (1997)
\bibitem {Shapiro}
 C. Leichtle, W.P. Schleich, I. Averbukh and M. Shapiro,
Phys. Rev. Lett. {\bf 80} 1418 (1998)
\bibitem {Buchsbaum}
T.C. Weinacht, J. Ahn and P.H. Buchsbaum, Phys. Rev. Lett. {\bf 80} 5508
(1998)
\bibitem {Nogues}
 G. Nogues, A. Rauschenbeutel, S. Osnaghi, M. Brune,
J.M. Raimond and S. Haroche, Nature {\bf 400} 239 (1999)
\bibitem{GRW}
 G.C.Ghirardi, A.Rimini and T.Weber, Phys. Rev. D{\bf 34} 470(1986)
\bibitem {Averin}
 D.V. Averin, Nature {\bf 398} 748 (1999)
\bibitem {Grover}
 L. Grover, The Sciences {\bf 39} 27 (1999, No.2)
\bibitem {Ramsey}
 N.F. Ramsey, {\it Molecular Beams} (Oxford University Press,
 New York, 1985)
\bibitem {NoelS}
 M.W Noel and C.R. Stroud, Jr., Phys. Rev. Letters {\bf 75}1252 (1995)
\bibitem {Goldstein}
H.Goldstein, {\it Classical Mechanics} (Addison-Wesley, Reading Mass. 1959)
\bibitem {Nieto}
M.M. Nieto, Physica Scripta T {\bf 48} 5 (1993)
\bibitem {Mandel}
L. Mandel and E. Wolf, {\it Optical Coherence and Quantum Optics} (University
Press, Cambridge, 1995) section 3.1
\bibitem {EYB1}
R. Englman, A.Yahalom and M. Baer, J. Chem. Phys.{109} 6550 (1998)
 (1999)
\bibitem {EYB2}
R. Englman, A. Yahalom and M. Baer, Phys. Lett. A {\bf 251} 223 (1999)
\bibitem {EY1}
R. Englman and A. Yahalom, Phys. Rev. A {\bf 60} 1802 (1999)
\bibitem {EYB3}
R. Englman, A.Yahalom and M. Baer, Eur. Phys. J. D {\bf 8} 1 (2000)
\bibitem {EY2}
R. Englman and A.Yahalom, Phys. Rev. B {\bf 61} 2716 (2000)
\bibitem {EY3}
R. Englman and A.Yahalom, Found. Phys. Lett. {\bf 13} 329 (2000)
\bibitem {EY4}
R. Englman and A.Yahalom, {\it The Jahn Teller Effect: A Permanent Presence
in the Frontiers of Science} in M.D. Kaplan and G. Zimmerman (editors),
{\it Proceedings of the NATO Advanced Research
Workshop, Boston, Sep. 2000}  (Kluwer, Dordrecht, 2001)
\bibitem {BE2}
M. Baer and R. Englman, Chem. Phys. Lett. {\bf 335} 85 (2001)
\bibitem {MBEY}
A. Mebel, M. Baer, R. Englman and A. Yahalom, J.Chem. Phys. {\bf 115} 3673 (2001)
\bibitem {EYBM4MCI}
R. Englman, A.Yahalom and M. Baer, Int. J. Quant. Chem. (In press)
\bibitem {Khalfin}
L.A. Khalfin, Soviet Phys. JETP {\bf 8} 1053 (1958) [J. Exp.Teor. Fyz.
(USSR) {\bf 33} 1371 (1957)]
\bibitem {PE}
M.E. Perel'man and R. Englman, Modern Phys. Lett. B {\bf 14} 907 (2000)
\bibitem {ScullyZubairy}

M.N. Scully and M.S. Zubairy, {\it Quantum Optics} (University Press, Cambridge,
1997)
\bibitem {Miller71}
W.H. Miller, J. Chem. Phys. {\bf 55} 3146 (1971)
\bibitem {HwangPechukas}
J.-T. Hwang and P. Pechukas, J. Chem. Phys. {\bf 67} 4640 (1977)
\bibitem {Nikitin}
E.E. Nikitin, J. Chem. Phys. {\bf 102} 6768 (1997)
\bibitem {ZhuNN}
C. Zhou, E.E. Nikitin and H. Nakamura, J. Chem. Phys. {\bf 104} 7059 (1998)
\bibitem {Bersuker1}
I.B. Bersuker, Chem. Rev.  {\bf 101} 1067 (2001)
\bibitem {BersukerPolinger}
I.B. Bersuker and V.Z. Polinger, {\it Vibronic Interactions in Molecules
 and Crystals} (Springer-Verlag, New York, 1989)
\bibitem {Baer75}
M. Baer, Chem. Phys. Lett. {\bf 35} 112 (1975)
\bibitem{Sidis}
V. Sidis in M. Baer and C.Y. Ng (editors), {\it State Selected and State-to-State
 Ion-Molecule Reaction Dynamics, Part 2, Theory} (Wiley, New York, 1992);
 Adv. Chem. Phys. {\bf 82} 73 (2000)
\bibitem{PCK}
T. Pacher, L.S. Cederbaum and H. Koppel, Adv. Chem. Phys. {\bf 84}, 293 (1993)
\bibitem{DRY1}
D.R. Yarkony, Rev. Mod. Phys. {\bf 68}, 985 (1996);
 Adv. At. Mol. Phys. {\bf 31}, 511 (1998)
\bibitem {Thummel}
H. Thummel, M. Peric, S.D. Peyerimhoff amd R.J. Buenker, Z. Phys. D {\bf 13}
307 (1989)
\bibitem {KoppelMeiswinkel}
H. Koppel and R. Meiswinkel, Z. Phys. D {\bf 32} 153 (1994)
\bibitem {DRY2}
D.R. Yarkony, Acc. Chem. Res. {\bf 31} 511 (1998)
\bibitem {KoizumiBersuker}
H. Koizumi and I.B. Bersuker, Phys. Rev. Lett. {\bf 83} 3009 (1999)
\bibitem {MebelBLin1}
A. M. Mebel, M. Baer and S.H. Lin, J. Chem. Phys. {\bf 112} 10703 (2000)
\bibitem {MebelBLin2}
A. M. Mebel, M. Baer and S.H. Lin, J. Chem. Phys. {\bf 114} 5109 (2001)
\bibitem {AvoryBaerBilling}
J. Avery, M. Baer and G.D. Billing (2001, to be published)
\bibitem {ChildRev}
M.S. Child, Adv. Chem. Phys. (This volume)
\bibitem {BEV}
M. Baer, R. Englman and A.J.C. Varandas,  Mol. Phys. {\bf 97} 1185 (1999)
\bibitem {BVE}
M. Baer, A.J.C. Varandas and R. Englman,  J. Chem. Phys. {\bf 111}
9493 (1999)
\bibitem {YE1}
A.Yahalom and R. Englman, Phys. Lett. A {\bf 272} 166 (2000)
\bibitem {BO}
M. Born and R.J. Oppenheimer, Ann. Physik (Leipzig) {\bf 89} 457 (1927)
\bibitem  {BH}
M. Born and K. Huang, {\it  Dynamical Theory of Crystal Lattices} (Clarendon
 Press, Oxford, 1951) Appendix VII
\bibitem {BatesMcCarroll}
D.R. Bates and R. McCarroll , Proc. Roy. Soc. (London) A {\bf 245} 175 (1958)
\bibitem {MottMassey}
N.F. Mott and H.S.W. Massey, {\it The Theory of Atomic Collisions} (Oxford
University Press, London, 1965)
\bibitem {Smith}
F.T. Smith, Phys. Rev. {\bf 115} 349 (1960); {\it ibid} {\bf 179} 111 (1969)
\bibitem {YM}
C.N. Yang and R. Mills, Phys. Rev. {\bf 96} 191 (1954)
\bibitem {Jackiw}
R. Jackiw, Rev. Mod. Phys. {\bf 52} 661 (1980)
\bibitem {ItzyksonGlauber}
C. Itzykson and  J.-B. Zuber, {\it Quantum Field Theories} (McGraw Hill,
New York, 1980)
\bibitem {PeskinSchroeder}

M.E. Peskin and D.V. Schroeder, {\it An Introduction to Quantum Field Theory}
 (Perseus Books, Reading Mass., 1995) Chapter 14
\bibitem {Weinberg}
S. Weinberg, {\it The Quantum Theory of Fields} (University Press, Cambridge,
1996) Vol. 2 Chapter 15
\bibitem {EYBMathieu}
R. Englman, A. Yahalom and M. Baer, Intern. J. Quantum Chem.
\bibitem {Mead87}
C.A. Mead, Phys. Rev. Lett. {\bf 59} 161 (1987)
\bibitem {Zygelman1}
B. Zygelman, Phys. Lett. A {\bf 125} 476 (1987)
\bibitem{MeadTruhlar79}
C.A. Mead and D.G. Truhlar, J.Chem. Phys. {\bf 70} 2284 (1979)
\bibitem {MoodySW1}
J. Moody, A. Shapere and F. Wilczek, Phys. Rev. Lett. {\bf 56} 893 (1986)
\bibitem {ShapereW}
 A. Shapere and F. Wilczek (editors),
{\it Geometrical Phases in Physics} (World Scientific, Singapore, 1990)
\bibitem {MoodySW2}
J. Moody, A. Shapere and F. Wilczek in A. Shapere and F. Wilczek (editors),
{\it Geometrical Phases in Physics} (World Scientific, Singapore, 1990)
\bibitem {Zygelman2}
B. Zygelman, Phys. Rev. Lett. {\bf 64}  256 (1990)
\bibitem {AharonovBRPR}
Y. Aharonov, E. Ben-Reuven, S. Popescu and D. Rohrlich, Nucl. Phys. B
{\bf 350} 818 (1991)
\bibitem {BerryShWi90}
M.V. Berry in A. Shapere and F. Wilczek (editors),
{\it Geometrical Phases in Physics} (World Scientific, Singapore, 1990) p.7
\bibitem {BaerMagnetic}
M. Baer, Chem. Phys. Lett. (to appear)
\bibitem{BrumerS}
P. Brumer and M. Shapiro, Chem. Phys. Lett. {\bf 12} 541 (1986)
\bibitem{ShapiroB87}
M. Shapiro and P. Brumer, Faraday Disc. Chem. Soc. {\bf 82} 177 (1987)
\bibitem {ShapiroB89}
M. Shapiro and P. Brumer, J. Chem. Phys. {\bf 90} 6179 (1989)
\bibitem{Shapiro93}
M. Shapiro, J. Phys. Chem. {\bf 97} 7396 (1993)
\bibitem {DunnWM}
T.J. Dunn, I.A. Walmsley and S. Mukamel, Phys. Rev. Lett. {\bf 74} 884 (1995)
\bibitem {UbernaKWPG}
R. Uberna, M. Khalil, R.M. Williams, J.M. Papanikolas and S. George, J. Chem.
Phys. {\bf 108} 9259  (1998)
\bibitem {LeichtleSAS}
C. Leichtle, W.P. Schleich, I.Sh. Averbukh and M. Shapiro, Phys. Rev. Lett.
 {\bf 80} 1418 (1998)
\bibitem{Shapiro98}
M. Shapiro, J. Phys. Chem. {\bf 102} 9570 (1998)
\bibitem {ZucchettiVWW}
A. Zucchetti, W. Vogel, D.-G. Welsch and I.A. Walmsley, Phys. Rev. A {\bf 60}
2716 (1999)
\bibitem {ShapiroFB}
M. Shapiro, E. Frishman and P. Brumer, Phys. Rev. Lett. {\bf 84} 1669 (2000)
\bibitem {BlanchetNBG}
V. Blanchet, C. Nicole, M.-A. Bouchene and B. Girard, Phys. Rev. Lett. {\bf
78}  2716 (1997)
\bibitem {BuchsbaumNature}        
T.C. Weinacht, J. Ahn and P.H. Buchsbaum, Nature {\bf 397} 233 (1999)
\bibitem {AraujoWS}
L.E.E. de Araujo, I.A. Walmsley and C.R. Stroud, Phys. Rev. Lett. {\bf 81}
955 (1998)
\bibitem {SchleichNature}
W.P. Schleich, Nature {\bf 397} 207 (1999)
\bibitem {FeshbachKL}
H. Feshbach, K. Kerman and R.H. Lemmer, Ann. Phys. (NY) {\bf 41} 230 (1967)
\bibitem {JortnerM}
J. Jortner and S. Mukamel in A.D. Buckingham and C.A. Coulson
(editors), Intern. Rev. Sci. Theoret. Chem., Phys. Chem. Ser. 2, Vol.1
(Butterworth, London, 1975) p. 327
\bibitem {Englman79}
R. Englman, {\it Non-Radiative Decay of Ions and Molecules in Solids} (North-
Holland, Amsterdam, 1979) p. 155
\bibitem {Zeh}
H.D. Zeh, {\it The Physical Basis of the Direction of Time} (Springer-Verlag,
Berlin, 1992)
\bibitem {Savitt}
S.F. Savitt (editor), {\it Time's Arrow Today, Recent Physical and
Philosophical Work on the Direction of Time} (University Press, Cambridge,
1995)
\bibitem {Weyl}
H. Weyl, {\it Space, Time, Matter} (Dover Books, New York, 1950);
 {\it The Theory of Groups and Quantum Mechanics} (Dover Books, New York,
1950)

\bibitem {Pauli}
W. Pauli, {\it General Principles of Quantum Mechanics} (Springer-Verlag, Berlin,
1980)
\bibitem {Lamb}
 W.E. Lamb, Physics Today {\bf 22} 23 (April 1969)
\bibitem {GaleGT}
W. Gale , E. Guth and G.T. Trammel, Phys. Rev. {\bf 165} 1434 (1968)
\bibitem {Royer}
 A. Royer, Found. Physics {\bf 19} 3 (1989) 3
\bibitem {BE1}
M. Baer and R. Englman, Chem. Phys. Lett. {\bf 265} 105 (1997)
\bibitem {EB1}
R. Englman and M. Baer, J. Phys.:Condensed Matter {\bf 11} 1059 (1999)
\bibitem {GuimareBVM}
Y. Guimare, B. Baseia, C.J. Villas-Boas and M.H.Y. Moussa, Phys. Lett.
 A {\bf 268} 260 (2000)
\bibitem {Rayleigh}
J.W.S. Rayleigh  {\it The Theory of Sound}  Volume II (Dover, New York, 1945)
section 282
\bibitem {Klyshko}
D.N. Klyshko, Physics - Uspekhi {\bf 36} 1005-19 (1993)
\bibitem {Pancharatnam}
S. Pancharatnam, Proc. Ind. Acad. Sci. A {\bf 44} 247 (1956)
\bibitem {RamaRama}
G.N. Ramachandran and S. Ramaseshan in S. Flugge (editor),
 {\it Handbuch de Physik} Vol. XXV.1  (Springer, Berlin, 1961) p.1
\bibitem {SchmitzerKD1}
H. Schmitzer, S. Klein and W. Dultz, Physica B {\bf175} 148 (1991)
\bibitem {SchmitzerKD2}
H. Schmitzer, S. Klein and W. Dultz, Phys. Rev. Lett. {\bf 71} 1530 (1993)
\bibitem {Berry85}
M.V. Berry, J. Phys. A: Math. Gen. {\bf 18 } 15 (1985)
\bibitem {GiavariniGRT}
G. Giavarini, E. Gozzi, D. Rohrlich and W.D. Thacker, J. Phys. A: Math. Gen.
{\bf 22} 3513 (1989)                    
\bibitem {BruknerZ}
C. Brukner and A. Zeilinger, Phys. Rev. Lett. {\bf 79} 2599 (1997)
\bibitem {LawandeLJ}
Q.V. Lawande, S.V. Lawande and A. Joshi, Phys. Lett. A {\bf 251} 164 (1999)
\bibitem {ShapiroShepard}
J.H. Shapiro and S.R. Shepard, Phys. Rev. {\bf A 43} 3795 (1991)
\bibitem {BalisteriKKH}
M.L.M. Balisteri, J.P. Korterik, L. Kuipers and N.F. van Hulst, Phys.
 Rev. Lett. {\bf 85} 294 (2000)
\bibitem {WaghRFI}
A. G. Wagh, V.C. Rakhecha, P. Fischer and A. Ioffe, Phys. Rev. Lett. {\bf
81} 1992 (1998)
\bibitem {WaghBRBS}
A. G. Wagh, G. Badurek, V.C. Rakhecha, R.J. Buchelt and A. Schricker, Phys.
Lett. A {\bf 268} 209 (2000)
\bibitem {NugentPG}
K.A. Nugent, A. Paganin and T.E. Gureyev, Physics Today {\bf 54} No. 8, 27 (2001)
\bibitem  {GrangierLP}
 P. Grangier, J.A. Levenson and J-P. Poizat, Nature {\bf 396} 537
(1998)
\bibitem {KohlerM}
D. Kohler and L. Mandel in L. Mandel and E. Wolf (editors), {\it Coherence
and Quantum Optics} (Plenum, New York, 1973) p.387
\bibitem {VijayWB}
A. Vijay, R.E. Wyatt and G.D. Billing, J. Chem. Phys. {\bf 111} 10 794 (1999)
\bibitem {ReckZBB}
M. Reck, A. Zeilinger, H.J. Bernstein and P. Bertani, Phys. Rev. Lett. {\bf 73}
58 (1994)
\bibitem {NohFM}
J.W. Noh, A. Fougeres  and L. Mandel, Physica Scripta T {\bf 48} 29 (1993)
\bibitem {Stenholm}
S. Stenholm, Physica Scripta  T {\bf 48} 77 (1993)
\bibitem {Vourdas}
A Vourdas, Physica Scripta T {\bf 48} 84 (1993)
\bibitem {Loudon}
R. Loudon, {\it The Quantum Theory of Light} (Clarendon Press, Oxford, 1983)
p.140
\bibitem {Dirac1927}
P.A. M. Dirac, Proc. Roy. Soc. (London) A {\bf 114} 243 (1927)
\bibitem {London}
F. London, Z. Physik, {\bf 37} 915 (1926); {\bf 40} 193 (1927)
\bibitem {Louisell}
W.H. Louisell, {\it Quantum Statistical Properties of Radiation}
 (Wiley, London, 1973)
\bibitem {SusskindG}
L. Susskind and J. Glogower, Physics {\bf 1} 49 (1964)
\bibitem {PeggB}
D.T. Pegg and S.M. Barnett, Europhys. Lett. {\bf 8} 463 (1988)
\bibitem {GronbechCR}
N. Gronbech-Jensen, P.L. Christiansen and P.S. Ramanujam, J. Opt. Soc.
Am. {\bf 6} 2423 (1989)
\bibitem {VorontsovR1}
Yu. I . Vorontsov and Yu. A. Rembovsky,  Phys. Lett. A {\bf 254} 7 (1999)
\bibitem {VaccaroPB}
J.A. Vaccaro, D.T. Pegg and S.M. Barnett, Phys. Lett. A {\bf 262} 483 (2000)
\bibitem {VorontsovR2}
Yu. I . Vorontsov and Yu. A. Rembovsky,  Phys. Lett. A {\bf 262} 486 (2000)
\bibitem {AMuller}
A. Muller, Phys. Rev. A {\bf 57} 731 (1998)
\bibitem {AharonovS1967}
Y. Aharonov and L. Susskind, Phys. Rev. {\bf 158} 1237 (1967)
\bibitem {AharonovV2000}
Y. Aharonov and L. Vaidman, Phys. Rev. A {\bf 61} 052108 (2000)
\bibitem {Floissart}
M. Floissart in E.P. Wigner (editor) {\it Dispersion Relations and their
 Connection with Causality} (Academic Press, New York, 1964) p. 1.
\bibitem {PeiponenVA}
K.E. Peiponen, E.M. Vertiainen and T. Asakura {\it Dispersion, Complex
Analysis and Optical Spectroscopy. (Classical theory)}
 (Springer-Verlag, Berlin, 1999)
\bibitem {Toll}
J.S. Toll, Phys Rev. {\bf 104} 1760 (1956)
\bibitem {BurgeFGR}
R.E. Burge, M.A. Fiddy, A.H. Greenaway and G. Ross, Proc. Roy. Soc.
(London) A {\bf 350} 191 (1976)
\bibitem {Peat}
F.D. Peat, {\it Infinite Potential, The Life and Times of David Bohm} (Addison-
Wesley, Reading Mass., 1997) p. 192
\bibitem {EhrenbergS}
W. Ehrenburg and R.E. Siday, Proc. Roy. Soc. (London) B {\bf 62} 8 (1949)
\bibitem {PeshkinTT}
M. Peshkin, I. Talmi and L.J. Tassie, Ann. Phys. (NY) {\bf 12} 426 (1960)
\bibitem {Simon}
B. Simon, Phys. Rev. Lett. {\bf 51} 2167 (1983)
\bibitem {PeshkinT}
M. Peshkin and A. Tonomura, {\it The Aharonov Bohm Effect} (Springer-Verlag,
 Berlin, 1989)
\bibitem {WY}
T.T. Wu and C.N. Yang, Phys. Rev. D {\bf 12} 3845 (1975)
\bibitem {Englman72}
R. Englman, {\it The Jahn-Teller Effect in Molecules and Crystals} (Wiley,
Chichester, 1972)
\bibitem {Longuet}
H.C. Longuet-Higgins, Proc. Roy. Soc. (London) A {\bf 344} 147 (1975)
\bibitem {Stone}
A.J. Stone, Proc. Roy. Soc. (London) A {\bf 351} 141 (1976)
\bibitem {MeadT79}
C.A. Mead and D.G. Truhlar, J. Chem. Phys. {\bf 70} 2284 (1979)
\bibitem {Mead80}
C.A. Mead, Chem. Phys. {\bf 49} 23, 33 (1980)
\bibitem {Berry90}
M.V. Berry, Physics Today {\bf 43} 34 (1990)
\bibitem {Ham87}
F.S. Ham, Phys. Rev. Lett. {\bf 58} 725 (1987)
\bibitem {ChanceyO}
C.C. Chancey and M.C.M. O'Brien, {\it The Jahn-Teller Effect in $C_{60}$ and other
Icosahedral Complexes} (University Press, Princeton, 1997)
\bibitem {Anandan90}
J. Anandan, Phys. Lett. A {\bf 147} 3 (1990)
\bibitem {Kohler98}
B. Kohler, Phys. Lett. A {\bf 237} 195 (1998)
\bibitem {YangY}
L. Yang and F. Yan, Phys. Lett. A {\bf 265} 326 (2000)
\bibitem {ManiniP}
N. Manini and F. Pistolesi,  Phys. Rev. Lett. {\bf 85} 3067 (2000)
\bibitem {ManolopoulosC}
D. E. Manolopoulos and M.S. Child, Phys. Rev. Lett. {\bf 82} 2223 (1999)
\bibitem {HasegawaLBBR}
Y. Hasegawa, R. Loidl, M. Baron, G. Badurek and M. Rauch, Phys. Rev. Lett.
 {\bf 87} 070401 (2001)
\bibitem {GarrisonW}
J.C. Garrison and E.M. Wright, Phys. Lett. A {\bf 128} 17781 (1988)
\bibitem {MiniaturaSBB}
Ch. Miniatura, C. Sire, J. Baudon and J. Belissard, Europhysics Lett.  {\bf
13} 199 (1990)
\bibitem {GeC1}
Y.C. Ge and M.S. Child, Phys. Rev. Lett. {\bf 78} 2507 (1997)
\bibitem {GeC2}
Y.C. Ge and M.S. Child, Phys. Rev. A {\bf 58 } 872 (1998)
\bibitem {BaiCG}
Z.-M. Bai, G.-Z. Chen and M.-L. Ge,  Phys. Lett. A {\bf 262} 137 (1999)
\bibitem {WhitneyG}
R. S. Whitney and Y. Gefen, (Preprint, 2001)
\bibitem {AharonovR}
Y. Aharonov and B. Reznik, Phys. Rev. Lett. {\bf 84}  490 (2000)
\bibitem {Cervero}
J. M. Cervero, Int. J. Theor. Phys. {\bf 38} 2095 (1999)
\bibitem {FelkerZ}
P.M. Felker and A.H. Zewail, Adv. Chem. Phys. {\bf 70} 265 (1988)
\bibitem {GaspardB}
P. Gaspard and I. Burghardt (editors) {\it Chemical Reactions and their Control on the
Femtosecond Time Scale} in Adv. Chem. Phys. {\bf 101} (Wiley, New York, 1997)
\bibitem {MeshulachS}
D. Meshulach and Y. Silberberg, Nature {\bf 396} 239 (1998);  Phys. Rev. A {\bf 60}
 1287 (1999)
\bibitem {RabitzVMK}
 H. Rabitz, R. de Vivie-Riedle, M. Motzkus and K.-L.Kompa, Science {\bf 259} 1581
(1993)
\bibitem {ZeidlerFKM}
D. Zeidler, S. Frey, K.-L.Kompa and M. Motzkus, Phys. Rev. A {\bf 64} 023420 (2001)
\bibitem {SchererRDF}
F. Scherer, A.J. Ruggiero, M. Du, G.R. Fleming, J. Chem. Phys. {\bf 93} 856
 (1990)
\bibitem {Tannor}
D. J. Tannor, {\it Design of Femtosecond Optical Pulse Sequences to Control
Photochemical Products}  in  A.D. Bandrark (editor), {\it Molecules in Laser
Fields} (Dekker, New York, 1994)
\bibitem {DixonHYHLY}
R.N. Dixon, D.W. Hwang, X.F. Yang, S. Harich, J.J. Lin and X. Yang, Science
{\bf 285} 1249 (1999)
\bibitem {Clary}
D. C. Clary, Science {\bf 285} 1218 (199)
\bibitem {AvronB}
J.E. Avron and J. Berger, Chem. Phys. Lett. {\bf 294} 13 (1998)
\bibitem {AdhikariB97}
S. Adhikari and G. Billing, J. Chem. Phys. {\bf 107} 6213 (1997)
\bibitem {AdhikariB98}
S. Adhikari and G. Billing, Chem. Phys. Lett. {\bf 284} 31 (1998)
\bibitem {AdhikariB99}
S. Adhikari and G. Billing, J. Chem. Phys. {\bf 111} 40 (1999)
\bibitem {ZilbergH}
S. Zilberg and Y. Haas, J. Phys. Chem. {\bf A 103} 2364 (1999).
\bibitem {BaerLAAB}
M. Baer, S.-H. Lin, A. Alijah, S. Adhikari and G.D. Billing, Phys. Rev. A
{\bf 62} 032506 (2000)
\bibitem {AdhikariBALB}
 S. Adhikari, G.D. Billing, A. Alijah, S.-H. Lin and M. Baer, Phys. Rev. A
{\bf 62} 032507 (2000)
\bibitem {BillingK}
G.D. Billing and A. Kuppermann, Chem. Phys. Lett. {\bf 294} 26 (1998)
\bibitem {Shapiro3}
M. Shapiro, J. Chem. Phys. {\bf 103} 1748 (1995); J. Phys. Chem. {\bf 100} 7859
(1996)
\bibitem {BromageS}
J. Bromage and C.R. Stroud, Jr., Phys. Rev. Lett. {\bf 83} 4963 (1999)
\bibitem {EberlySN}
J.H. Eberly, J.J. Sanchez-Mondragon and N.B. Narozhny, Phys. Rev. Lett.
{\bf 44} 1323 (1980)
\bibitem {AverbukhP2}
I. Sh. Averbukh and N.F. Perel'man, Phys. Lett. A {\bf 139} 449 (1989)
\bibitem {AverbukhVVS}
I. Sh. Averbukh, M.J.J. Vrakking, D.M. Villeneuve and A. Stolow, Phys. Rev.
Lett. {\bf 77} 3518  (1996)
\bibitem {Jarzynski}
J. Jarzynski,  Phys. Rev. Lett. {\bf 74} 1264 (1995)
\bibitem {Heller}
E.J. Heller, J. Chem. Phys. {\bf 62} 1544 (1975)
\bibitem {HermanK}
M.F. Herman and E. Kluk, Chem. Phys. {\bf 91} 27 (1984)
\bibitem {HuberH}
D. Huber and E.J. Heller , J. Chem. Phys. {\bf 87} 5302 (1987)
\bibitem {Kay}
K.G. Kay, Phys. Rev. Lett. {\bf 83} 5190 (1999)
\bibitem {Percival}
I. C. Percival, {\it Semiclassical Theory of Bound States} in Adv. Chem. Phys.
{\bf 36} 1 (1977)
\bibitem {ZhangFG}
W.M. Zhang, D.H. Feng and R. Gilmore, Rev. Mod. Phys. {\bf 62} 867-927
(1990)
\bibitem {Childbook}
M.S. Child, {\it Semiclassical Mechanics with Molecular Applications} (Clarendon
Press, Oxford, 1991)
\bibitem {MaslovF}
V.P. Maslov and M.V. Fedoriuk, {\it Semi-classical Approximations in Quantum
Mechanics} (Reidel, Boston, 1981)
\bibitem {Davydov}
A.S. Davydov, {\it Quantum Mechanics} (Pergamon Press, Oxford 1965)
Section 108
\bibitem {Levine}
R.D. Levine {\it Quantum Mechanics of Molecular Rate Processes} (Clarendon
Press. Oxford, 1969) section 2.8.1
\bibitem {TYWu}
T.Y. Wu, Am. J. Phys., {\bf 26} 568 (1958)
\bibitem {Popper}
K.R. Popper, Nature {bf\ 177} 538 (1956) ; {\it ibid} {\bf 178} 382 (1956);
 {\it ibid} {\bf 179} 1293 (1957)
\bibitem {HillG}
E.L. Hill and A. Grunbaum, Nature {\bf 179} 1292 (1957)
\bibitem {Costa}
O. Costa de Beauregard, {\it La Second Principe de la Science du Temps} (De
Seuil, Paris, 1963)
\bibitem {Lebowitz}
J.L. Lebowitz, Physics Today {\bf Sep. 1993} 32 (1993) [Various responses in
  Physics Today {\bf Nov. 1994} 11  (1994)]
\bibitem {HalliwellPZ}
J.J. Halliwell, J. Perez-Mercader and W.H. Zurek (editors), {\it Physical
 Origin of Time Asymmetry} (University Press, Cambridge, 1994)
\bibitem {Schulman}
 L.S. Schulman, {\it Time Arrows and Quantum Measurement} (University
Press, Cambridge, 1997)
\bibitem {BanksSP}
T. Banks, L. Susskind and M.E. Peskin, Nucl. Phys. B {\bf 244} 125 (1984)
\bibitem {BerettaGPH}
G.P. Beretta, E.P. Gyftopoulos, J.L. Park and G.N. Hatsopoulos. Nuovo. Cim.
 {\bf 82B} 169 (1984)
\bibitem {Gheorghiu}
S. Gheorghiu- Svirshevski, Phys. Rev. A {\bf 63} 022105 (2001);
 {\bf63} 054102 (2001)
\bibitem {ABohmAK}
A. Bohm, I. Antoniou and P. Kielanowski, Phys. Letters,  A {\bf 189} 442
(1994)
\bibitem {ABohm}
A. Bohm, Phys. Rev. {\bf A60} 861 (1999)
\bibitem {DBohm}
D. Bohm, {\it Quantum Theory} (Prentice Hall, New York, 1952)
\bibitem {Wigner}
E.P. Wigner, Phys. Rev. {\bf 98} 145 (1955)
\bibitem {PollakM}
E. Pollak and W.H. Miller, Phys. Rev. Lett. {\bf 53} 115 (1984)
\bibitem {Pollak}
E. Pollak. J. Chem. Phys. {\bf 83 } 1111 (1985)
\bibitem {Perel'man}
M.E. Perel'man, {\it Kinetical Quantum Theory of Optical Dispersion} (Mezniereba,
Tblisi, 1989) (In Russian)
\bibitem {Perel'man2}
 M.E. Perel'man, Soviet Phys. JETP {\bf 23} 407 (1966); [J. Eksp. Teoret.
Fyz.  {\bf 50} 613 (1966)]; Soviet Physics Doklady {\bf 14} 772 (1970) [DAN
 SSSR {\bf 187} 781-3 (1969)]
\bibitem {LandauerM}
R. Landauer and Th. Martin, Rev. Mod. Phys. {\bf 66} 217 (1994)
\bibitem {ChauvatEBL}
 D. Chauvat, O. Emile, F. Bretenaker and A. Le Floch, Phys. Rev. Lett. {\bf
84} 71 (1999)
\bibitem {Debu}
P. Debu , Europhysics News {\bf 31} 5 (May/June 2000)
\bibitem {AharonovAu}
Y. Aharonov and C.K. Au, Phys. Lett. A {\bf 86 } 269 (1981)
\bibitem {FeuchtwangKGC}
T.E. Feuchtwang, E. Kazes, H. Grotch and P.H. Cutler, Physics Letters A {\bf
93} 4 (1982)
\bibitem {Kaempffer}
 F.A. Kaempffer, {\it  Concepts in Quantum Mechanics} (Academic, New York
1965) p. 169
\bibitem {OzimbaM}
P.A. Ozimba and A.S. Msezane, Chem. Phys. {\bf 246} 87 (1999)
\bibitem {Heuss}
 W.D. Heuss, Eur. Phys. J. D {\bf 7} 1 (1999)
\bibitem {Schiff}
 L.I. Schiff, {\it Quantum Mechanics} (McGraw-Hill, New York, 1955)
 Section 12
\bibitem {SelloniCCP}
A. Selloni, P. Carnevali, R. Car and M. Parinello, Phys. Rev. Lett.
{\bf 59} 823 (1987)
\bibitem {AncilottiT}
 F. Ancilotti and F. Toigo, Phys. Rev. A {\bf 45} 4015 (1992)
\bibitem {Resta}
 R. Resta, Phys. Rev. Lett. {\bf 80} 1800 (1998)
\bibitem {SternAI}
 A. Stern, Y. Aharonov and Y. Imry, Phys Rev. A {\bf 41} 3436 (1990)
\bibitem {Neumann}
J. von Neumann, {\it The Mathematical Foundation of Quantum Mechanics} (Dover,
New York, 1959)
\bibitem {Perelman69}
M.E. Perel'man, Soviet Phys. Doklady {\bf 14} 772 (1969)[DAN SSR {\bf 187} 781
(1969)]
\bibitem {Perelman66}
M.E. Perel'man, Soviet Phys. JETP {\bf 23} 407 (1966) [Zh. Eksp. Teor. Fiz. {\bf
50} 613 (1966)]
\bibitem {Titchmarsh1}
 E.C. Titchmarsh, {\it Introduction to the Theory of Fourier Integrals}
 (Clarendon Press, Oxford, 1948) Chapter V
\bibitem {Caratheodory}
 C. Caratheodory, {\it Theory of Functions} (Chelsea, New York, 1958) Vol. I,
 Chapter 3.
\bibitem {Resta94}
 R. Resta, Rev. Mod. Phys. {\bf 66} 899 (1994)
\bibitem {PaleyW}
R.A.E.C. Paley and N. Wiener, {\it Fourier Transforms in the Complex Domain}
(American Physical Soc., New York, 1934)
\bibitem {Davison}
 B. Davison, {\it Neutron
Transport Theory}  (Clarendon Press, Oxford, 1957) Chapter VI , Section 1
\bibitem {BornF}
 M. Born and V. Fock,  Zeitschrifte fur Physik {\bf 51} 165 (1928)
\bibitem {Kato}
 T. Kato, Progr. Theor. Phys. {\bf 5} 435 (1950)
\bibitem {Messiah}
 A.  Messiah, {\it Quantum Mechanics}  Vol. II
(North Holland, Amsterdam, 1961) Chapter XVII
\bibitem {LiuHL}
 J. Liu, B. Hu and B. Li, Phys. Rev. Lett. {\bf 81} 1749 (1998)
\bibitem {Zigmund}
 A. Zigmund, {\it Trigonometrical Series} (Dover, New York, 1955) Chapter III.

\bibitem {NeumannW}
 J. von Neumann and E.P. Wigner, Physik. Zeits. {\bf 30} 467 (1929)
\bibitem {ZwanzigerG}
 J.W. Zwanziger and E.R. Grant, J. Chem. Phys. {\bf 87} 2954 (1987)
\bibitem {Kosloff}
R. Kosloff, Ann. Rev. Phys. Chem. {\bf 45} 145 (1994)
\bibitem {Tomonaga}
 S.-I. Tomonaga, {\it Quantum Mechanics}, Vol II (North Holland, Amsterdam
 1966), Sections 41, 61
\bibitem {Heller81}
 E.J. Heller, J. Chem. Phys. {\bf 75} 2923 (1981)
\bibitem {BerryK}
 M.V. Berry and J. Klein, J. Phys. A: Math. and General {\bf 17} 1805 (1984)
\bibitem {DodonovMN}
 V.V. Dodonov, V. I. Manko and D.E. Nikonov, Phys. Lett. A {\bf 162} 359 (1992)
\bibitem {Grosche}
C. Grosche, Phys. Lett. A {\bf 182} 28 (1993)
\bibitem {AharonovKPR}
 Y. Aharonov, T. Kaufherr, S. Popescu and B. Reznik, Phys. Rev.
    Lett. {\bf 80} 2023 (1998)
\bibitem {ZwanzigerRC}
 J.W. Zwanziger, S.P. Rucker and G.C. Chingas, Phys. Rev. A {\bf 43} 3232 (1991)
\bibitem {Titchmarsh2}
 E.C. Titchmarsh, {\it The Theory of Functions} (Clarendon Press, Oxford, 1932)
 sections 3.42, 3.45, 7.8, 8.1.
\bibitem {Shvartsman}
 N. Shvartsman and I. Freund, Phys. Rev. Lett. {\bf 72} 1008 (1994)
\bibitem {JahnT}
A.H. Jahn, and E. Teller,  Proc. Roy. Soc. (London) A {\bf 161} 220 (1937)
\bibitem {O'Brien}
M.C.M. O'Brien, Proc. Roy. Soc. (London) A {\bf 281} 323 (1964)
 \bibitem {Ham71}
F.S.  Ham, {\it Jahn-Teller Effects in Electron Paramagnetic Spectra} ( Plenum Press  New York, 1971)
\bibitem {Yarkony99a}
 D. Yarkony,  J. Chem. Phys. {\bf 111} 4906  (1999)
\bibitem  {VarandasYX}
D.J.A. Varandas, H.G. Yu and Z.R.  Xu,  Mol. Phys. {\bf 96} 1193 (1999)
\bibitem  {VarandasX}
D.J.A. Varandas  and Z.R.  Xu, Int. J. Quant. Chem.{\bf 75} 89 (1999)
\bibitem {Yarkony99b}
 D. Yarkony,  J. Chem. Phys. {\bf 110} 701  (1999)
\bibitem {HerzbergL}
G. Herzberg and H.C. Longuet-Higgins,  Discuss. Faraday Soc. {\bf 35} 77 (1963)
\bibitem{KoizumiBBP}
H. Koizumi, I.B. Bersuker, J. Boggs and V.Z.
Polinger, J. Chem. Phys. {\bf 112}  8470  (2000)
\bibitem {MoateODBLP}
M.C.P. Moate, M.C.O. O'Brien, J.I. Dunn, C.A. Bates, Y.M. Liu and V.Z. Polinger,  Phys. Rev. Lett. {\bf 77} 4362 (1996)
\bibitem {SaxeLY}
P. Saxe,  B.H. Lengsfield and D.H. Yarkony, Chem. Phys. Lett. {\bf 113} 19 (1985)
\bibitem {ThummelPPB}
H. Thummel, M. Peric, S.D. Peyerimhoff and R.J. Buenker,  Z. Phys. D
{\bf 13} 307 (1989)
\bibitem {RadazosRBO}
 I.N. Radazos, M.A. Robb, M. A. Bernardi and M. Olivucci, Chem. Phys. Lett.
 {\bf 192} (1992) 217
\bibitem {WaschewskyKMKB}
G.C.G. Waschewsky, P. W. Kash, T.L. Myers, D.C. Kitchen and L.J. Butler,
 J. Chem. Soc. Faraday Trans. {\bf 90} 1581 (1994)
\bibitem {MebelBL2000}
A.M.  Mebel, M Baer and S.H. Lin,  J. Chem. Phys. {\bf 112} 10 (2000) 10 703
\bibitem {MebelBL2001}
A.M.  Mebel, M Baer and S.H. Lin, Mol. Phys. (in press)
\bibitem {Yarkoni98}
D. Yarkony, Acc. Chem. Res. {\bf 31} 511 (1998)
\bibitem {Herzberg}
G. Herzberg, {\it Molecular Spectra and Molecular Structure}
  (Van Nostrand: Princeton, 1966) Vol.3
\bibitem {ChabanGY}
G. Chaban, M.S. Gordon and D. Yarkony, J. Phys. Chem. {\bf 43 } 7953 (1997)
\bibitem {HobeyM}
W.D. Hobey and A.D. McLachlan, J. Chem. Phys. {\bf 33} 1695 (1961)
\bibitem {McLachlan}
A.D. McLachlan, Mol. Phys. {\bf 4} 417 (1961)
\bibitem {LepetitK}
B. Lepetit and A. Kuppermann, Chem. Phys. Lett. {\bf 166} 581 (1990)
\bibitem {WuK}
Y.-S. M. Wu and A. Kuppermann, Chem. Phys. Lett. {\bf 201} 178 (1993);
 {\it ibid} {\bf 235}  105 (1995)
\bibitem {Baer76 }
M. Baer, Chem. Phys. {\bf 15} 49 (1976)
\bibitem { Baer80 }
M. Baer, Mol. Phys. {\bf 40} 1011 (1980)
\bibitem { Baer85 }
M. Baer, in M. Baer (editor) {\it Theory of Chemical Reaction Dynamics}
 Vol. II (CRC Press, Boca Raton, 1985) Chapter 4
\bibitem {Baer 97}
M. Baer, J. Chem. Phys. {\bf 107}  2694, 10662 (1997)
\bibitem {ChapuisatND}
X. Chapuisat, A. Nauts and D. Hehaureg-Dao, Chem. Phys. Lett. {\bf 95} 139 (1983)
\bibitem {Hehaureg CLGR}
D. Hehaureg, X. Chapuisat, J.C. Lorquet, G. Galloy and G. Raseev, J. Chem. Phys. {\bf 78} 1246 (1983)
\bibitem {CederbaumKD}
L.S. Cederbaum, H. Koppel and W. Domcke, Int. J. Quant. Chem. Symp. {\bf 15} 251 (1981)
\bibitem {PacherCK2 }
T. Pacher, L.S. Cederbaum and H. Koppel, J.Chem. Phys. {\bf 84} 293 (1993)
\bibitem {BaerA}
M. Baer and A. Alijah, Chem. Phys. Lett. {\bf 319} 489 (2000)
\bibitem {Baer2000JPC}
M. Baer, J. Phys. Chem. A{\bf 104} 3181 (2000)
\bibitem {BaerCKB}
R. Baer, D. Charutz, R. Kosloff and M. Baer, J. Chem. Phys. {\bf 105} 9141 (1996)
\bibitem {CharutzBB}
D. Charutz, R. Baer and M. Baer, Chem. Phys. Lett. {\bf 265} 629 (1997)
\bibitem {ThompsonM}
T.C. Thompson and C.A. Mead, J. Chem. Phys. {\bf 82} 2408 (1985)
\bibitem {Griffith62}
J.S. Griffith, {\it The Irreducible Tensor Method for Molecular Symmetry Groups} (Prentice Hall, Englewood Cliffs, NJ, 1962) p. 20
\bibitem {Griffith64}
J.S. Griffith, {\it The Theory of Transition Metal Ions} (University Press, Cambridge, 1964)
\bibitem {BevilacquaMP}
G. Bevilacqua, L. Martinelli and G.P. Parravicini, (Preprint 2001)
\bibitem {BitterD}
T. Bitter and D.  Dubbens, Phys. Rev. Lett. {\bf 59} 251 (1988)
\bibitem {SuterMP}
D. Suter, K.T. Mueller and  A. Pines, Phys. Rev. Lett. {\bf  60} 1218 (1988)
\bibitem {SimonKS}
R. Simon, H.J. Kimble and E.C.G.  Sudarshan,  Phys. Rev. Lett. {\bf 61} 19 (1988)
\bibitem {Morpurgo HKVB}
A.F. Morpurgo, J.P. Heida, T.M. Klapwijk, B.J. Van Wees and G. Borghs, Phys. Rev. Lett. {\bf 80} 1050 (1998)
\bibitem {VonBuschDEKWDSM}
H.  Von Busch, V.  Dev, H.-E.  Eckel,  S. Kasahara, J. Wang, W. Demtroder, P.  Sebald and W.  Meyer, Phys. Rev. Lett. {\ 81} 4584 (1998)
\bibitem { LossSG}
D. Loss, H. Schoeller and P.M. Goldbart,  Phys. Rev. B {\bf 59} 13 328 (1999)
\bibitem {FuentesBV}
I.  Fuentes-Guri, S. Bose and V. Vedral,  Phys. Rev. Lett. {\bf 85} 5018 (2000)
\bibitem {Sjoquist}
E. Sj\"{o}quist, Phys. Lett. A {\bf 286} 4 (2001)
\bibitem {PothierLUED}
H. Pothier, P. Lafarge, C. Urbina, D.  Esteve and M.H. Devoret, Europhys.
Lett. {\bf 17} 1183 (1992)
\bibitem {LiuDBH}
C. Liu, Z. Dutton, C. Behroozi and L.V. Hau,  Nature {\bf 409} 490 (2001)
\bibitem {PhillipsFMWL}
D.F. Phillips, A.  Fleischhauer, A. Mair,  R.L. Walsworth and M.D. Lukin, Phys. Rev. Lett. {\bf 86} 783 (2001)
\bibitem {DivicenzoT}
D. DiVicenzo and B. Terhal,  Physics World, {\bf March 1998}  53 (1998)
\bibitem {Mead92}
C.A. Mead,  Rev. Mod. Phys. {\bf 64} 51 (1992)
\bibitem {ThielK}
A. Thiel and H. Koppel, J. Chem. Phys. {\bf 110} 9371 (1999)
\bibitem{Bohm} D. Bohm, {\it Quantum Theory} (Prentice Hall, New York, 1966)
 section 12.6
\bibitem{Schrodinger} E. Schr\"{o}dinger, Ann. d. Phys. {\bf 81} 109 (1926).
 English translation appears in E. Schr\"{o}dinger, {\it Collected Papers in Wave
 Mechanics} (Blackie and Sons, London, 1928) p. 102
\bibitem{Gordon1}
 W. Gordon, Z. Physik {\bf 41} 117 (1926).
\bibitem{Greiner2}
W. Greiner, {\it Relativistic Quantum Mechanics: Wave Equations} (Springer-Verlag, Berlin, 1997)
\bibitem{Bogoliubov} N.N. Bogoliubov and D.V. Shirkov, {\it Introduction
to the Theory of Quantized Fields} (Interscience, New York, 1959).
\bibitem{Gordon2} W. Gordon, Z. Physik {\bf 50} 630 (1928).
\bibitem{Holland}
P.R. Holland {\it The Quantum Theory of Motion}
(Cambridge University Press, Cambridge, 1993)
\bibitem{Seliger}
 R. L. Seliger and G. B. Whitham, Proc. Roy. Soc. (London) A {\bf A 305} 1 (1968)
\bibitem{Madelung}
E. Madelung, Z. Phys., {\bf 40} 322 (1926)
\bibitem{Talmi}
I. Talmi, {\it Simple Models of Complex Nuclei} (Harcourt Academic, Chur, 1993)
\end{thebibliography}
\end{document}